\documentclass[review, preprint,12pt]{elsarticle} %

\usepackage{amssymb}
\usepackage{amsmath,amsfonts,amssymb,bm,accents}
\usepackage{hyperref}
\usepackage{mathrsfs}
\usepackage{graphicx}
\usepackage{epstopdf}
\usepackage{float}
\usepackage{caption}
\usepackage{subcaption}
\usepackage{bbm}
\usepackage{mathrsfs}
\usepackage{cleveref}
\usepackage{soul}
\usepackage{accents}
\usepackage{graphicx}
\usepackage{xcolor}
\usepackage{courier} 
\usepackage{listings} 
\usepackage{tabu} 
\usepackage{booktabs}
\usepackage{longtable}
\usepackage{changepage} 
\usepackage[margin=2cm]{geometry}
\biboptions{sort&compress} 
\usepackage[section]{placeins}
\usepackage[font={small}]{caption}

\usepackage{siunitx}
\usepackage[export]{adjustbox}

\newcommand{\beq}{\begin{equation}}
\newcommand{\eeq}{\end{equation}}
\newcommand{\bea}{\begin{eqnarray}}
\newcommand{\eea}{\end{eqnarray}}

\newcommand{\dd}{\,{\rm d}}

\usepackage{lineno}

\journal{Construction and Building Materials}

\makeatletter
\def\@author#1{\g@addto@macro\elsauthors{\normalsize%
    \def\baselinestretch{1}%
    \upshape\authorsep#1\unskip\textsuperscript{%
      \ifx\@fnmark\@empty\else\unskip\sep\@fnmark\let\sep=,\fi
      \ifx\@corref\@empty\else\unskip\sep\@corref\let\sep=,\fi
      }%
    \def\authorsep{\unskip,\space}%
    \global\let\@fnmark\@empty
    \global\let\@corref\@empty  
    \global\let\sep\@empty}%
    \@eadauthor={#1}
}
\makeatother

\begin{document}

\begin{frontmatter}



\title{A phase-field chemo-mechanical model for corrosion-induced cracking in reinforced concrete}


\author[IC]{Ev\v{z}en Korec}

\author[CTU]{Milan Jir\'{a}sek}

\author[IC]{Hong S. Wong}

\author[IC]{Emilio Mart\'{\i}nez-Pa\~neda\corref{cor1}}
\ead{e.martinez-paneda@imperial.ac.uk}

\address[IC]{Department of Civil and Environmental Engineering, Imperial College London, London SW7 2AZ, UK}

\address[CTU]{Department of Mechanics, Faculty of Civil Engineering, Czech Technical University in Prague, Th\'{a}kurova 7, Prague - 6, 166 29, Czech Republic}

\cortext[cor1]{Corresponding author.}

\begin{abstract}
We present a new mechanistic framework for corrosion-induced cracking in reinforced concrete that resolves the underlying chemo-mechanical processes. The framework combines, for the first time, (i) a model for reactive transport and precipitation of dissolved $\mathrm{Fe}^{2+}$ and $\mathrm{Fe}^{3+}$ ions in the concrete pore space, (ii) a precipitation eigenstrain model for the pressure caused by the accumulation of precipitates (rusts) under pore confinement conditions, (iii) a phase-field model calibrated for the quasi-brittle fracture behaviour of concrete, and (iv) a damage-dependent diffusivity tensor. Finite element model predictions show good agreement with experimental data from impressed current tests under natural-like corrosion current densities.  \\
\end{abstract}

\begin{keyword}

Reinforced concrete \sep Corrosion-induced cracking \sep Phase-field fracture \sep \texttt{PF-CZM} (Phase-field cohesive zone model) \sep Reactive transport in porous media \sep Precipitation eigenstrain 



\end{keyword}

\end{frontmatter}


\section{Introduction}
\label{Introduction}

This study is concerned with the modelling of crack nucleation and growth in reinforced concrete resulting from the electrochemical corrosion of embedded steel reinforcement. Corrosion of steel in concrete is responsible for the premature deterioration of 70-90\% of concrete structures \cite{Gehlen2011-za, British_Cement_Association_BCA1997-jj} and thus induces considerable repair costs \citep{Koch}. Especially in industrialized countries, these are becoming pressing issues as most of the reinforced concrete bridges have been built in the 1960s and are coming to the end of their predicted service life. Only in the Netherlands, the number of bridges in the need of repair is estimated to rise by a factor of 2 - 4 in the next 20 years and by a factor of 3 - 6 in the next 40 years \citep{Polder2012}. Since direct replacement of all degraded reinforced-concrete infrastructure in a short period is unfeasible, advances in long-term corrosion durability predictions are needed to accurately assess the remaining service life. Many models have been proposed to estimate corrosion-induced cracking, the majority of which are of empirical nature (see Ref. \cite{Jamali2013a} for a review). Phenomenological models generally require extensive calibration and have a regime of applicability that is typically limited to scenarios resembling the calibration schemes; predictive modelling requires explicitly resolving the physical processes at play. \\ 

Recent years have seen promising progress in the development of mechanistic models for corrosion-induced cracking of reinforced concrete. The aim is to model the physical phenomena underlying the formation of a rust layer and how this rust layer triggers cracking. However, these phenomena are complex and hence have only been accounted for to a certain extent. For example, mechanical models have been presented that provide a mechanistic description of the fracture process but assume a given (often non-uniform) rust layer \cite{Zhao2011a,Xi2018b,Zhao2020b,Hu2022,Wu2020}. Instead of simulating changes in the corrosion current density or the transport of involved chemical species, corrosion-induced fracture predictions are obtained based on a prescribed non-uniform rust distribution described by Gaussian \citep{Zhao2011a}, Von Mises \citep{Zhao2020b,Xi2018b} or semi-elliptical \citep{Hu2022} functions. Within prescribed rust layer approach, even mesoscale fracture of concrete both in 2D \cite{Wu2020} and 3D \cite{Jin2020} has been simulated. Other models go one step further and estimate the rust layer thickness from a given corrosion current density. \citet{Molina1993} and Grassl and co-workers \cite{Fahy2017,Aldellaa2022,Grassl2011} studied corrosion-induced cracking caused  by uniform corrosion. \citet{Fahy2017} considered the pressure-driven transport of corrosion products, simplified to be an incompressible fluid, into cracked porous concrete. \citet{Tran2011} investigated the case of non-uniform corrosion and allowed for the accommodation of corrosion products in cracks, which results in a reduction of the corrosion-induced pressure. The impact of the mechanical properties of rust \citet{Tang2022}, steel-concrete interface \cite{Zhao2016b} and stirrups \cite{Zhao2021} have also been considered. A third class of mechanistic models are those that predict the thickness of the rust layer, often considering the chemo-mechanical nature of the problem. \citet{Wei2021} assumed uniform corrosion and related the expansion of the rust layer to the flux of dissolved oxygen. \citet{Nossoni2014} predicted the corrosion current density from the oxygen concentration, assuming it to be the limiting factor, and considered a detailed structure of a rust layer composed of ferrous and ferric rusts. More comprehensive approaches have also been employed, whereby the transport of water and certain chemical species such as oxygen or chlorides is solved for, the corrosion current density is calculated from the distribution of electrochemical potential, and the rust layer thickness is then related to the predicted corrosion current density \cite{Ozbolt2010,Ozbolt2011,Ozbolt2012,Ozbolt2014,Ozbolt2016,Ozbolt2017,Sola2019,KusterMaric2020,Solgaard2013,Flint2014,Michel2014,Thybo2017,Geiker2017, Zhu2017}. O{\v{z}}bolt, Balabani{\'{c}} and co-workers \citep{Ozbolt2010,Ozbolt2011,Ozbolt2012,Ozbolt2014,Ozbolt2016,Ozbolt2017,Sola2019,KusterMaric2020} employed contact finite elements on the boundary of the rebars to simulate the corrosion-induced inelastic strain. Also, their model simulated the transport of corrosion products into porosity and cracks as a convective diffusion problem. Michel, Geiker and co-workers \citep{Solgaard2013,Flint2014,Michel2014,Thybo2017,Geiker2017} simulated the corrosion-induced pressure of the rust layer with a thermal analogy and considered a corrosion accommodation region (CAR), where corrosion products can accumulate stress-free until the CAR is filled. Other examples of this group of models are those by \citet{Bazant1979,bazant1979physical} and \citet{Chen2019b,Chen2020}.  \\

These mechanistic models have paved the way to a better understanding of corrosion-induced cracking in reinforced concrete but the predictive abilities of existing models are still limited \cite{Angst2018a,Geiker2019}, as also highlighted in the critical review by \citet{Jamali2013a}. It has been argued that key chemo-mechanical phenomena must be included into the modelling to improve predictive capabilities \cite{Angst2018a,Angst2019a,Geiker2019}. In particular, the aforementioned models do not explicitly simulate: (i) the reactive transport of iron species in the pore space of concrete, (ii) the subsequent precipitation of iron species into rusts that blocks the pore space, and (iii) the precipitation-induced pressure caused by the accumulation of rust under confined conditions. The reactive transport of dissolved iron species and their precipitation to rusts was modelled by \citet{stefanoni_kinetic_2018,Furcas2022} and \citet{Zhang2021}, but the corrosion-induced pressure and subsequent fracture have not yet been investigated within this framework. 

In this work, a mechanistic chemo-mechanical model for corrosion-induced cracking in reinforced concrete is presented to fill this important gap. That is, unlike previous corrosion-induced cracking models, the growth of precipitates from the steel surface is not implicitly assumed but explicitly resolved by a simplified model of reactive transport and precipitation of dissolved iron species in the concrete pore space. Moreover, an eigenstrain model is adopted to capture the pressure buildup resulting from the accumulation of precipitates and this is coupled with a phase-field description of quasi-brittle fracture that incorporates the role of cracks in enhancing the transport of iron ions. The theoretical chemo-mechanical framework is implemented using the finite element method and the predictive capabilities of the coupled numerical model are demonstrated by simulating corrosion-induced cracking under mildly accelerated conditions. 


\section{Theory and computational framework}
\label{Sec:Theory}

We shall begin the formulation of our theory by describing the key chemo-mechanical mechanisms underlying corrosion-induced cracking, including a discussion on existing approaches for capturing such phenomena and on how these phenomena are incorporated into our model (Section \ref{Sec:MechanismsIntro}). Then, in Section \ref{subSec:ReTransMod}, the reactive transport model for dissolved $\mathrm{Fe}^{2+} $ and $ \mathrm{Fe}^{3+}$ ions is presented. This is followed by a description of the precipitation eigenstrain model (Section \ref{sec:mechanics_eigenstrain}). Section \ref{SubSec:fractureWu} presents the coupled phase-field fracture formulation, including the definition of the damage-dependent diffusivity tensor. And this part of the manuscript concludes with an overview of the governing equations of the theory and a brief description of the numerical implementation (Section \ref{Sec:govEq}).\\ 

\noindent \emph{Notation}. Scalar quantities are denoted by light-faced italic letters, e.g. $\phi$, Cartesian vectors by upright bold letters, e.g. $\mathbf{u}$, and Cartesian second- and higher-order tensors by bold italic letters, e.g. $\bm{\sigma}$. The symbol $ \bm{1} $ represents the second-order identity tensor while $ \bm{I} $ corresponds to the fourth-order identity tensor. Inner products are denoted by a number of vertically stacked dots, where the number of dots corresponds to the number of indices over which summation takes place, such that $ \bm{\sigma}:\bm{\varepsilon} = \sigma_{ij} \varepsilon_{ij}$. Finally, $ \bm{\nabla} c $ denotes gradient of scalar variable $c$ with respect to spatial coordinate \textbf{x} and $ \bm{\nabla} \cdot \bm{\sigma}$ is the divergence of tensor $\bm{\sigma}$.

\subsection{Mechanism of corrosion-induced cracking and its current modelling}
\label{Sec:MechanismsIntro}

As shown in Fig. \ref{Fig1}, after the initiation of corrosion, the corrosion-induced cracking process of reinforced concrete structures is divided into three main stages: (a) the transport of iron species and their precipitation, (b) the fracture of concrete due to precipitation-induced pressure, and (c) the widening of cracks, which results in enhanced ionic transport through the crack network, and eventually leads to the delamination/spalling of the concrete cover. Our modelling approach to each of these stages is discussed below.

\subsubsection{Transport of $\mathrm{Fe}^{2+}$ and $\mathrm{Fe}^{3+}$ through the concrete matrix and their precipitation}

Concrete provides a protective cover to the embedded steel. There are two levels of protection. Firstly, concrete provides a natural barrier to substances causing the corrosion of embedded steel. Secondly, the concrete environment is strongly alkaline with a pH between 12.5 and 13.5 \citep{Poursaee2016a}, which leads to the formation of a nanometre-thick protective semiconductive layer around the steel surface. This protective layer can be disrupted by various phenomena -- typically (i) carbonation caused by carbon dioxide gradually penetrating concrete, changing the pH in the process, and (ii) chloride ions causing a localised breakdown of the passive layer. In both cases, a sufficiently high content of the relevant species needs to be transported to the steel surface before corrosion begins. 

\begin{figure}[htp]
\begin{center}
\includegraphics[scale=0.81]{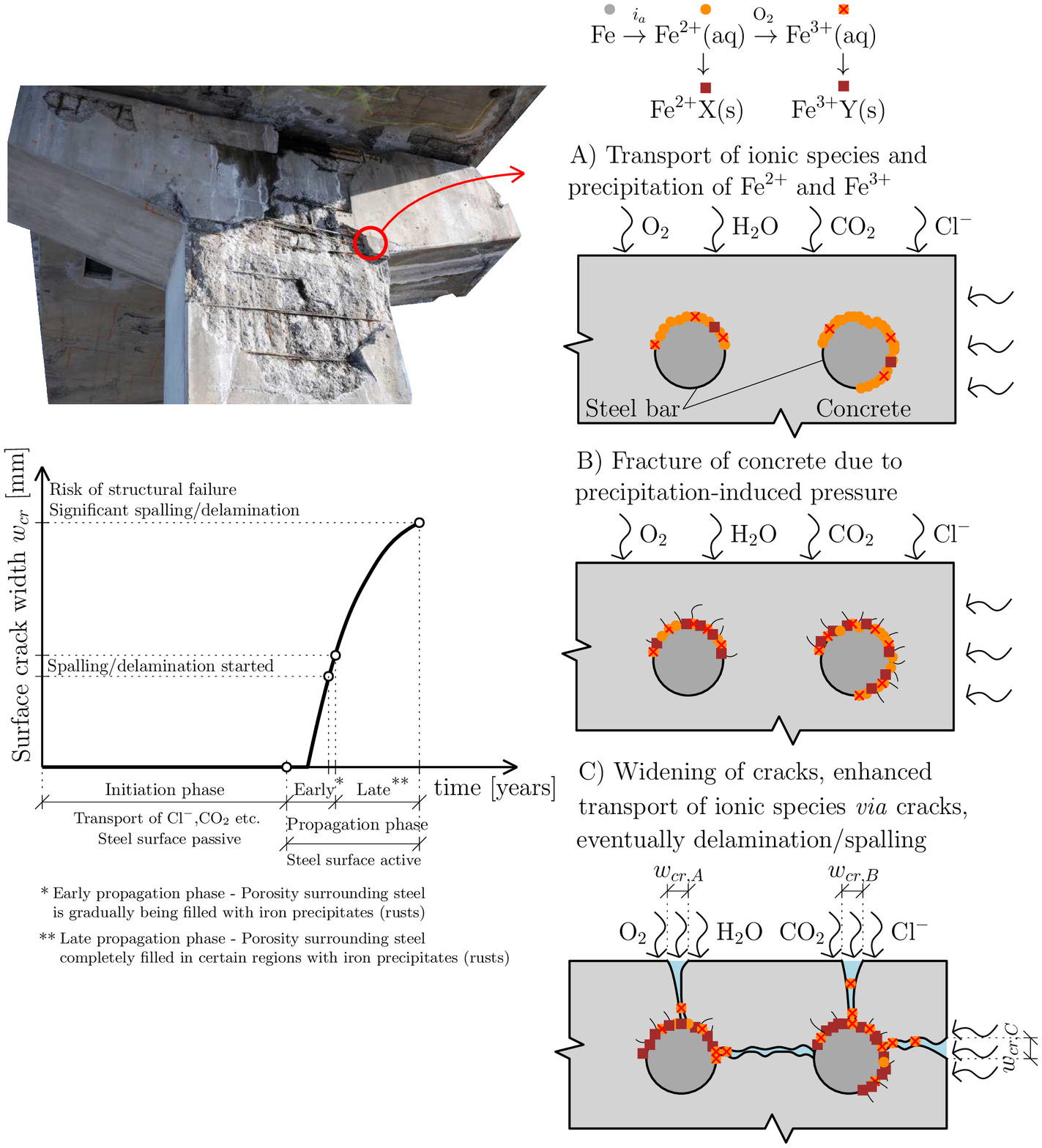}
\caption{Schematic illustration of the various stages and phenomena involved in the evolution of corrosion-induced cracking in reinforced concrete.}
\label{Fig1}
\end{center} 
\end{figure}

This period is traditionally referred to as the initiation phase of corrosion, which is followed by a propagation phase, where an anodic corrosion current density appears on the anodic sites of the steel surface. The electro-chemical process during the propagation stage is quite complex, involving a number of intermediate products, see for instance \citet{Furcas2022} for further information and reference. Nevertheless, some key processes can be identified. On anodic sites, atoms of iron are oxidised to $ \mathrm{Fe}^{2+} $ ions and released from the steel surface. Then, ferrous ions are transported deeper into the concrete pore space (not purely as charged particles but in the form of complex chemical substances). The transport of $ \mathrm{Fe}^{2+} $ is facilitated by diffusion, electromagnetic migration and advection. $ \mathrm{Fe}^{2+} $ ions precipitate to ferrous rusts (denoted as $ \mathrm{Fe}^{2+}X $ in Fig. \ref{Fig1}) after reaching a saturation concentration, and they are oxidised to $ \mathrm{Fe}^{3+} $ if there is sufficient oxygen supply. $ \mathrm{Fe}^{3+} $ ions are also transported in the form of complex chemical substances and precipitate to ferric rusts ($ \mathrm{Fe}^{3+}Y $ in Fig. \ref{Fig1}) after reaching the saturation concentration. For a pH of 8.5, the solubilities of $ \mathrm{Fe}^{2+} $ and $ \mathrm{Fe}^{3+} $ are respectively $ 10^{-3} $ M and $ 10^{-10} $ M \cite{Angst2019a}, where M denotes molarity (moles per L of solution); i.e., the precipitation of $ \mathrm{Fe}^{3+} $ ions is favoured relative to that of $\mathrm{Fe}^{2+} $. Also, the concentration of $ \mathrm{Fe}^{2+} $ is constantly diluted by the oxidation to $ \mathrm{Fe}^{3+} $. The ratio of dissolved $ \mathrm{Fe}^{2+} $ and $ \mathrm{Fe}^{3+} $ and the distribution of their precipitates were found to depend on the corrosion current density \citep{stefanoni_kinetic_2018}. The computational studies of \citet{stefanoni_kinetic_2018} and \citet{Zhang2021} showed that in chloride-free carbonated concrete under natural corrosion currents, concentrations of both dissolved and precipitated $ \mathrm{Fe}^{2+} $ are significantly lower than of $ \mathrm{Fe}^{3+} $ and that dissolved iron species can travel for significant distances (millimetres) before precipitating. However, under the high corrosion currents typical of accelerated corrosion tests, the rate of released $ \mathrm{Fe}^{2+} $ is so high that it quickly reaches the saturation concentration in the close vicinity of the steel surface, blocking the transport of $ \mathrm{Fe}^{2+} $ further into the concrete pore space. These theoretical results are also supported by experimental study of \citet{Zhang2019c} who found out that the total content of well-oxidised ferric hydroxyl oxides (specifically $\alpha-$,$\beta-$ and $\gamma-\mathrm{FeO(OH)}$) significantly decreases with increasing applied corrosion current density compared to the content of poorly oxidised $\mathrm{FeO}$ and $\mathrm{Fe_{3}O_{4}}$.

Existing models generally assume that rust grows directly from the steel surface. This is a sensible assumption when modelling impressed current tests, where corrosion is accelerated by applying a very high corrosion current density, often in the order of hundreds of \unit{\micro\ampere\per\centi\metre^2} \citep{stefanoni_kinetic_2018, Andrade2016a}. However, under the low current densities relevant to natural conditions, corrosion products can form in the pore space of concrete as far as several mm away from the steel surface \citep{stefanoni_kinetic_2018}. Some of the currently available models (e.g., \citet{Michel2014}) attempt to compensate for this discrepancy by introducing a corrosion accommodating region (CAR) located in close vicinity of steel, which can accommodate initially formed corrosion products in a stress-free state until its saturation, thus delaying the build-up of the rust layer. Other authors have simulated the transport of corrosion products into porosity as a convective diffusion problem \citep{Ozbolt2014} or as the flow of incompressible fluid \citep{Fahy2017}. Even though the transport and evolution of $ \mathrm{Fe}^{2+} $ and $ \mathrm{Fe}^{3+} $ in concrete has been simulated by \citet{stefanoni_kinetic_2018}, to the best of our knowledge, the modelling of these phenomena has not yet been considered in any corrosion-induced cracking model. As described below, we build upon the standard model for reactive transport in porous media (see, e.g., Ref. \cite{Marchand2016}) to explicitly resolve: (i) the transport of $\mathrm{Fe}^{2+}$ through the concrete matrix, (ii) the oxidation of $\mathrm{Fe}^{2+}$ to $\mathrm{Fe}^{3+}$, (iii) the transport of $\mathrm{Fe}^{3+}$ through the concrete matrix, (iv) the precipitation of $\mathrm{Fe}^{2+}$ and $\mathrm{Fe}^{3+}$ in the concrete pore space, and (v) the clogging of the pore space by precipitates (rusts).

\subsubsection{Fracture of concrete due to precipitation-induced pressure}

When $ \mathrm{Fe}^{2+} $ ions and $ \mathrm{Fe}^{3+} $ ions precipitate, the resulting rusts have a significantly lower density than original iron, typically by 3 - 6 times \citep{Angst2019a}, which allows them to quickly fill concrete pores. Since precipitates are forced to grow under confined conditions in the pore space, they exert pressure on the surrounding concrete matrix. This gradually increasing pressure is arguably the key driving mechanism of corrosion-induced cracking in its early stages \cite{Angst2019a, Angst2018a}. This mechanism is likely similar to the one governing salt damage in porous materials \cite{Scherer1999, Flatt2014, Flatt2017a, Coussy2006, Castellazzi2013, Koniorczyk2012, Espinosa2008}. The precise governing mechanism in later stages, when the majority of the surrounding pore space is filled with rusts, remains unclear. It is likely influenced by the porosity of the cement paste and cracks, which can facilitate the transport of $ \mathrm{Fe}^{2+} $ and $ \mathrm{Fe}^{3+} $ away from the steel surface, reducing the local amount of rusts accumulated at the steel-concrete interface in the process.\\

Existing models typically consider an expansion coefficient to account for the density mismatch between rusts and steel. A fixed value is commonly assumed, regardless of the corrosion current density, and this value is obtained by fitting data from highly accelerated corrosion tests. However, this approach compromises the predictive capabilities of the models in the regime of natural corrosion currents, as the composition and distribution of rusts change with the corrosion current. The pressure on concrete induced by accumulating rusts is commonly accounted for by either expanding the rebar (e.g., via a thermal expansion analogy \cite{Michel2014,Chernin2011}) or by prescribing on the rebar boundary a suitable displacement field \cite{Chen2020, Chen2019b, Tran2011, Roshan2020} or internal pressure \cite{Du2006}. However, all these strategies implicitly assume that the precipitates grow directly from the steel surface and that the fracture is driven by the growth of the corrosion layer surrounding the steel rebar. This contrasts with the fact that corrosion products can precipitate even millimetres away from the steel surface \citep{stefanoni_kinetic_2018} and that the initial corrosion-induced fracture is supposedly driven by the accumulation of precipitates in pores surrounding steel. Even if a corrosion accommodating region (CAR) is considered, the accumulation of rusts in CAR is considered stress-free. The proposed model aims at overcoming these shortcomings and developing a framework capable of predicting corrosion-induced cracking for natural corrosion conditions. This is achieved by incorporating two key elements into the modelling: (i) the pressure build-up due to the precipitation of ferrous and ferric ions, and (ii) the nucleation and growth of cracks due to the resulting stress state. The former is introduced into our theory by means of an eigenstrain. This strategy has been successfully employed by \citet{Krajcinovic1992} and \citet{Basista2008,Basista2009} to simulate sulphate attack on concrete. The eigenstrain approach has also been used by \citet{Evans2018,Evans2020} to describe reaction-driven cracking in rocks due to olivine hydration/carbonation. Thus, an eigenstrain-based model is formulated for the precipitation of corrosion products, which should provide a sound description of the underlying physics of the early propagation phase of corrosion-induced cracking (see Fig. \ref{Fig1}). Regarding the fracture process, this is captured by means of a phase-field fracture model \cite{Bourdin2000,Amor2009,Miehe2010}. Phase-field models have gained remarkable popularity in recent years due to their numerical robustness and ability to capture complex cracking phenomena of arbitrary complexity (see, e.g. \cite{Wei2021,Wu2018b,Kristensen2020a,Tan2021} and Refs. therein). In particular, given the quasi-brittle nature of concrete \cite{Bazant1991}, we build our model upon the phase-field fracture model of Wu and co-workers \cite{Wu2017,Wu2018} (see \ref{Sec:A0} for a numerical comparison with other approaches, illustrating its ability to predict quasi-brittle behaviour).

\subsubsection{Sustained cracking and spalling, with enhanced transport of chemical species through cracks}
 
When cracks appear, they become a preferred pathway for the transport of new $ \mathrm{Fe}^{2+}$ ions and are thus filled preferentially instead of the local pore space, which reduces the build-up of precipitation pressure in these regions. Once cracks reach the concrete surface, depending on the humidity of surrounding environment, the corrosion process can be accelerated because the necessary substances such as oxygen, moisture or chlorides can be easily transported to the steel bar from the concrete surface through the cracks (see Fig. \ref{Fig1}). However, cracks also get gradually blocked by precipitates. \\

Some models have aimed at incorporating the influence of cracks on the distribution of precipitates and induced pressure by modelling the transport of rusts from the steel surface. In the model of \citet{Fahy2017}, the transport of corrosion products away from the steel surface is assumed to be analogous to the flow of viscous fluid. \citet{Ozbolt2014} consider the transport of rust to be a convective diffusion problem. \citet{Tran2011} assume that cracks can accommodate a certain volume of corrosion products so that the corrosion-induced pressure is reduced. However, to our best knowledge, explicit modelling of transport and precipitation of $ \mathrm{Fe}^{2+} $ and $ \mathrm{Fe}^{3+} $, and their interplay with cracking, have not yet been accounted for in existing models. The enhancement of diffusivity facilitated by micro-cracks and cracks is here captured by adopting a damage-dependent diffusivity tensor, building upon the work by \citet{Wu2016}. Other strategies have also been presented to capture the diffusion-fracture interplay, such as penalty approaches \citep{Martinez-Paneda2020}.

\subsection{Reactive transport model}
\label{subSec:ReTransMod}

\subsubsection{Representative volume element (RVE) and primary fields}

\begin{figure}[htp]
\begin{center}
\includegraphics[scale=1.0]{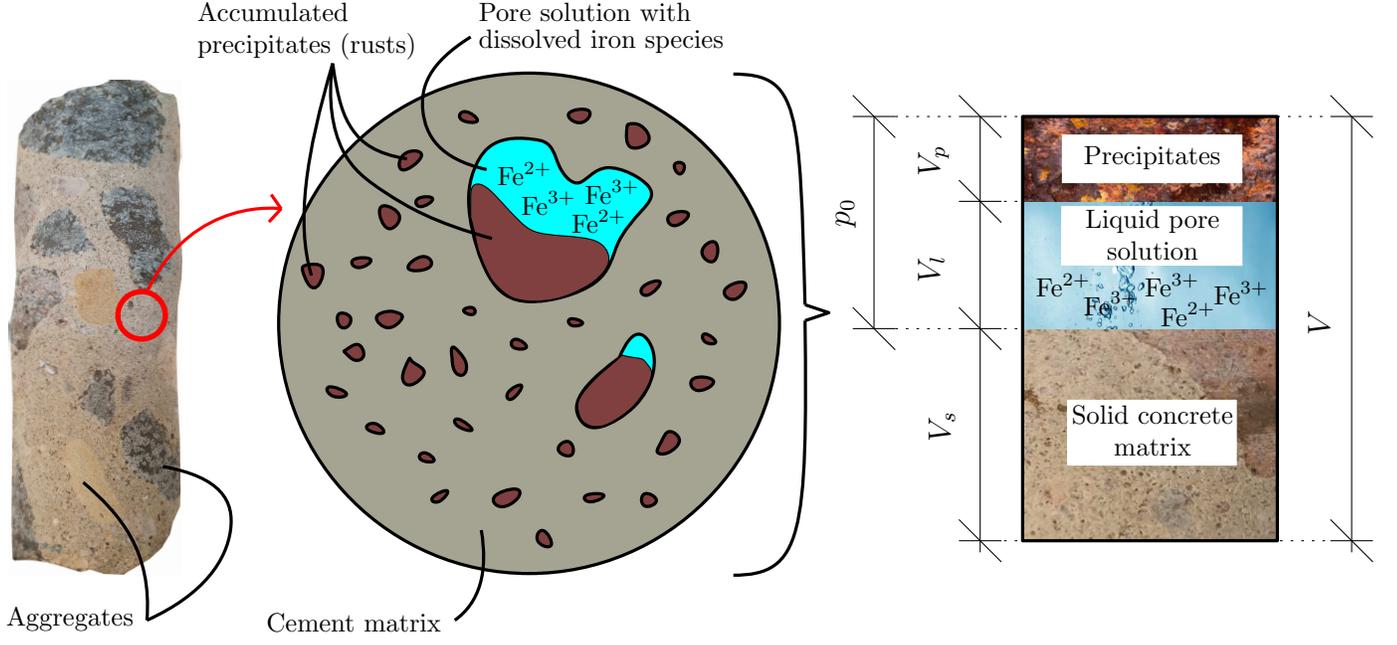}
\caption{Schematic illustration of a Representative Volume Element (RVE) of concrete at the vicinity of corroding steel, highlighting the relevant phases.}
\label{Fig2}
\end{center} 
\end{figure}

Consider the representative volume element (RVE) of concrete (at the vicinity of corroding steel) depicted in Fig. \ref{Fig2}. The RVE of concrete consists of a volume $V_{s}$ occupied by solid matrix with porosity $ p_{0}$. Concrete is idealised to be fully saturated with its porosity $ p_{0} = (V-V_{s})/V $ divided between liquid pore solution and iron precipitates (rusts), which are generated by the sequence of chemical reactions from $\mathrm{Fe}^{2+}$ and $\mathrm{Fe}^{3+}$ ions. 

The distribution of liquid pore solution and rusts is described by the liquid volume fraction $\theta_{l} = V_{l}/V$ and the precipitate volume fraction $\theta_{p} = V_{p}/V$, respectively, while the distribution of $\mathrm{Fe}^{2+}$ and $\mathrm{Fe}^{3+}$ is described in terms of their concentrations $ c_{II} $ and $ c_{III} $, with units of mol per cubic meter of liquid pore solution. It is assumed that the volume of the solid concrete matrix is not changing in time and thus the available porosity $p_{0}$ remains constant too. This facilitates the definition of a precipitate saturation ratio as $ S_{p} = \theta_{p}/p_{0} $.\\

\begin{figure}[!htb]
\begin{center}
\includegraphics[width=0.5\textwidth]{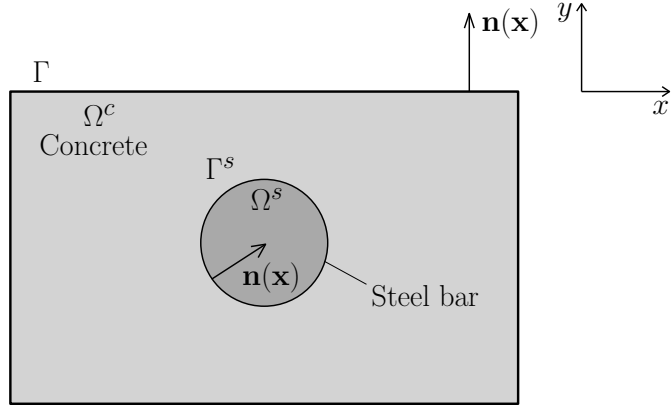}
\caption{Graphical illustration of the domain and relevant variables for the chemo-mechanical problem.}
\label{Fig3}
\end{center} 
\end{figure}

The reactive transport problem is solved on a concrete domain $\Omega^{c}$, depicted in Fig. \ref{Fig3}. The boundaries of the concrete domain are the outer boundary $\Gamma$ and the boundary of steel rebar $\Gamma^{s}$. Here, $\mathbf{n}(\mathbf{x})$ denotes an outward-pointing vector normal to $ \Gamma \cup \Gamma^{s} $. The primary fields for the reactive transport model are
\begin{equation}\label{setCII}
c_{II}(\mathbf{x},t) \in \mathbb{A} = \lbrace \forall t \geq 0: c_{II}(\mathbf{x},t) \in W^{1,2}(\Omega^{c}) \rbrace
\end{equation}
\begin{equation}\label{setCIII}
c_{III}(\mathbf{x},t) \in \mathbb{B} = \lbrace \forall t \geq 0: c_{III}(\mathbf{x},t) \in W^{1,2}(\Omega^{c})\rbrace
\end{equation}
\begin{equation}\label{setPr}
\theta_{p}(\mathbf{x},t) \in \mathbb{C} = \lbrace \forall t \geq 0: \theta_{p}(\mathbf{x},t) \in W^{1,2}(\Omega^{c}) \rbrace
\end{equation}
where $ W^{1,2}(\Omega^{c}) $ is the Sobolev space consisting of functions with square-integrable weak derivatives. For $ c_{II}(\mathbf{x},t) $, in addition to a Dirichlet type boundary condition, a Neumann type boundary condition describing the flux of $\mathrm{Fe}^{2+}$ from the corroding steel surface is prescribed on the steel boundary $ \Gamma^{s}$. The formulation of this boundary condition is introduced further in this section. Let us remark that the pore space of concrete is simplified here to be fully saturated with water, at least in the close vicinity of the rebar. This assumption is presumed to be particularly sensible for well-cured concrete samples in laboratory conditions, which are constantly kept wet. It should be emphasised that the analysis of in-situ structures would require reflecting the variable water saturation caused by periodic wetting and drying cycles, which is the goal of an ongoing research. Assuming  full saturation, the volume fraction of pore solution $ \theta_{l} $ can be calculated directly from the precipitate volume fraction $ \theta_{p} $ as $\theta_{l} = p_{0} - \theta_{p}$. This means that growing precipitates are presumed to push the required liquid volume fraction of pore solution out of the pores immediately.  

\subsubsection{Governing equations of reactive transport}
Assuming small deformations, we can neglect the velocity of the solid concrete matrix and derive the transport equations for $\mathrm{Fe}^{2+}$, $\mathrm{Fe}^{3+}$ and precipitates (rusts) from the condition of mass conservation. The resulting transport equation reads
\begin{equation}\label{total_mass_derivative_N_1}
\frac{\partial \widetilde{\rho}_{\alpha}}{\partial t} + \bm{\nabla} \cdot \left(\widetilde{\rho}_{\alpha} \mathbf{v}_{\alpha} \right) =  M_{\alpha} \theta_{l} R_{\alpha} \,\,\,\,\,\,\, \text{ in } \,\, \Omega^{c}, \quad \alpha = II,III,p  
\end{equation}
where $\widetilde{\rho}_{\alpha} $ is the averaged density of species $ \alpha $ calculated with respect to the whole RVE, $\mathbf{v}_{\alpha}$ is the velocity of species $\alpha$ and $M_{\alpha}$ is the molar mass of species $\alpha$. $ R_{\alpha} $ describes the resulting rate of the chemical transformation of $\alpha$ summing the production and consumption effects of all the chemical reactions that $\alpha$ is undergoing. In Eq. (\ref{total_mass_derivative_N_1}), the term $ \widetilde{\rho}_{\alpha} \mathbf{v}_{\alpha} $ can also be expressed as  
\begin{equation}\label{totalFlux1}
\widetilde{\rho}_{\alpha} \mathbf{v}_{\alpha} = \widetilde{\rho}_{\alpha} \mathbf{v}_{l} + \mathbf{J}_{\alpha}
\end{equation}
where $\widetilde{\rho}_{\alpha} \mathbf{v}_{l}$ is the flux caused by the flow of $\alpha$ with liquid (advection) and $ \mathbf{J}_{\alpha}$ is the additional flux caused by the transport of $\alpha$. Substituting (\ref{totalFlux1}) into (\ref{total_mass_derivative_N_1}), we get
\begin{equation}\label{total_mass_derivative_N_2}
\frac{\partial \widetilde{\rho}_{\alpha}}{\partial t} + \bm{\nabla} \cdot \left(\widetilde{\rho}_{\alpha} \mathbf{v}_{l} + \mathbf{J}_{\alpha}\right) =  M_{\alpha} \theta_{l} R_{\alpha}\,\,\,\,\,\,\, \text{ in } \,\, \Omega^{c}
\end{equation}
The averaged density $ \widetilde{\rho}_{\alpha} $ can be expressed based either on the volume fraction $ \theta_{\alpha} $ or concentration $c_{\alpha}$ as
\begin{equation}\label{avr_mass_density_forms_N_1}
\widetilde{\rho}_{\alpha} = \theta_{\alpha} \rho_{\alpha} = \theta_{l} M_{\alpha} c_{\alpha}
\end{equation}
with $ \rho_{\alpha} $ being the intrinsic density of species $ \alpha $ with respect to its volume in RVE. Since the precipitates are considered to be immobile (i.e., their total flux is zero) and $ \rho_{c} $ is assumed to be constant both in time and space, we can substitute (\ref{avr_mass_density_forms_N_1}) into (\ref{total_mass_derivative_N_2}) and obtain
\begin{equation}\label{total_mass_derivative_crystals_N_2}
\frac{\partial \theta_{p}}{\partial t} =  \dfrac{M_{p}}{\rho_{p}} \theta_{l} R_{p} \,\,\,\,\,\,\, \text{ in } \,\, \Omega^{c}
\end{equation} 
Contrarily to immobile precipitates, $ \mathrm{Fe}^{2+} $ and $ \mathrm{Fe}^{3+} $ ions can be transported through the pore space of concrete by advection, diffusion, chemical activity-related movement and electromagnetic migration. Diffusion is assumed to be the dominant phenomenon in the transport of iron ions (as in Refs. \cite{stefanoni_kinetic_2018,Zhang2021}) and thus the one modelled here. In pore solution, Fick's law reads 
\begin{equation}\label{diffFluxFickLaws}
\mathbf{J}_{\alpha} = -M_{\alpha}\bm{D}_{\alpha}\cdot\nabla c_{\alpha}, \quad \alpha = II,III
\end{equation}
where $ \bm{D}_{\alpha} $ is the second-order diffusivity tensor of species $\alpha$. However, since we consider transport in a porous medium and not only the pore solution, the diffusion-driven flux term (\ref{diffFluxFickLaws}) needs to be scaled adequately. This problem has been the subject of many studies, which approached it in different ways. For example \citet{Baroghel-Bouny2011} proposed to introduce a saturation ratio-dependent diffusivity which needs to be fitted experimentally. Alternatively, \citet{Marchand2016} adopted an integral averaging procedure \cite{Bear1990}, which is based on the averaging of the transport equation in pore solution over the RVE of the porous material. Thus, as \citet{Marchand2016}, we write the resulting diffusion term as
\begin{equation}\label{diffFluxMarchand}
\mathbf{J}_{\alpha} = -M_{\alpha}\theta_{l}\bm{D}_{\alpha}\cdot\nabla c_{\alpha}, \quad \alpha = II,III   
\end{equation}
The effects of the geometry of the pore space and variation of the cross-section of pores can also be considered by multiplying (\ref{diffFluxMarchand}) by additional factors reflecting constrictivity and tortuosity of the pore space, as done by \citet{Zhang2021}. Here, the effects of constrictivity and tortuosity are introduced via $\bm{D}_{\alpha}$. We will adopt (\ref{diffFluxMarchand}) as the expression for diffusion-driven flux in porous media and substitute it into (\ref{total_mass_derivative_N_2}). If the velocity of liquid $ \mathbf{v}_{l} $ and thus the flux related to advection is neglected, we obtain 
\begin{equation}\label{total_mass_derivative_ionic_N_2}
\frac{\partial \left(\theta_{l}c_{\alpha}\right)}{\partial t} - \bm{\nabla} \cdot \left(\theta_{l}\bm{D}_{\alpha}\cdot\nabla c_{\alpha} \right) =  \theta_{l} R_{\alpha} \,\,\,\,\,\,\, \text{ in } \,\, \Omega^{c}, \quad \alpha = II,III
\end{equation}    

\subsubsection{Considered chemical reactions}
In order to evaluate reaction terms $ R_{II} $, $ R_{III} $ and $ R_{p} $ and the boundary condition describing the rate of production of the ferrous ions on the steel surface, let us consider the following system of chemical reactions: 
\begin{equation}\label{reaction_Fe2}
2 \mathrm{Fe}+\mathrm{O}_{2}+2 \mathrm{H}_{2} \mathrm{O} \rightarrow 2 \mathrm{Fe}^{2+}+4 \mathrm{OH}^{-}
\end{equation}
\begin{equation}\label{reaction_Fe3}
4 \mathrm{Fe}^{2+} +\mathrm{O}_{2}+2 \mathrm{H}_{2} \mathrm{O} \rightarrow 4 \mathrm{Fe}^{3+}+4 \mathrm{OH}^{-}
\end{equation}
\begin{equation}\label{reaction_FeOOH}
\mathrm{Fe}^{3+} + 3 \mathrm{OH}^{-} \rightarrow \mathrm{FeO(OH)}+\mathrm{H}_{2}\mathrm{O}
\end{equation}
Firstly, $ \mathrm{Fe}^{2+} $ ions are released from the corroding steel surface because of the electrochemical reaction, as described by (\ref{reaction_Fe2}). Corrosion is assumed to be uniform along the steel surface, such that cathodic and anodic reactions proceed over the entire surface with the same corrosion current density $ i_{a} $. Since the focus of this study is on the modelling of the corrosion-induced cracking and not predicting the corrosion current density, the latter is taken as an input parameter to the model. Then, by reaction (\ref{reaction_Fe3}), $\mathrm{Fe}^{2+}$ ions are oxidised to $\mathrm{Fe}^{3+}$ ions. The possibility and rate of this reaction depend on the availability of oxygen. Its transport is not simulated in this model and a constant oxygen concentration is assumed instead. This simplification is introduced merely to reduce the complexity of the model at the current stage of research. A parametric study on the influence of oxygen concentration (discussed in \ref{Sec:C}) suggests that unless oxygen is nearly depleted (so that the reaction stops), its concentration does not significantly affect the reaction rate and the resulting concrete damage. It seems reasonable to assume that oxygen will not be entirely depleted during the early propagation phase as there is experimental evidence that even in water-submerged conditions, oxygen in concrete is not entirely depleted in years or even decades \cite{AngstNature2019}.\\

Finally, reaction (\ref{reaction_FeOOH}) leads to the precipitation of $ \mathrm{Fe}(\mathrm{OH})_{3} $ or otherwise $ \mathrm{FeO(OH)}+\mathrm{H}_{2}\mathrm{O} $, which is assumed to be the final corrosion product in this model. Since the focus is on natural corrosion conditions, the precipitation of $\mathrm{Fe}^{2+}$ is neglected. This assumption is supported by the results of \citet{stefanoni_kinetic_2018} and \citet{Zhang2021}, who have shown that in carbonated chloride-free concrete under natural corrosion currents, the rate of precipitation of $\mathrm{Fe}^{2+}$ is significantly slower than that of $\mathrm{Fe}^{3+}$. Moreover, it is known that the presence of chloride ions seriously hinders the precipitation of $\mathrm{Fe}^{2+}$ \citep{Sagoe-Crentsil1993, Furcas2022}. In addition, it should be emphasised that reactions (\ref{reaction_Fe2})-(\ref{reaction_FeOOH}) constitute a simplification of a very complex system of reactions, see for instance \citet{Furcas2022}. 

\subsubsection{Reaction rates}

The rate law for reactions (\ref{reaction_Fe3})-(\ref{reaction_FeOOH}) is often considered to be of the form $ R = k_{r}c^{X_{1}}_{\alpha_{1}}c^{X_{2}}_{\alpha_{2}}...c^{X_{N}}_{\alpha_{1}} $, where $ c_{\alpha_{1}}, c_{\alpha_{2}}...c_{\alpha_{N}} $ are the concentrations of reactants and $k_{r}$ is the rate constant, which has to be measured experimentally, as is the case with the exponents $X_i$ \cite{Atkins2018}. Here, the rate law for reaction (\ref{reaction_Fe3}) is defined based on the work by \citet{stefanoni_kinetic_2018} and \citet{Zhang2021}. Thus, for a pH higher than approximately 8, the reaction rate is independent of pH \citep{Morgan2007} and can be written as
\begin{equation}\label{rateLaw1}
R_{II} = -k^{II \rightarrow III}_{r} c_{II}c_{ox}
\end{equation}
where $c_{ox}$ is the concentration of oxygen dissolved in the pore solution. Since the rate law for the precipitation of $\mathrm{Fe}^{3+}$ (reaction (\ref{reaction_FeOOH})) is unknown, we estimate   
\begin{equation}\label{rateLaw2}
R_{p}= k^{III \rightarrow p}_{r} c_{III}
\end{equation} 
Based on (\ref{rateLaw1}) and (\ref{rateLaw2}), the rate of reaction (\ref{reaction_Fe3}) can be calculated as 
\begin{equation}\label{rateLaw3}
R_{III} = -R_{II} - R_{p} = k^{II \rightarrow III}_{r} c_{II}c_{ox} - k^{III \rightarrow p}_{r} c_{III}
\end{equation} 
The boundary condition on $ \Gamma^{s} $ describing the influx of $ \mathrm{Fe}^{2+} $ released from the steel surface according to reaction (\ref{reaction_Fe2}) reads
\begin{equation}\label{FarLaw_1}
J_{II} = -\mathbf{n} \cdot \left(-\boldsymbol{D}_{II} \cdot \nabla c_{II} \right)
\end{equation}
The inward flux $ J_{II} $ can also be related to Faraday's law as  
\begin{equation}\label{new_FarLaw}
J_{II} = \frac{2i_{a}}{z_{a}F}
\end{equation}
where $ i_{a} $ is the corrosion current density, $F$ is the Faraday constant and $ z_{a} = 2 $ stands for the number of electrons exchanged in anodic reaction (\ref{reaction_Fe2}) per one atom of iron. The factor of two multiplying the right-hand side of (\ref{new_FarLaw}) results from the stoichiometry of the corrosion reaction (\ref{reaction_Fe2}).

\subsection{Precipitation eigenstrain}
\label{sec:mechanics_eigenstrain}

Due to its simplicity and numerical robustness, the macroscopic stress resulting from the precipitation products is incorporated through an eigenstrain $ \bm{\varepsilon}_{\star} $, which is derived from micromechanical considerations \cite{Krajcinovic1992,mura1987}. In the absence of damage, the small strain tensor $\bm{\varepsilon} = \nabla_{s}\mathbf{u} = (\nabla \mathbf{u} + (\nabla \mathbf{u})^{T})/2$ can be additively decomposed into the elastic part $ \bm{\varepsilon}_{e} $ and the precipitation eigenstrain $ \bm{\varepsilon}_{\star} $, such that 
\begin{equation}\label{smallStrainTensor1}
\bm{\varepsilon} = \bm{\varepsilon}_{e} + \bm{\varepsilon}_{\star}
\end{equation}
The Cauchy stress can then be readily estimated using the fourth-order elastic stiffness tensor $\mathcal{\bm{C}}_{e}$ as
\begin{equation}\label{CauchyStressTensor}
\bm{\sigma} = \mathcal{\bm{C}}_{e}:(\bm{\varepsilon} - \bm{\varepsilon}_{\star}) 
\end{equation} 

Inspired by the analytical result of \citet{Coussy2006} for isothermal conditions, a linear dependence of $ \bm{\varepsilon}_{\star} $ on the precipitate saturation ratio $ S_{p} = \theta_{p}/p_{0}$ is assumed, such that 
\begin{equation}\label{crystEigStr}
\bm{\varepsilon}_{\star} = \varepsilon_{\star} \bm{1} = f(S_{p},\phi,...)\bm{1} \approx C S_{p} \bm{1}
\end{equation}
\noindent where $C$ is a positive constant, to be defined. More complex eigenstrain functions $ f(S_{p},\phi,...)$ can be considered; for example, to capture the influence of pore size or to define a threshold for $S_p$, below which rust can be accommodated in the pores stress-free. Regarding $C$, we assume that it is proportional to the volumetric strain $\varepsilon_{v}$ resulting from the geometrically unconstrained precipitation of dissolved $ \mathrm{Fe}^{3+} $. This strain can be calculated as 
\begin{equation}\label{volStrainCryst}
\varepsilon_{v} = \dfrac{\upsilon_{p}}{(1-r_{0})\upsilon_{III}} - 1 = \dfrac{\rho_{III} M_{p}}{(1-r_{0}) \rho_{p} M_{III}} - 1 
\end{equation} 
where $\upsilon_{\alpha}  = M_{\alpha}/\rho_{\alpha}$ is the molar volume and $r_{0}$ is the porosity of rusts. Now, let us firstly calculate the eigenstrain $ \bm{\varepsilon}_{\star\star} = \varepsilon_{\star\star} \bm{1}$ assuming that rusts have the same mechanical properties as the surrounding concrete matrix. Since concrete is considered to be isotropic, we can write 
\begin{equation}\label{crystEigStr1}
\bm{\varepsilon}_{\star\star} = \dfrac{1}{3}\varepsilon_{v}S_{p}\bm{1} 
\end{equation}   
We then follow \citet{Krajcinovic1992} and incorporate a correction factor in Eq. (\ref{crystEigStr1}) to account for the material property mismatch between rust and concrete. The approach follows the equivalent inclusion analysis by \citet{mura1987}, and in particular the case of the spherical inclusion. Thus, assuming that the inhomogeneity is sufficiently far from the boundary to be unaffected by surface tractions $ \mathbf{t}$,  the precipitation eigenstrain reads
\begin{equation}\label{formulaMura}
\bm{\varepsilon}_{\star} = \dfrac{3(1-\nu)K_{p}}{(1+\nu)K_{p}+(2-4\nu)K}\bm{\varepsilon}_{\star\star}
\end{equation}
where $ K $ is the bulk modulus of concrete and and $ K_{p} $ is the bulk modulus of iron precipitates (rusts). $ K $ and $ K_{p} $ are calculated as $K_i=E_i/(3(1-2\nu_i))$, where $E_{p}$ and $\nu_{p}$ respectively denote the the Young's modulus and the Poisson's ratio of iron precipitates (rusts), and $E$ and $\nu$ are the associated counterparts for a rust-filled concrete. It remains to consider that the pores in which rusts can accumulate represent a significant part of the total volume of concrete. Thus, the mechanical properties of rust-filled regions of concrete must be a function of the precipitate volume fraction $ \theta_{p} $, and the mechanical properties of rust-free concrete ($E_c$, $\nu_c$) and rust ($E_p$, $\nu_p$). Making use of the rule of mixtures,  
\begin{equation}\label{rustFillConcProp}
E = (1-\theta_{p})E_{c} + \theta_{p}E_{p},\quad \nu = (1-\theta_{p})\nu_{c} + \theta_{p}\nu_{p}
\end{equation}
\noindent and thus the precipitation eigenstrain $\bm{\varepsilon}_{\star}$ is given by
\begin{equation}\label{crystEigStr2}
\bm{\varepsilon}_{\star} = C S_{p}\bm{1}, \text{ with } C = \dfrac{(1-\nu)K_{p}}{(1+\nu)K_{p}+(2-4\nu)K}\left(\dfrac{\rho_{III} M_{p}}{(1-r_{0}) \rho_{p} M_{III}} - 1 \right)
\end{equation}

\subsection{Phase-field description of precipitation-induced cracks}
\label{SubSec:fractureWu}

\subsubsection{A generalised structure}

A general formulation of phase-field fracture accounting for precipitation eigenstrain is first presented. Of interest is the nucleation and growth of cracks in a concrete domain $ \Omega^{c} \subset \mathbb{R}^{d}, ~ d = 2,3 $, where $ d $ is the geometrical dimension of the problem. The domain $\Omega^{s} \subset \mathbb{R}^{d} $ corresponds to the steel rebars. The boundary of the concrete domain $ \Omega^{c} $ is $ \Gamma \cup \Gamma^{s}$, where $ \Gamma^{s} $ is the boundary with steel rebars and $ \Gamma $ is an outer concrete boundary. The boundary $ \Gamma $ can be decomposed into $\Gamma = \Gamma^{u} \cup \Gamma^{t}$, where $ \Gamma^{u} $ is the portion of boundary with prescribed displacements  $\overline{\mathbf{u}}(\mathbf{x}) $ and $ \Gamma^{t} $ is the portion of boundary where tractions $\overline{\mathbf{t}}(\mathbf{x}) $ are prescribed. The outward-pointing normal vector to $ \Gamma \cup \Gamma^{s} $ is $\mathbf{n}(\mathbf{x})$ and $\overline{\mathbf{b}}(\mathbf{x}) $ denotes a prescribed volume force vector acting on $\Omega = \Omega^{c} + \Omega^{s}$.  
\begin{figure}[!htb]
    \centering
    \includegraphics[width=0.5\textwidth]{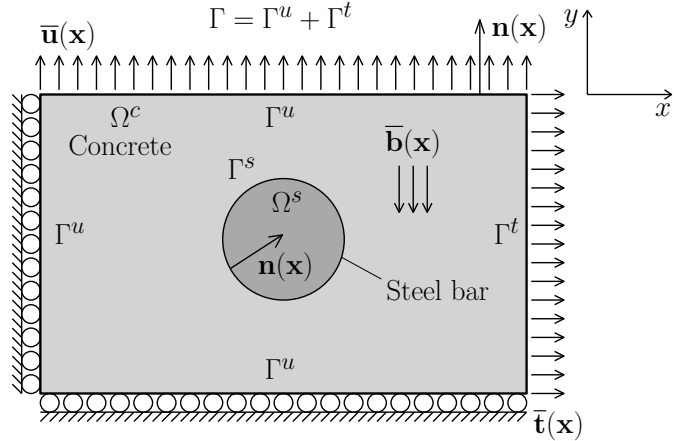}
    \caption{Graphical illustration of the domain and relevant variables for the deformation-fracture problem.}
    \label{FigNotation}    
\end{figure}
The primary (unknown) fields are the displacement components
\begin{equation}\label{setDispl}
\mathbf{u}(\mathbf{x},t) \in \mathbb{U} = \lbrace \forall t \geq 0: \mathbf{u}(\mathbf{x},t) \in W^{1,2}(\Omega)^{d}; \mathbf{u}(\mathbf{x},t) = \overline{\mathbf{u}}(\mathbf{x},t)\text{ in }\Gamma^{u} \rbrace
\end{equation} 
and the phase-field variable 
\begin{equation}\label{setPF}
\phi(\mathbf{x},t) \in \mathbb{P} = \lbrace \forall t \geq 0: \phi(\mathbf{x},t) \in W^{1,2}(\Omega^{c}); 0 \leq \phi(\mathbf{x},t) \leq 1 \text { in } \Omega^{c}; t_{1} \leq t_{2} \Longrightarrow \phi(\textbf{x},t_{1}) \leq \phi(\textbf{x},t_{2}) \rbrace
\end{equation} 
where $ W^{1,2}(\Omega^{c})^{d} $ is the Cartesian product of $ d $ Sobolev spaces $ W^{1,2}(\Omega^{c}) $ consisting of functions with square-integrable weak derivatives. The phase-field variable characterises the current state of damage and can be transformed into the degradation function  $ g(\phi) \in \mathbb{S} = \lbrace g(\phi) \in W^{1,2}(\Omega^{c}); 0 \leq g(\phi) \leq 1 \text { in } \Omega^{c} \rbrace $, which reflects the remaining integrity of the material. The degradation function $ g $ is expected to be decreasing from 1 for $ \phi = 0 $, which corresponds to the initial undamaged material, to 0 for $\phi=1$, which corresponds to the fully cracked material. The steel domain $ \Omega^{s} $ is expected to remain elastic.\\ 

For elasto-damaged material, quasi-static conditions are assumed, such that the small strain tensor reads  $\bm{\varepsilon} = \nabla_{s}\mathbf{u} = (\nabla \mathbf{u} + (\nabla \mathbf{u})^{T})/2 $ and the Cauchy stress tensor $ \bm{\sigma} $ is given by
\begin{equation}\label{CauchyStressTensor2}
\bm{\sigma} = g(\phi)\mathcal{\bm{C}}_{e}:(\bm{\varepsilon} - \bm{\varepsilon}_{\star}) \end{equation} 

Now, let us adopt the minimum total energy principle and formulate the total energy functional $ \Pi $ of our phase-field fracture problem. The total energy functional reads 
\begin{subequations}\label{engyFun1}
\begin{align}
\Pi(\mathbf{u}, \phi) &= E_{std} (\mathbf{u}, \phi) + E_{reg} (\phi) + D (\phi_{n},\phi) + B (\mathbf{u})+ T(\mathbf{u}) \\ 
E_{std}(\mathbf{u}, \phi) &=\frac{1}{2} \int_{\Omega^{c}}g(\phi) (\nabla_{s} \mathbf{u} - \bm{\varepsilon}_{\star}): \mathcal{\bm{C}}^{c}_{e}: (\nabla_{s} \mathbf{u} - \bm{\varepsilon}_{\star}) \mathrm{d} V + \frac{1}{2} \int_{\Omega^{s}}\nabla_{s} \mathbf{u}:\mathcal{\bm{C}}^{s}_{e}:\nabla_{s} \mathbf{u} \dd V \\
E_{reg}(\phi) &= \int_{\Omega^{c}}\varepsilon_{reg}(\phi) \mathrm{d} V =  \int_{\Omega^{c}} \mathcal{D}(1) \ell^{2}\|\nabla \phi\|^{2} \mathrm{d} V \\ \label{engyFun1a}
D(\phi_{n},\phi) &= \begin{cases}\displaystyle\int_{\Omega^{c}} \left(\mathcal{D}\left(\phi\right)-\mathcal{D}\left(\phi_{n}\right)\right) \dd V \quad &\text { if } \phi \geq \phi_{n} \text { in } \Omega^{c} \\ +\infty & \text { otherwise }\end{cases}\\
B(\mathbf{u}) &= -\int_{\Omega} \overline{\mathbf{b}} \cdot \mathbf{u} \dd V \\
T(\mathbf{u}) &= -\int_{\Gamma} \overline{\mathbf{t}} \cdot \mathbf{u} \dd S 
\end{align}
\end{subequations}
In (\ref{engyFun1}), $\mathcal{\bm{C}}^{c}_{e}$ and $\mathcal{\bm{C}}^{s}_{e}$ are the fourth-order elastic stiffness tensor of concrete and steel respectively. $\mathcal{\bm{C}}^{c}_{e}$ and $\mathcal{\bm{C}}^{s}_{e}$ are calculated as $\mathcal{\bm{C}}^{i}_{e}=\lambda^{i} \bm{1} \otimes \bm{1} +2\mu^{i} \bm{I} $, where $ \lambda^{i} $ and $\mu^{i}$ are the Lamé constants of steel or concrete. $ D(\phi_{n},\phi) $ is the dissipation distance evaluating dissipation caused by the cracking process changing phase-field from $\phi_{n}$ to $\phi$, where $\phi_{n}$ is the previously reached phase-field distribution. In terms of the numerical solution, $\phi_{n}$ is the phase-field distribution in the previous time step. $B(\mathbf{u})$ is the potential energy related to prescribed body forces and $T(\mathbf{u})$ is the potential energy related to prescribed surface tractions. The energy stored in the system is given by $ E_{std}(\mathbf{u}, \phi) + E_{reg}(\phi) $, where $  E_{std}(\mathbf{u}, \phi) $ is a standard term which is commonly used in damage mechanics and $ E_{reg}(\phi) $ is a regularization term. The purpose of the regularization term $ E_{reg}(\phi) $ in (\ref{engyFun1}) is to ensure that the phase-field variable $ \phi $ is smooth over the process zone adjacent to the crack and to act as a localisation limiter. For this reason, $ E_{reg}(\phi) $ is constructed to depend on the Cartesian norm of $ \nabla \phi $ and damage distributions with high gradients are thus disadvantageous for the considered physical system, as it strives to minimize its total energy. In $ E_{reg}(\phi) $, $ \ell $ is a characteristic phase field length scale that governs the size of the process zone \cite{Kristensen2020c}. From a numerical point of view, the implications of the non-locality of the model is that mesh objective results can be attained if the characteristic element length $h$ in the process zone is sufficiently small (5-7 times smaller than $\ell$ \cite{Kristensen2021}). The parameter $\mathcal{D}(1)$ [J m$^{-3}$] in $ E_{reg}(\phi) $ in (\ref{engyFun1}) is the density of energy dissipated by the complete failure process and it should be understood as a scaling parameter of $ E_{reg}(\phi) $. The dissipation density function $ \mathcal{D}(\phi) $ also determines the dissipation distance $ D(\phi_{n},\phi) $ (\ref{engyFun1a}) and quantifies the irreversible energy density consumed by the damage process. The infinity term in (\ref{engyFun1a}) enforces the
irreversibility of damage evolution, i.e., ensures that the phase-field variable cannot decrease in time. \\

According to the minimum total energy principle, we search for the global minimum of the functional $\Pi$, defined in (\ref{engyFun1}), if such a global minimum exists. From functional analysis, the optimality condition requires the first variation of $ \Pi(\mathbf{u},\phi) $ to be non-negative for all admissible variations $ \delta \mathbf{u} $ and $\delta \phi$, where
\begin{equation}\label{setVarDispl}
\delta\mathbf{u}(\mathbf{x},t) \in \mathbb{V} = \lbrace \forall t \geq 0: \delta\mathbf{u}(\mathbf{x},t) \in W^{1,2}(\Omega)^{d}; \delta\mathbf{u}(\mathbf{x},t) = \bm{0}\text{ in }\Gamma^{u} \rbrace
\end{equation} 
\begin{equation}\label{setVarPF}
\delta\phi(\mathbf{x},t) \in \mathbb{W} = \lbrace \forall t \geq 0: \delta\phi(\mathbf{x},t) \in W^{1,2}(\Omega^{c}) \rbrace 
\end{equation} 
The weak formulation of our problem of mechanical fracture thus reads
\begin{equation}\label{VarEngyFun1}
\begin{aligned}
\delta \Pi(\mathbf{u}, \phi, \delta \mathbf{u}, \delta \phi) &= 
\int_{V}g(\phi) (\bm{\nabla}_{s} \mathbf{u} - \bm{\varepsilon}_{\star}): \mathcal{\bm{C}}_{e}: \bm{\nabla}_{s} \delta \mathbf{u} \dd V \\
&+\frac{1}{2} \int_{V} \frac{\dd g(\phi)}{\dd \phi} \delta \phi (\bm{\nabla}_{s} \mathbf{u} - \bm{\varepsilon}_{\star}): \mathcal{\bm{C}}_{e}: (\bm{\nabla}_{s} \mathbf{u} - \bm{\varepsilon}_{\star}) \dd V \\
&+2\int_{V} \mathcal{D}(1) \ell^{2} \bm{\nabla} \phi \cdot \bm{\nabla} \delta \phi \dd V+\int_{V} \frac{\dd \mathcal{D}(\phi)}{\dd \phi} \delta \phi \dd V \\
&-\int_{V} \overline{\mathbf{b}} \cdot \delta \mathbf{u} \dd V - \int_{V} \overline{\mathbf{t}} \cdot \delta \mathbf{u} \dd V \geq 0 \,\,\,\,\,\,\,\,\, \forall \delta \mathbf{u} \in \mathbb{V}, \delta \phi \in \mathbb{W}
\end{aligned}
\end{equation}
It can be shown that if the solution ($\mathbf{u}$, $\phi$) that satisfies variational inequality (\ref{VarEngyFun1}) is sufficiently regular/smooth, then it satisfies the set of equations and inequalities in a strong form, which reads  
\begin{subequations}\label{govEqFr1}
\begin{align}
\bm{\nabla} \cdot\left( g(\phi) \mathcal{\bm{C}}_{e}: (\bm{\nabla}_{s} \mathbf{u} - \bm{\varepsilon}_{\star})\right)+\overline{\mathbf{b}} &=0 \,\,\,\,\,\,\, \text{ in } \,\, \Omega^{c}\\
-\frac{\dd g(\phi)}{\mathrm{~d} \phi} \frac{1}{2} (\bm{\nabla}_{s} \mathbf{u} - \bm{\varepsilon}_{\star}): \mathcal{\bm{C}}_{e}: (\bm{\nabla}_{s} \mathbf{u} - \bm{\varepsilon}_{\star})+2 \mathcal{D}(1) \ell^{2} \nabla^{2} \phi-\frac{\dd \mathcal{D}(\phi)}{\dd \phi} &=0 \,\,\,\,\,\,\, \text{ in } \,\,\Omega^{\phi}(t) \\
-\frac{\dd g(\phi)}{\mathrm{~d} \phi} \frac{1}{2} (\bm{\nabla}_{s} \mathbf{u} - \bm{\varepsilon}_{\star}): \mathcal{\bm{C}}_{e}: (\bm{\nabla}_{s} \mathbf{u} - \bm{\varepsilon}_{\star})+2 \mathcal{D}(1) \ell^{2} \nabla^{2} \phi-\frac{\dd \mathcal{D}(\phi)}{\dd \phi} & < 0 \,\,\,\,\,\,\, \text{ in } \,\, \Omega^{c} \backslash \Omega^{\phi}(t)
\end{align}
\end{subequations}
where $ \nabla ^{2} \phi $ is the Laplacian operator and $ \Omega^{\phi} $ is the active part of the damage zone, i.e., the part in which damage is growing, such that
\begin{equation}\label{setActPF}
\Omega^{\phi}(t) = \left\lbrace \mathbf{x} \in \Omega^{c}; \dfrac{\dd \phi(\mathbf{x},t)}{\dd t} > 0 \right\rbrace  
\end{equation}
The governing equations (\ref{govEqFr1}) are accompanied by the following boundary conditions
\begin{subequations}\label{govBC1}
\begin{align}
\mathbf{u} &= \overline{\mathbf{u}}\,\,\,\,\,\,\, \text{ in } \,\,\Gamma^{u} \\
\bm{\sigma} \cdot \mathbf{n} &= \overline{\mathbf{t}}\,\,\,\,\,\,\, \text{ in } \,\,\Gamma^{t} \\
\nabla \phi \cdot \mathbf{n} &= 0\,\,\,\,\,\,\, \text{ in } \,\,\Gamma \cup \Gamma^{s} \\
\nabla \phi \cdot \mathbf{n} &\geq 0\,\,\,\,\,\,\, \text{ in } \,\,\Gamma^{\phi} 
\end{align}
\end{subequations}
with $ \Gamma^{\phi} $ being the boundary of $ \Omega^{\phi} $.\\

The inequality (\ref{govEqFr1}c) can be reformulated as a variational equality to solve the system of equations without using `ad hoc' solvers \cite{Wu2017, Wu2018}. This is achieved by replacing the so-called crack driving force $ H = (\bm{\nabla}_{s} \mathbf{u} - \bm{\varepsilon}_{\star}): \mathcal{\bm{C}}_{e}: (\bm{\nabla}_{s} \mathbf{u} - \bm{\varepsilon}_{\star})/2 $ in (\ref{VarEngyFun1}) and (\ref{govEqFr1}) with 
\begin{equation}\label{historyFunc}
\mathcal{H}(t) = \displaystyle\max_{t\in\langle 0, T \rangle} \left(\widetilde{H}, H(t)\right)
\end{equation}
where the crack driving force history function $ \mathcal{H}(t) $ calculates the maximum value of the crack driving force $ H(t) $ that has been reached during the loading process, with $ \widetilde{H} $ being the threshold for damage nucleation. Eq. (\ref{historyFunc}) enforces damage irreversibility. Thus, the resulting weak formulation can be stated as   
\begin{equation}\label{VarEngyFun2}
\begin{aligned}
\delta \Pi(\mathbf{u}, \phi, \delta \mathbf{u}, \delta \phi) &= 
\int_{V} g(\phi) (\bm{\nabla}_{s} \mathbf{u} - \bm{\varepsilon}_{\star}): \mathcal{\bm{C}}_{e}: \bm{\nabla}_{s} \delta \mathbf{u} \dd V+\int_{V} \frac{\dd g(\phi)}{\dd \phi} \delta \phi \mathcal{H}(t) \dd V \\
&+\int_{V} 2 \mathcal{D}(1) \ell^{2} \bm{\nabla} \phi \cdot \bm{\nabla} \delta \phi \dd V+\int_{V} \frac{\dd \mathcal{D}(\phi)}{\dd \phi} \delta \phi \dd V \\
&-\int_{V} \overline{\mathbf{b}} \cdot \delta \mathbf{u} \dd V - \int_{V} \overline{\mathbf{t}} \cdot \delta \mathbf{u} \dd V = 0 \,\,\,\,\,\, ~ \forall \delta \mathbf{u} \in \mathbb{V}, \delta \phi \in \mathbb{W}
\end{aligned}
\end{equation}
with the corresponding strong formulation being
\begin{subequations}\label{govEqFr2}
\begin{align}
\bm{\nabla} \cdot\left(g(\phi) \mathcal{\bm{C}}_{e}: (\bm{\nabla}_{s} \mathbf{u} - \bm{\varepsilon}_{\star}\right)+\overline{\mathbf{b}} &=0 \,\,\,\,\,\,\, \text{ in } \,\, \Omega^{c}\\
-\frac{\dd g(\phi)}{\mathrm{~d} \phi} \mathcal{H}(t)+2 \mathcal{D}(1) \ell^{2} \nabla^{2} \phi-\frac{\dd \mathcal{D}(\phi)}{\dd \phi} &= 0 \,\,\,\,\,\,\, \text{ in } \,\, \Omega^{c}
\end{align}
\end{subequations} 
which is complemented by the boundary conditions (\ref{govBC1}a)-(\ref{govBC1}c). The presented scheme can be generalised to existing phase-field fracture models upon appropriate choices of dissipation density $ \mathcal{D} $, degradation function $ g(\phi) $ and crack driving force history function $ H(t) $.

\subsubsection{Particularisation to quasi-brittle fracture behaviour}

Of interest here is to particularise the generalised formulation presented in the previous section to a phase-field fracture model that mimics the quasi-brittle behaviour of concrete. As showcased in \ref{Sec:A0}, the so-called phase-field cohesive zone model (\texttt{PF-CZM}) by Wu and co-workers \cite{Wu2017,Wu2018} can capture the softening behaviour typical of concrete-like materials. Thus, appropriate constitutive choices are made to particularise the generalised formulation presented above to a \texttt{PF-CZM}-based model for precipitation-induced cracking. The dissipation function is chosen to be \cite{Wu2017}:
\begin{equation}\label{dissDenWu2}
\mathcal{D}(\phi) = \mathcal{D}(1) \alpha(\phi) =\mathcal{D}(1)\left(2 \phi-\phi^{2}\right)
\end{equation}
For a one-dimensional problem, the final distribution of $\phi$ in the completely cracked state can be solved analytically and its spatial distribution then determines the spatial distribution of dissipated energy densities $\mathcal{D}(\phi)$ and $ \varepsilon_{reg}(\phi)$ from (\ref{engyFun1}). The integral of $\mathcal{D}(\phi) + \varepsilon_{reg}(\phi)$ over the damage zone then determines the energy consumed by the complete failure process and could be thus interpreted as fracture energy $G_{f}$. Based on these ideas, the fracture energy could be expressed as
\begin{equation}\label{fracEnWu1}
G_{f} = \pi D(1) \ell
\end{equation}
The crack driving force $H(t)$ and its threshold $\widetilde{H}$ in (\ref{historyFunc}) are then defined as follows \cite{Wu2018} 
\begin{equation}\label{historyFunc1}
H(t)=\frac{\left\langle \bar{\sigma}_{1}\right\rangle^{2}}{2 \widetilde{E}}, \quad
\widetilde{H}=\frac{f^{2}_{t}}{2 \widetilde{E}}
\end{equation}
where $ f_{t} $ is the tensile strength and $ \bar{\sigma}_{1} $ is the maximum principal value of the effective stress tensor $\bar{\bm{\sigma}} = \mathcal{\bm{C}}_{e}:(\bm{\varepsilon} - \bm{\varepsilon}_{\star})$. The positive part function is defined as $ \langle x \rangle = (x + |x|)/2$ and $ \widetilde{E} = \lambda + 2\mu$ is the elongation modulus. Eq. (\ref{historyFunc1}) means that the onset of damage is controlled by the Rankine criterion based on the maximum principal stress. Here, one should note that before the onset of damage, the effective stress $\bar{\bm{\sigma}}$ is equal to the nominal stress $\bm{\sigma}$, since $g(0)=1$. Only tensile failure is considered here, hence the positive part operator. By taking the square of the stress divided by the elastic modulus and multiplied by 1/2, we convert the stress into the density of elastically stored energy, which is the original definition of the crack driving force. However, this equivalence holds only under uniaxial tension, if we use the Young modulus in this conversion. \citet{Wu2018} used the elongation (oedometric) modulus, and so the equivalence would hold under uniaxial strain.

Finally, the degradation function $g(\phi)$ is chosen in such a way so as to match the required initial slope of the softening curve of the considered stress-strain response and the required mode-I crack opening at complete failure, the latter resulting from a well-defined cohesive zone model with the chosen softening curve \cite{Wu2017,Lorentz2011}. This considers fracture as the process of separation of the material in the cohesive zone located at the crack tip, with the crack opening resisted by cohesive tractions. Based on these considerations, \citet{Wu2017} set 
\begin{equation}\label{damFunWu1}
g(\phi)=\frac{(1-\phi)^{p}}{(1-\phi)^{p}+a_{1} \phi (1+a_{2} \phi+a_{3}  \phi^{2})}
\end{equation}
in which parameters $ p \geq 2 $, $ a_{1} > 0 $, $ a_{2} $ and $ a_{3} $ allow calibration of the model. Substituting (\ref{dissDenWu2})--(\ref{damFunWu1}) into (\ref{govEqFr2}), the governing equations of the deformation-fracture problem read
\begin{subequations}\label{govEqFr3}
\begin{align}
\bm{\nabla} \cdot\left(g(\phi) \mathcal{\bm{C}}_{e}: (\bm{\nabla}_{s} \mathbf{u} - \bm{\varepsilon}_{\star}\right)+\overline{\mathbf{b}} &=0 \,\,\,\,\,\,\, \text{ in } \,\, \Omega^{c}\\
-\dfrac{1}{2}\frac{\dd g(\phi)}{\mathrm{~d} \phi} \mathcal{H}(t)+\dfrac{\ell}{\pi} G_{f} \nabla^{2} \phi-\dfrac{G_{f}}{\pi \ell}(1-\phi) &=0 \,\,\,\,\,\,\, \text{ in } \,\, \Omega^{c}
\end{align}
\end{subequations} 
The governing equations are accompanied by the set of boundary conditions (\ref{govBC1}a)-(\ref{govBC1}c). It remains to calibrate $g(\phi)$ by appropriately choosing $ p $, $ a_{1} $, $ a_{2} $ and $ a_{3} $ in (\ref{damFunWu1}). For example, a particular choice allows to recover the brittle phase-field fracture model of \citet{Francfort2008}. Other choices lead to linear, hyperbolic or exponential softening laws \cite{Wu2017}. For a quasi-brittle concrete considered in this model, commonly used softening curve by \citet{Cornelissen1962} is employed. \citet{Wu2017} showed that the Hordijk-Cornelissen softening curve can be approximated with the following choice of the parameters of the degradation function $g(\phi)$:  
\begin{equation}\label{a1a2a3}
a_{1}=\frac{4}{\pi} \frac{\ell_{irw}}{\ell}, \quad a_{2}=2 \beta_{k}^{2 / 3}-p-\frac{1}{2}, \quad a_{3}= \begin{cases} 1/2 \beta_{w}^{2}-a_{2}-1 & \text{if} \quad p=2 \\ 0 & \text{if} \quad p>2  \end{cases}
\end{equation}    
In (\ref{a1a2a3}), $ \ell_{irw} = \widetilde{E}G_{f}/f^{2}_{t} $ is the Irwin internal length. The constants $ \beta_{w} $ and $ \beta_{k} $ are given by
\begin{equation}  
\beta_{w} = \dfrac{w_{c}}{w_{c,lin}}, \quad w_{c,lin} = \dfrac{2G_{f}}{f_{t}}
\end{equation}
\begin{equation}
\beta_{k} = \dfrac{k_{0}}{k_{0,lin}} \geq 1, \quad k_{0,lin} = -\frac{f_{t}^{2}}{2G_{f}} 
\end{equation}
where $ w_{c} $ is the limit crack opening given by the chosen softening curve and $k_{0}$ is the initial slope of the selected softening curve. Ratios $ \beta_{w} $ and $\beta_{k} $ then compare $ w_{c} $ and $k_{0}$ with the values of the parameters governing the linear softening curve, $ w_{c,lin} $ and $k_{0,lin}$, respectively. Finally, the Hordijk-Cornelissen softening curve is recovered by setting 
\begin{equation}\label{CornelissenSoft}
p = 2, \quad w_{c}=5.1361 \frac{G_{f}}{f_{t}}, \quad k_{0}=-1.3546 \frac{f_{t}^{2}}{G_{f}}
\end{equation}     
\noindent These choices deliver a length-insensitive phase-field fracture model, provided that the phase field length scale chosen is sufficiently small so as to ensure that $\ell \leq \text{min} \left( 8 \ell_{irw} / 3 \pi, \, L/100 \sim L/50 \right)$, where $L$ is the characteristic length of the structure in the sense of its dominant geometric proportions. 

\subsubsection{Damage--dependent diffusivity tensor}

The role of fracture networks in enhancing the transport of ionic species is captured by means of a second-order diffusivity tensor $ \boldsymbol{D}_{\alpha} $, see Eq. (\ref{total_mass_derivative_ionic_N_2}). Cracks filled with pore solution facilitate diffusion relative to the surrounding concrete matrix and thus become preferable transport pathways for dissolved iron species. This directly influences the distribution of precipitates because their accumulation is locally slowed down if iron species can escape through adjacent micro-cracks or cracks. Inspired by the work of \citet{Wu2016}, we calculate the product of liquid volume fraction and diffusivity tensor in (\ref{total_mass_derivative_ionic_N_2}) as 
\begin{equation}\label{diffusivity_tensor_2}
\theta_{l}\boldsymbol{D}_{\alpha}=\theta_{l}(1-\phi)D_{m,\alpha}\boldsymbol{1} + \phi D_{c,\alpha} \boldsymbol{1}, \quad \alpha = II, III
\end{equation}    
where $ D_{m,\alpha} $ is an isotropic diffusivity of the considered species in concrete. The parameter $D_{c,\alpha}$ controls the anisotropic diffusivity of the cracked material \citep{Wu2016}, and generally $ D_{c} \gg D_{m} $. It should be noted that the original formulation of \citet{Wu2016} leads to an anisotropic diffusivity tensor because the diffusivity is enhanced only in the direction of the crack, i.e. perpendicular to the gradient of $\phi$. However, in the proposed model, an isotropic diffusivity enhancement is considered because it was found to be more numerically robust and to better reflect the enhanced diffusivity of damaged concrete adjacent to steel rebar before the full localisation of cracks. 

\subsection{Overview of the governing equations}
\label{Sec:govEq}

We conclude the description of our theory by recalling its governing equations:

\begin{subequations}\label{summGovEq}
\begin{align}
\bm{\nabla} \cdot\left(g(\phi) \mathcal{\bm{C}}_{e}: (\bm{\nabla}_{s} \mathbf{u} - \bm{\varepsilon}_{\star}\right)+\overline{\mathbf{b}} &=0 \,\,\,\,\,\, \text{ in } \Omega^{c}\\
-\frac{\dd g(\phi)}{\mathrm{~d} \phi} \mathcal{H}(t)+\dfrac{2\ell}{\pi} G_{f} \nabla^{2} \phi-\dfrac{G_{f}}{\pi \ell}\frac{\dd \alpha(\phi)}{\dd \phi} &=0  \,\,\,\,\,\, \text{ in } \Omega^{c} \\
\frac{\partial \left(\theta_{l}c_{II}\right)}{\partial t} - \bm{\nabla} \cdot \left(\theta_{l}\bm{D}_{II}\cdot\nabla c_{II} \right) &=  \theta_{l} R_{II} \,\,\,\,\,\, \text{ in } \Omega^{c} \\
\frac{\partial \left(\theta_{l}c_{III}\right)}{\partial t} - \bm{\nabla} \cdot \left(\theta_{l}\bm{D}_{III}\cdot\nabla c_{III} \right) &=  \theta_{l} R_{III} \,\,\,\,\,\, \text{ in } \Omega^{c} \\ \label{summGovEq5}
\frac{\partial \theta_{p}}{\partial t} &=  \dfrac{M_{p}}{\rho_{p}} \theta_{l} R_{p} \,\,\,\,\,\, \text{ in } \Omega^{c}
\end{align}
\end{subequations}
for the five primary field variables -- displacement vector $\mathbf{u}$, phase-field variable $\phi$, $\mathrm{Fe}^{2+}$ ions concentration $c_{II}$, $\mathrm{Fe}^{3+}$ ions concentration $c_{III}$, and precipitate volume fraction $\theta_{p}$. The governing equations (\ref{summGovEq}) are accompanied by boundary conditions
\begin{subequations}\label{summGovBC}
\begin{align*} 
\tag{52a}
\mathbf{u} &= \overline{\mathbf{u}} \,\,\,\,\,\,\, \text{ in } \,\,\Gamma^{u} \\ 
\tag{52b}
\nabla \phi \cdot \mathbf{n} &= 0 \,\,\,\,\,\,\, \text{ in } \,\,\Gamma \cup \Gamma^{s}  \\
\tag{52c}
\mathbf{n} \cdot \left(\boldsymbol{D}_{II} \cdot \nabla c_{II} \right) &= 0\,\,\,\,\,\,\, \text{ in } \,\,\Gamma, \hspace{0.5cm} \hspace{0.5cm} \mathbf{n} \cdot \left(\boldsymbol{D}_{II} \cdot \nabla c_{II} \right) = \frac{2i_{a}}{z_{a}F}\,\,\,\,\,\,\, \text{ in } \,\,\Gamma^{s} \\ 
\tag{52d}
\mathbf{n} \cdot \left(\boldsymbol{D}_{III} \cdot \nabla c_{III} \right) &= 0\,\,\,\,\,\,\, \text{ in } \,\,\Gamma\cup\Gamma^{s}  \\ 
\end{align*}
\end{subequations}
Equation (\ref{summGovEq5}) does not require any boundary condition because it does not contain any space derivatives. We solve the resulting system of equations using the finite element method and discretising the domain $\Omega^{c} \cup \Omega^{s}$ with linear triangular elements. The phase-field variable is not solved in the steel domain $\Omega^{s}$ and steel is assumed to be linear elastic. A staggered solution scheme is used \cite{Zhou2018}. The model is numerically implemented using the finite element package COMSOL Multiphysics \footnote{The COMSOL model developed is made freely available at \url{https://www.imperial.ac.uk/mechanics-materials/codes}}.

\section{Results}
\label{Sec:Results}

The ability of the theoretical and computational framework presented to capture experimental data and simulate corrosion-induced cracking in complex scenarios is showcased by means of representative examples. First, the impressed current tests of \citet{Pedrosa2017} are modelled to benchmark surface crack width predictions against their measurements (Section \ref{Sec:Andrade}). Then, in Section \ref{Sec:ResultsDelamSpalling}, the analysis is extended to boundary value problems involving multiple rebars, to assess the model capabilities in capturing crack interaction and the associated gradual delamination/spalling of the concrete cover. Finally, a three-dimensional boundary value problem is simulated in Section \ref{Sec:ResultsThreeDim}, to showcase the ability of the framework to address large-scale studies of technological importance. 

\subsection{Model validation against the experiments by \citet{Pedrosa2017} and analysis}
\label{Sec:Andrade}

\subsubsection{Choice of model parameters}
\label{Sec:modelParam}

\begin{table}[htb!]
\begin{small}
\begin{longtable}[ht]{p{5cm} p{2.5cm} p{2cm} p{1.5cm}}\toprule
\multicolumn{4}{ c }{\textbf{Properties of concrete - 28 and 147 days cured samples}} \\
\toprule
\textbf{Parameter} & \textbf{Value} & \textbf{Unit} & \textbf{Source}\\
\toprule
\toprule
Compressive strength $f_{c,cube}$ & 37.5 \& 54.7 & MPa & \cite{Pedrosa2017}\\
\midrule
Tensile strength $f_{t}$ & 2.2 \& 3.9 & MPa &  \cite{Pedrosa2017} \\
\midrule
Young's modulus $E_{c}$ & 33 \& 36 & GPa & \cite{standard2004eurocode}  \\
\midrule
Poisson's ratio $\nu_{c}$ & 0.2 \& 0.2 & - & \cite{standard2004eurocode}  \\
\midrule
Fracture energy $G_{f}$ & 95 \& 114 & N m$^{-1}$ & \cite{Bazant2002} \\
\bottomrule
\caption{Model parameters: mechanical properties of concrete, based on the measurements by \citet{Pedrosa2017} and literature data.}  
\label{tab:tableMechRust1}
\end{longtable}
\end{small}
\end{table}

The modelling framework takes as input parameters that have a physical basis and can be independently measured. Thus, when possible, the magnitudes of the model parameters were taken from the measurements by \citet{Pedrosa2017}. For those properties for which measurements were not conducted, parameter values are taken from the literature. The mechanical properties of concrete are listed in Table \ref{tab:tableMechRust1}, for samples cured for both 28 and 147 days. The compressive strength is taken from the cube samples measurements by \citet{Pedrosa2017} and the fracture energy is determined assuming crushed aggregates in the formula of \citet{Bazant2002}. 

\begin{table}[htb!]
\begin{small}
\begin{longtable}{p{8cm} p{3cm} p{2cm} p{1.5cm}}
\toprule
\textbf{Parameter} & \textbf{Value} & \textbf{Unit} & \textbf{Source} \\
\toprule
\multicolumn{4}{ c }{\textbf{Properties of rust ($\bm{\mathrm{FeO(OH) + H_{2}O}}$)}} \\
\toprule
\toprule
Young's modulus $E_{p}$ & $ 440  $ & MPa & \cite{ZHAO201619} \\
\midrule
Poisson's ratio $\nu_{p}$ & $ 0.4  $ & - &  \cite{ZHAO201619} \\
\midrule
Porosity $r_{0}$ & $ 0.16 $ & - & \cite{Ansari2019} \\
\midrule
Molar mass of rust $M_{p}$ & $ 106.85  $ & g mol$^{-1}$ & \\
\midrule
Density of rust $\rho_{p}$ & $ 3560  $ & kg m$^{-3}$ & \cite{ZHAO201619} \\
\toprule
\multicolumn{4}{ c }{\textbf{Transport properties of concrete}} \\
\toprule
\toprule
Porosity $p_{0}$ & $ 0.26 $ & - & \cite{Powers1946} \\
\midrule
Initial concrete diffusivity $\theta_{l}D_{m,II}$ and $\theta_{l}D_{m,III}$ & $ 10^{-11} $ & m$^{2}$ s$^{-1}$ & \cite{stefanoni_kinetic_2018}  \\
\midrule
Cracked concrete diffusivity $ D_{c,II} $ and $ D_{c,III} $ & $ 7\cdot10^{-10}  $ & m$^{2}$ s$^{-1}$ &  \cite{stefanoni_kinetic_2018,Leupin2021} \\
\toprule
\multicolumn{4}{ c }{\textbf{Other chemical properties}} \\
\toprule
\toprule
Rate constant $k^{II \rightarrow III}_{r}$ & $ 0.1 $ & mol$^{-1}$m$^{3}$s$^{-1}$ & \cite{stefanoni_kinetic_2018}  \\
\midrule
Rate constant $k^{III \rightarrow p}_{r}$ & $ 2 \cdot 10^{-4}  $ & s$^{-1}$ & \cite{Leupin2021} \\
\midrule
Oxygen concentration $c_{ox}$ & $ 0.28  $ & mol m$^{-3}$ &  \cite{Zhang2021} \\
\bottomrule
\caption{Model parameters: properties of rust, transport properties of concrete and other relevant chemical properties.}
\label{tab:tableMechRust2}
\end{longtable}
\end{small}
\end{table}

The remaining model parameters are given in Table \ref{tab:tableMechRust2}. These are the properties of rust, the transport properties of concrete and other relevant chemical properties. There is some scatter within the mechanical properties of rust reported in the literature. Two intermediate values within the range reported by \citet{ZHAO201619} are assumed for the Young's modulus and Poisson's ratio of rust. The molar mass of rust is estimated as the molar mass of $\mathrm{FeO(OH) + H_{2}O}$, the rust density is taken as the density of $\beta$-FeO(OH) (Akaganeite), a commonly found rust in chloride contaminated concrete, and the rust porosity $r_0$ is taken to be an intermediate value from those reported by \citet{Ansari2019}. 

In regards to the transport properties of concrete, the initial (undamaged) concrete diffusivity ($\theta_{l}D_{m,II}$ and $\theta_{l}D_{m,III}$) is estimated from the known diffusivity of the species in water, assuming that the ratio between the diffusivity in water and concrete is approximately the same as for chloride ions. On the other side, the transport properties through cracks ($ D_{c,II} $ and $ D_{c,III} $) are estimated as the diffusivity in pore solution, assuming that cracks are filled with water. The porosity of concrete is estimated from the seminal work of \citet{Powers1946}, assuming that only the porosity of the cement paste is relevant in the close vicinity of steel rebars and a degree of hydration of 0.9, as the samples were well-cured and had high water to cement ratio \cite{Pedrosa2017}. Importantly, it is well-known that porosity at the steel-concrete interface (SCI) could be significantly larger than in the bulk concrete, providing considerable space for rust accumulation. In order to reflect this, a 0.2 mm layer around the rebar with a porosity twice as high as the rest of the sample is considered (i.e., $p_0=0.52$ within the SCI)\footnote{The precipitate saturation ratio in the SCI is calculated with respect to the basic bulk porosity}. Within sensible values \cite{Angst2017}, the thickness of SCI has been found to have a very small impact on the resulting surface crack width (see Fig. \ref{FigSweepSCIthc} in \ref{Sec:C}). The remaining chemical properties are taken from the literature or as an intermediate value of reported values. It should also be noted that the molar mass and the intrinsic density of $\mathrm{Fe}^{3+}$ are considered to be the same as for iron. For the steel rebar, a Young's modulus of 205 GPa and a Poisson's ratio of 0.28 are assumed, as these are common values for steel.\\

Triangular finite elements are used to discretise the concrete domain, with the mesh being finer in the region close to the rebar and, particularly, within the SCI. The maximum element size in the SCI is 0.1 mm and 0.6 mm in the remaining regions of interest. Since the phase field length scale is chosen to be $\ell = 3$ mm, five times larger than the maximum element size in the characteristic region, mesh objective results are expected \cite{Kristensen2021}. The boundary conditions are given in Fig. \ref{FigExpTestSetup}.

\FloatBarrier
\subsubsection{Validation of the model - the impressed current test of \citet{Pedrosa2017}}
\label{Sec:ResultsValidationAndrade}
\begin{figure}[!htb]
\begin{center}
\includegraphics[width=0.8\textwidth]{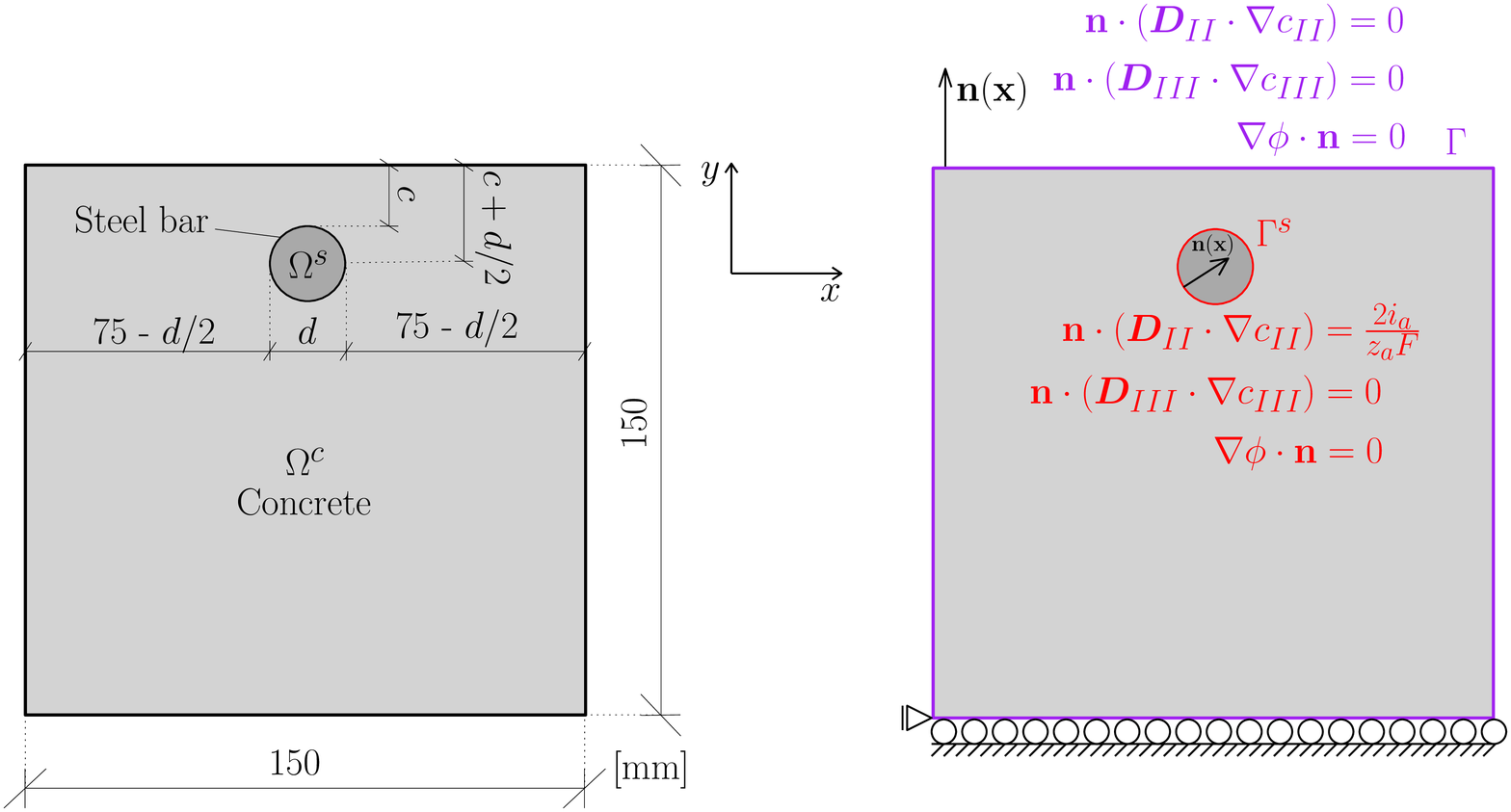}
\caption{Geometry and boundary conditions of the cross-section of concrete samples used for the simulation of the impressed current tests by \citet{Pedrosa2017}. Dimensions are given in mm.}
\label{FigExpTestSetup}
\end{center} 
\end{figure}

To validate the proposed model, corrosion-induced cracking simulations are conducted mimicking the conditions of the impressed current tests by \citet{Pedrosa2017}. The tested samples were 500 mm long concrete prisms, longitudinally reinforced with a single 600 mm long steel bar with a diameter of 16 mm. The concrete cover of the steel rebar was either 20 or 40 mm thick. Due to the relatively uniform distribution of corrosion current density along the steel surface, the simulated domain of the concrete samples is reduced to the two-dimensional cross-section depicted in Fig. \ref{FigExpTestSetup}. On such prepared samples, various corrosion current densities were applied for a certain period. In this study, we restrict our attention to the tests with applied corrosion current density lower than 10 \unit{\micro\ampere\per\centi\metre^2}, such that the assumption of neglected precipitation of $\mathrm{Fe}^{2+}$ is not violated. The three cases fulfilling this condition, henceforth referred to as tests 1--3, are characterised by the following values of corrosion current density $i_{a}$, the thickness of concrete cover $c$ and sample curing times: 
\begin{enumerate}
\item $i_{a} = 5$ \unit{\micro\ampere\per\centi\metre^2}, $c = 20$ mm, 28-day cured 
\item $i_{a} = 10$ \unit{\micro\ampere\per\centi\metre^2}, $c = 20$ mm, 147-day cured 
\item $i_{a} = 10$ \unit{\micro\ampere\per\centi\metre^2}, $c = 40$ mm, 28-day cured
\end{enumerate}

Relevant to the estimation of $G_f$ and $p_0$, \citet{Pedrosa2017} reported that the used concrete mixture had a maximum aggregate size of 12 mm and a water to cement ratio of 0.55. To cause immediate corrosion of the steel rebar, a high content of chlorides in the form of $ \mathrm{CaCl_{2}}\cdot2\mathrm{H_{2}O} $ was added to the mixture such that a 3$\%$ mass ratio of chlorides to cement was achieved. Under such a high chloride concentration, relatively uniform corrosion is expected.

\begin{figure}[!htb]
\centering
\includegraphics[scale=0.6]{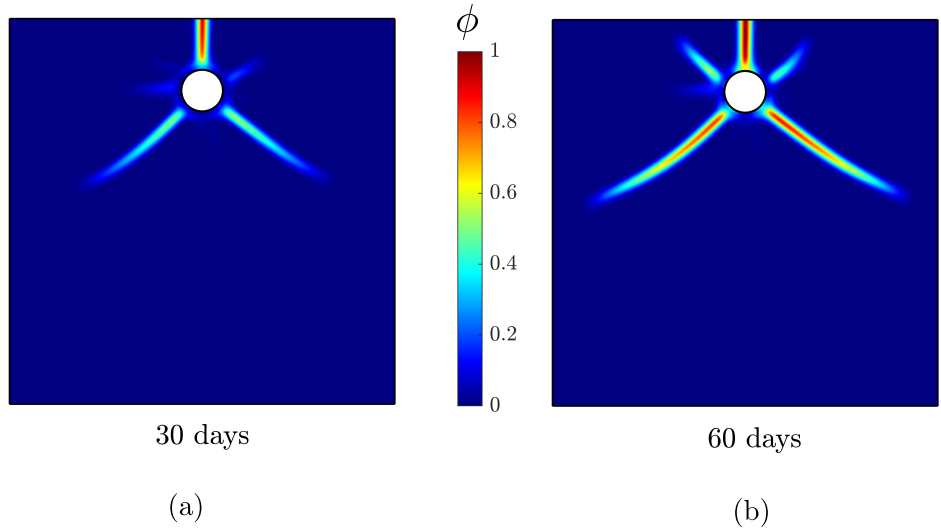}
\caption{Nucleation and growth of cracks as characterised by the contours of the phase field variable $\phi$. Results obtained for the conditions of test 2 ($i_{a} = 10$ \unit{\micro\ampere\per\centi\metre^2} and $c = 20$ mm) after (a) 30, and (b) 60 days.}
\label{Fig:PFcontours1}
\end{figure}

The cast samples were separated into two batches, which were subjected to curing for different periods of time, specifically 28 and 147 days, resulting in different mechanical properties (see Table \ref{tab:tableMechRust1}). The impressed current tests were carried out for 279 days for tests 1 and 3 and for 359 days for test 2. Here, the focus is on early-stage corrosion-induced cracking (assumed to last at least for the first 60 days), where the assumptions inherent to the model hold. Over this period, and for the corrosion current densities explored, the change in shape and size of the steel rebar can be neglected. Also, the samples were kept wet throughout the experiment. During the tests, the length of the crack opening at the upper concrete surface ($w$) was monitored. The crack width $w$ was measured in test 2 using a digital calliper while in tests 1 and 3 three polyester wire strain gauges were placed on the upper face of the concrete samples. Since the crack is perpendicular to the upper concrete surface, the crack width is estimated in the model as an integral over the x-component of the inelastic strain tensor $ \bm{\varepsilon}_{d}  = \bm{\varepsilon} - \bm{\varepsilon}_{e} - \bm{\varepsilon}_{\star} $ over the upper concrete surface $ \Gamma^{us} $ \cite{Navidtehrani2022}, such that 
\begin{equation}
w = \int_{\Gamma^{us}} (\bm{\varepsilon}_{d})_{x} \dd \Gamma =  \int_{\Gamma^{us}} (1-g(\phi))(\bm{\varepsilon}_{x}-(\bm{\varepsilon}_{\star})_{x}) \dd \Gamma
\end{equation}   
We proceed to present the finite element predictions. First, as shown in Fig. \ref{Fig:PFcontours1}, the model captures a key qualitative element of the experiments: the nucleation and growth of a vertical crack that connects the steel rebar and the closest (upper) concrete surface, followed by the formation of lateral cracks nucleating in the vicinity of the steel surface. The reasons for the slight offset of cracks from the rebar is explained in \ref{Sec:D}.

\begin{figure}[H]
     \centering
     \begin{subfigure}[!htb]{0.49\textwidth}
         \centering
         \includegraphics[width=\textwidth]{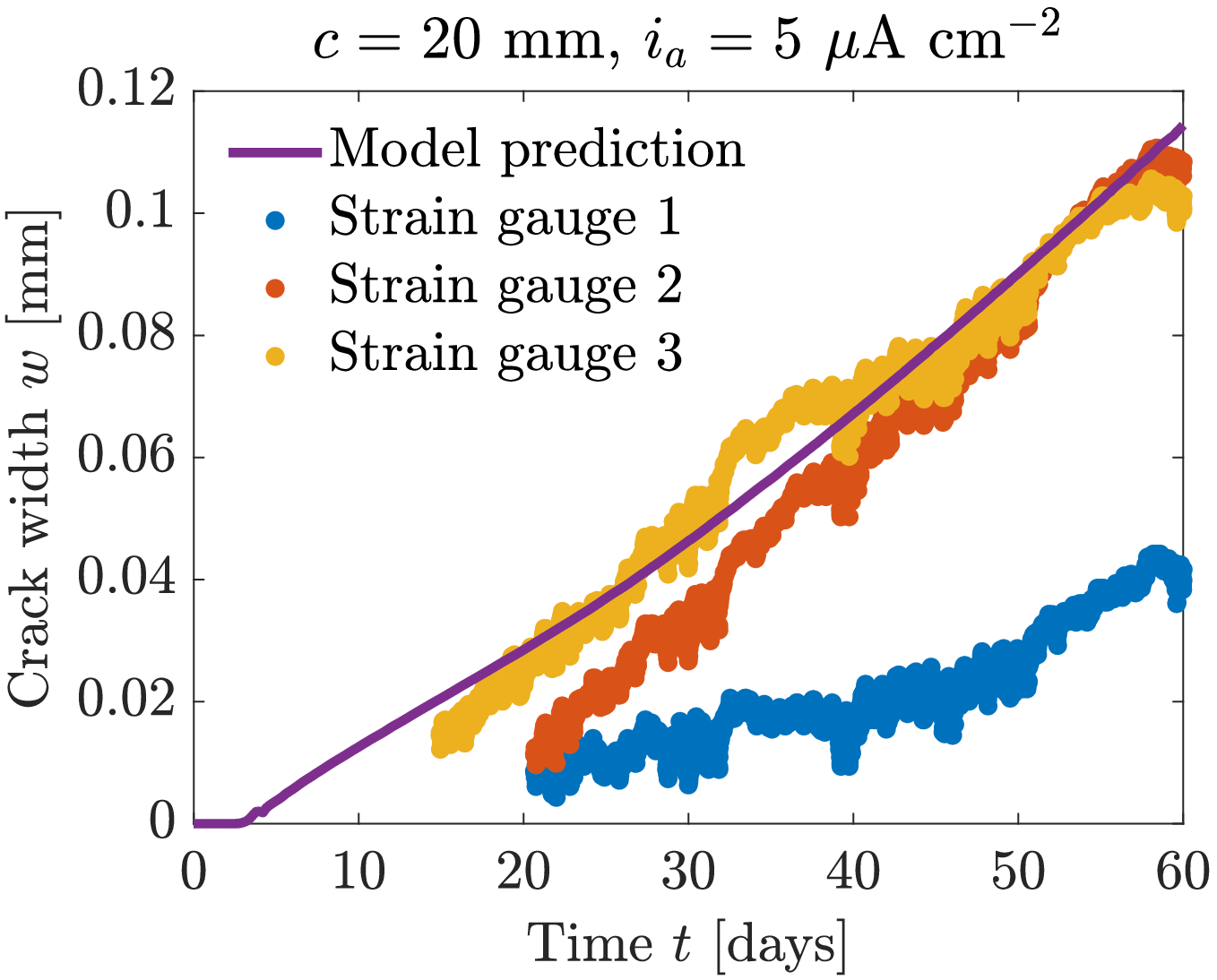}
         \caption{}
         \label{FitC20I5}
     \end{subfigure}
     \hfill
     \begin{subfigure}[!htb]{0.49\textwidth}
         \centering
         \includegraphics[width=\textwidth]{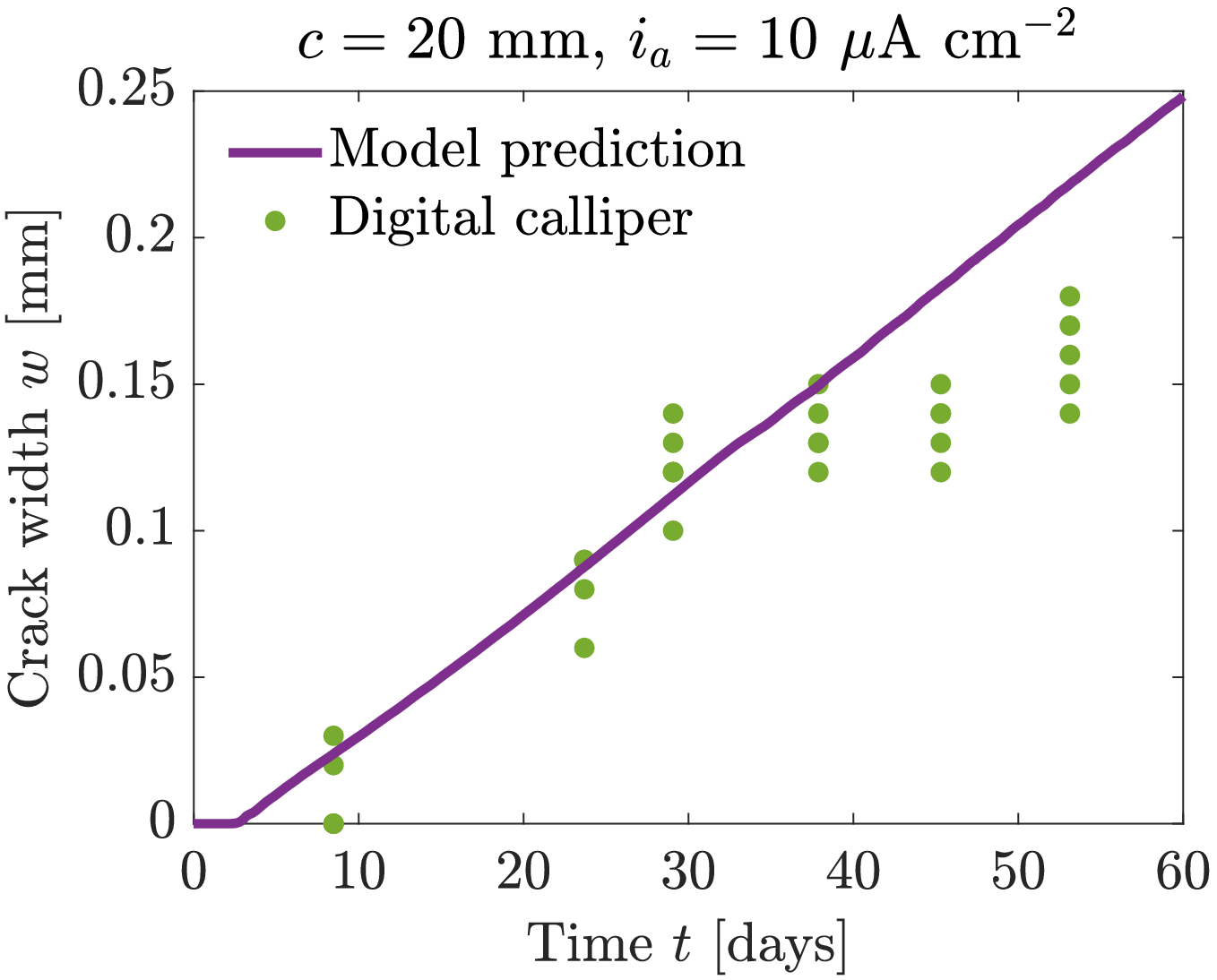}
         \caption{}
         \label{FitC20I10}
     \end{subfigure}
     \hfill
     \begin{subfigure}[!htb]{0.49\textwidth}
         \centering
         \includegraphics[width=\textwidth]{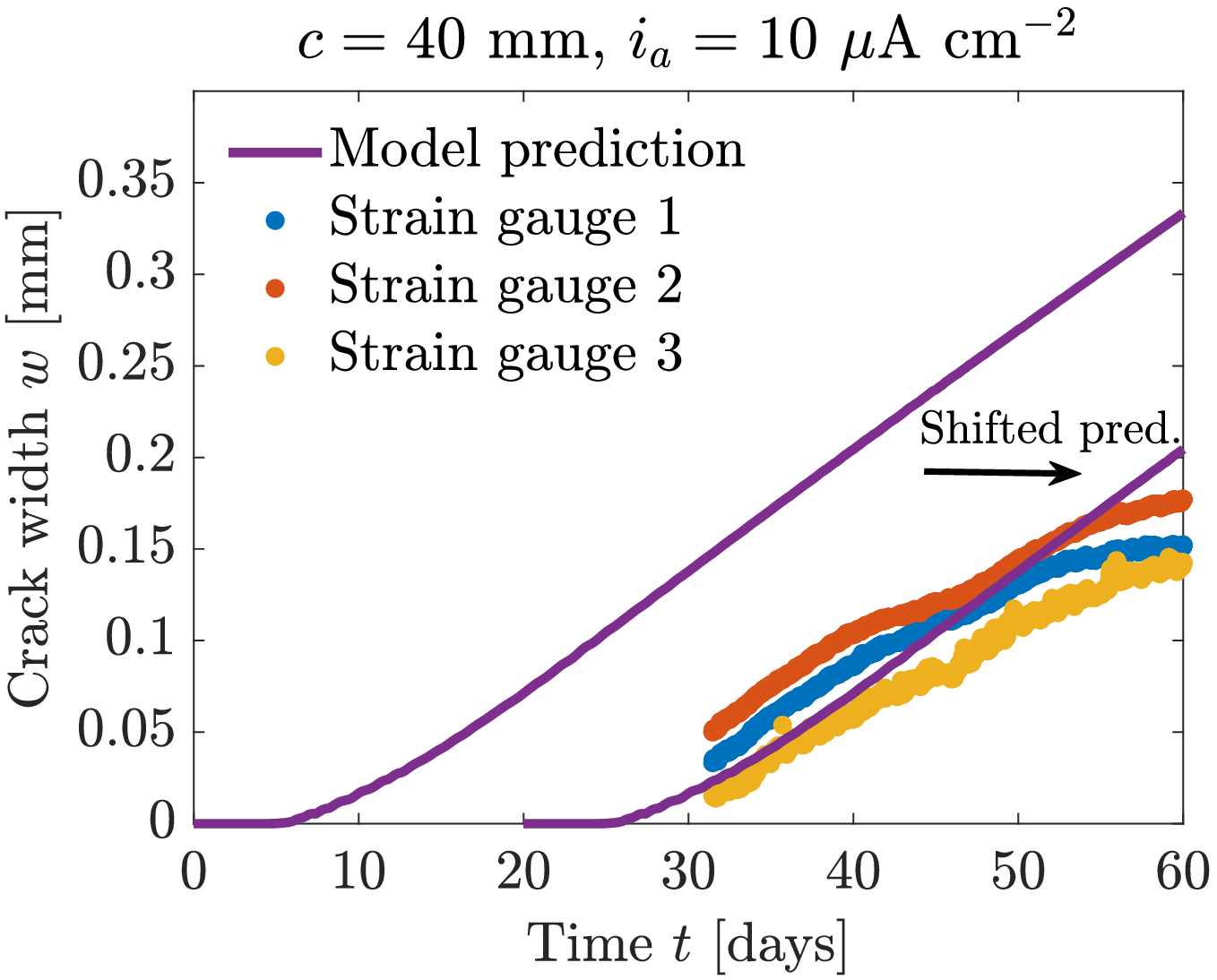}
         \caption{}
         \label{FitC40I10}
     \end{subfigure}
        \caption{Comparison of the evolution of simulated and experimentally-measured surface crack width for impressed current tests by \citet{Pedrosa2017} -- (a) test 1 ($i_{a} = 5$ \unit{\micro\ampere\per\centi\metre^2}, $c = 20$ mm), (b) test 2 ($i_{a} = 10$ \unit{\micro\ampere\per\centi\metre^2}, $c = 20$ mm), (c) test 3 ($i_{a} = 10$ \unit{\micro\ampere\per\centi\metre^2}, $c = 40$ mm). A good agreement with experiments is attained but, for test 3, this requires shifting the prediction in time. Such a shift could be due to the role that aggregates play in initiating secondary cracks that delay the growth of the monitored crack.}
\label{fig:three graphs}
\end{figure}

Second, quantitative predictions of crack width $w$ versus time are given in Fig. \ref{fig:three graphs}. Numerical predictions show a good agreement with experimental measurements in tests 1 and 2 (Figs. \ref{fig:three graphs}a and \ref{fig:three graphs}b, respectively), despite the differences in current density and sample curing times. This agreement is notable because, as discussed in Section \ref{Sec:ResultsParamStudy}, crack width estimates are very sensitive to the current density, which comes into play via Eq. (\ref{new_FarLaw}). In the case of test 3, the appropriate scaling (curve slope) is captured but quantitative agreement requires shifting the simulation result. This indicates that the model is able to accurately capture the growth of the main crack, once it has initiated, but that it fails in capturing the crack nucleation event for that particular experiment. This is not entirely surprising as the onset of crack growth in a heterogeneous material like concrete can be of stochastic nature. E.g., material heterogeneities such as aggregates could have triggered the nucleation of secondary cracks that relax the stress state and delay the formation of the measured crack. 
    
It should me emphasized here that the predicted and experimentally measured crack widths have been compared only for the first 60 days. However, \citet{Pedrosa2017} found out that in nearly all impressed current tests they performed, the rate of growth of surface crack width dropped significantly at a certain time after the initial stage of approximately linear evolution. For this reason, \citet{Pedrosa2017} characterised the surface crack width evolution as a two-stage process. The computational results indicate that the pore space is not entirely filled with precipitates during the investigated period, suggesting the validity of the assumptions of the model in the investigated regime. Thus, we hypothesise that the early stage is encompassed in the first stage described by \citet{Pedrosa2017}, which is also strengthened by the good agreement between the experimental and predicted surface crack width. We also hypothesise that the change in the slope of the crack width in time is related to the filling of the local porosity by accumulated rust.  It is possible that as iron ions can no longer escape into the porosity (with the exception of cracks), they accumulate in the layer in the corroded regions of the steel rebar. The change of the mechanism generating corrosion-induced pressure may result in the change of the slope of the crack width curve in time. An accurate prediction of the second phase described by \citet{Pedrosa2017} will be an objective of future studies.

\subsubsection{General aspects of the simulation results}
\label{sec:ResultsGeneralApects}

\begin{figure}[!htb]
    \centering
    \begin{subfigure}[!htb]{0.49\textwidth}
    \centering
    \includegraphics[width=\textwidth]{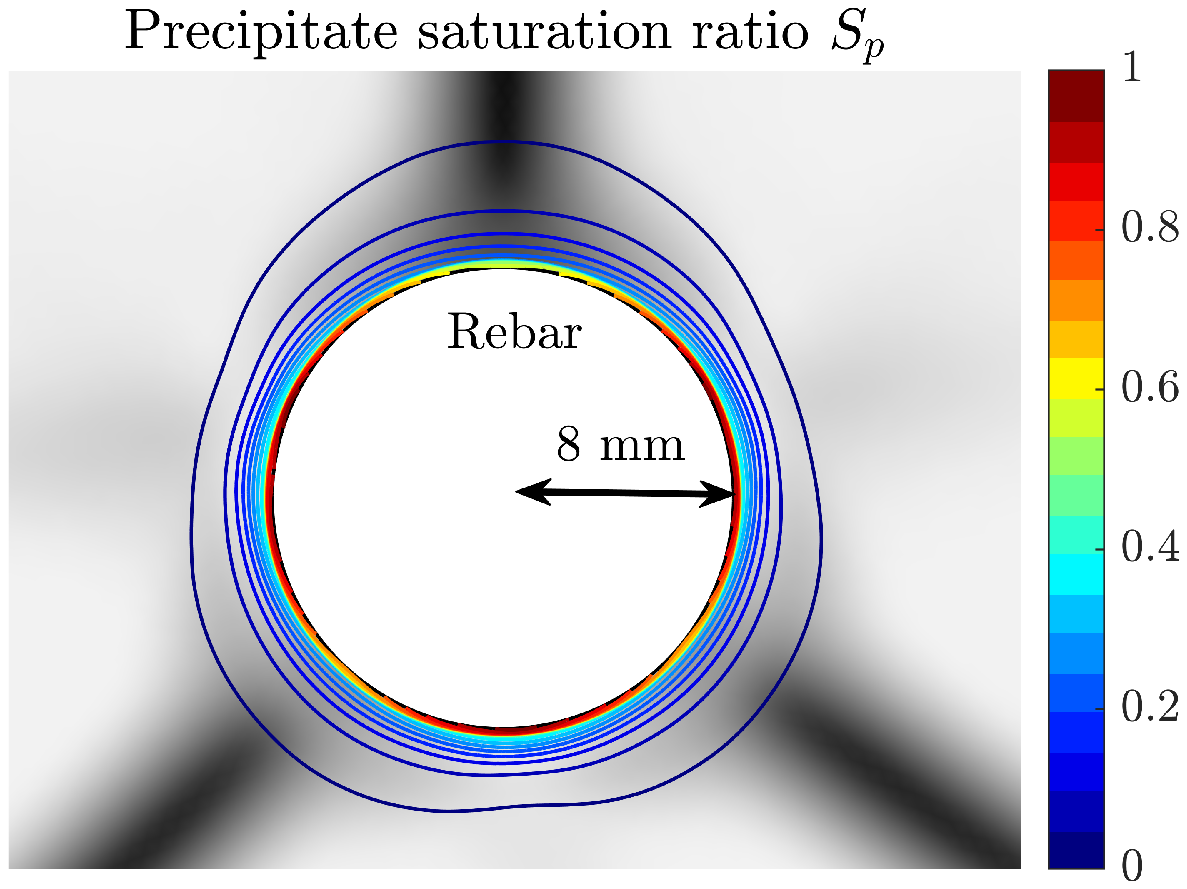}
    \caption{}
    \label{FigSpcont60d}    
    \end{subfigure}
    \hfill
    \begin{subfigure}[!htb]{0.49\textwidth}
    \centering
    \includegraphics[width=\textwidth]{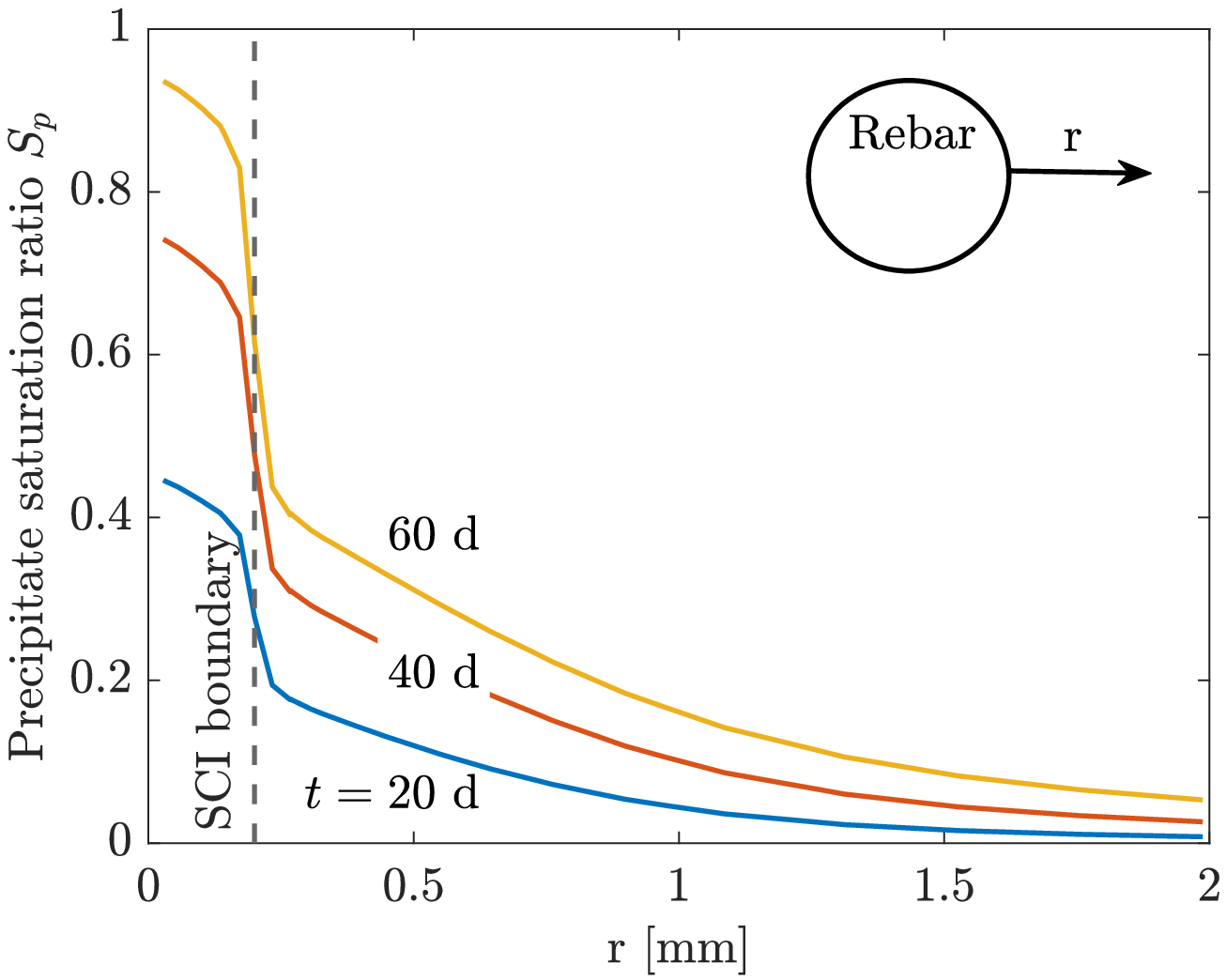}
    \caption{}
    \label{FigSectionSp} 
    \end{subfigure} 
\hfill
    \begin{subfigure}[!htb]{0.49\textwidth}
    \centering
    \includegraphics[width=\textwidth]{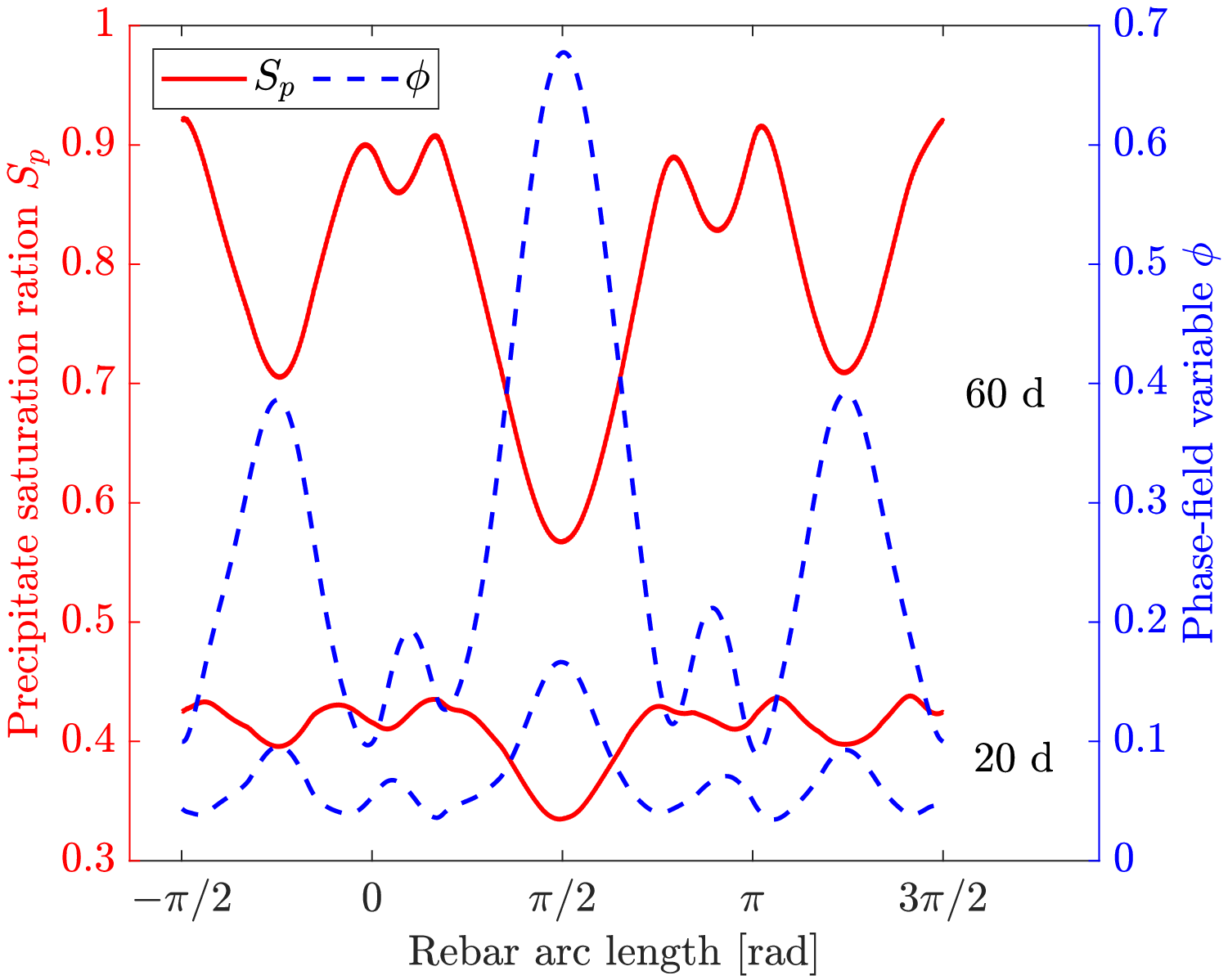}
    \caption{}
    \label{FigAlongRebarPFvsSp}    
    \end{subfigure} 
\caption{Precipitate saturation ratio $S_{p}$ for test 2 ($i_{a} = 10$ \unit{\micro\ampere\per\centi\metre^2}, $c = 20$ mm) -- (a) contours of $S_{p}$ in the vicinity of steel rebar at 60 days, with the phase-field variable $\phi$ shown in the back in a grey scale bar (0 -- white, 1 -- black); (b) evolution in time of $S_{p}$ in a radial direction from the rebar ($r, \theta=0^\circ$), $S_{p}$ increases rapidly in the vicinity of SCI boundary, due to the significantly higher porosity and thus diffusivity of $\mathrm{Fe}^{2+}$ and $\mathrm{Fe}^{3+}$ in SCI; (c) comparison of the distribution of $\phi$ and $S_{p}$ around the circumference of steel rebar at 20 and 60 days.}
\label{fig:SpResults}
\end{figure}

Let us examine the evolution of key variables to understand cracking predictions and gain insight into the model. Since the evolution of the distribution of rust in the concrete pore space determines corrosion-induced cracking, its accurate description was one of the main objectives of the proposed formulation. For this purpose, the reactive transport and precipitation of dissolved iron species were explicitly modelled. It is known that precipitates can form millimetres from the steel surface \citep{stefanoni_kinetic_2018}, which reduces the precipitation-induced pressure and thus delays the cracking process. As shown in Figs. \ref{FigSpcont60d} and \ref{FigSectionSp}, for the representative case of test 2 ($c = 20$ mm, $i_{a} = 10 $ \unit{\micro\ampere\per\centi\metre^2}), the proposed model is able to capture how precipitates spread over several mm away from the steel surface. Also, Fig. \ref{FigSectionSp} shows how the precipitate content dramatically increases in the close vicinity of the rebar, where a highly porous steel-concrete interface is located and thus where significantly more rust can accumulate. This `layer' of rust around the rebar is commonly observed in experimental studies \cite{Wong2010a,Michel2014}. The evolution of the distributions of $ \mathrm{Fe}^{2+} $ ions and $ \mathrm{Fe}^{3+} $ ions are discussed in more detail in \ref{Sec:A}. \\

Another objective of the proposed model was to simulate the volume expansion of rust compared to the original steel, going beyond the simplified concepts of rebar expansion and stress-free accumulation of precipitates in the concrete pore space. The abilities of the model in capturing the interplay between the growth of precipitates under confined conditions and corrosion-induced cracking are showcased in Figs. \ref{fig:three graphs} and \ref{fig:SpResults}. Moreover, the results in Fig. \ref{fig:SpResults} show that fracture proceeds even though the pores are still not entirely filled with precipitates, in contrast with the popular simplifying assumption of the existence of a region of stress-free accumulation of corrosion products around the steel rebar. In this regard, it has been argued that growing crystals in concrete can cause pressure before the pore space is completely filled, as the growing crystal exerts pressure on the pore walls \cite{Flatt2017a}. \\

A third physical phenomenon of particular interest is the preferential transport of dissolved iron species through cracks \citep{Wong2010a}, instead of through the concrete pore space, and the associated local reduction of pressure. This was incorporated into the proposed model by employing a damage-dependent formulation of the diffusivity tensor (\ref{diffusivity_tensor_2}). The ability of the model to capture the role of cracks on the precipitate saturation ratio $S_{p}$ is shown in Fig. \ref{FigAlongRebarPFvsSp}. The results show how in regions with a higher degree of damage (i.e., locally growing cracks), dissolved iron species are preferentially transported deeper into the cracks and precipitate there or further away in the concrete porosity rather than on the steel-concrete boundary. Thus, the precipitate saturation ratio at the steel-concrete boundary is locally reduced in the vicinity of cracks and is highest in the regions with low damage.    

\FloatBarrier
\subsubsection{Parametric studies}
\label{Sec:ResultsParamStudy}

\begin{figure}[htp!]
    \centering
    \begin{subfigure}[!htb]{0.49\textwidth}
    \centering
    \includegraphics[width=\textwidth]{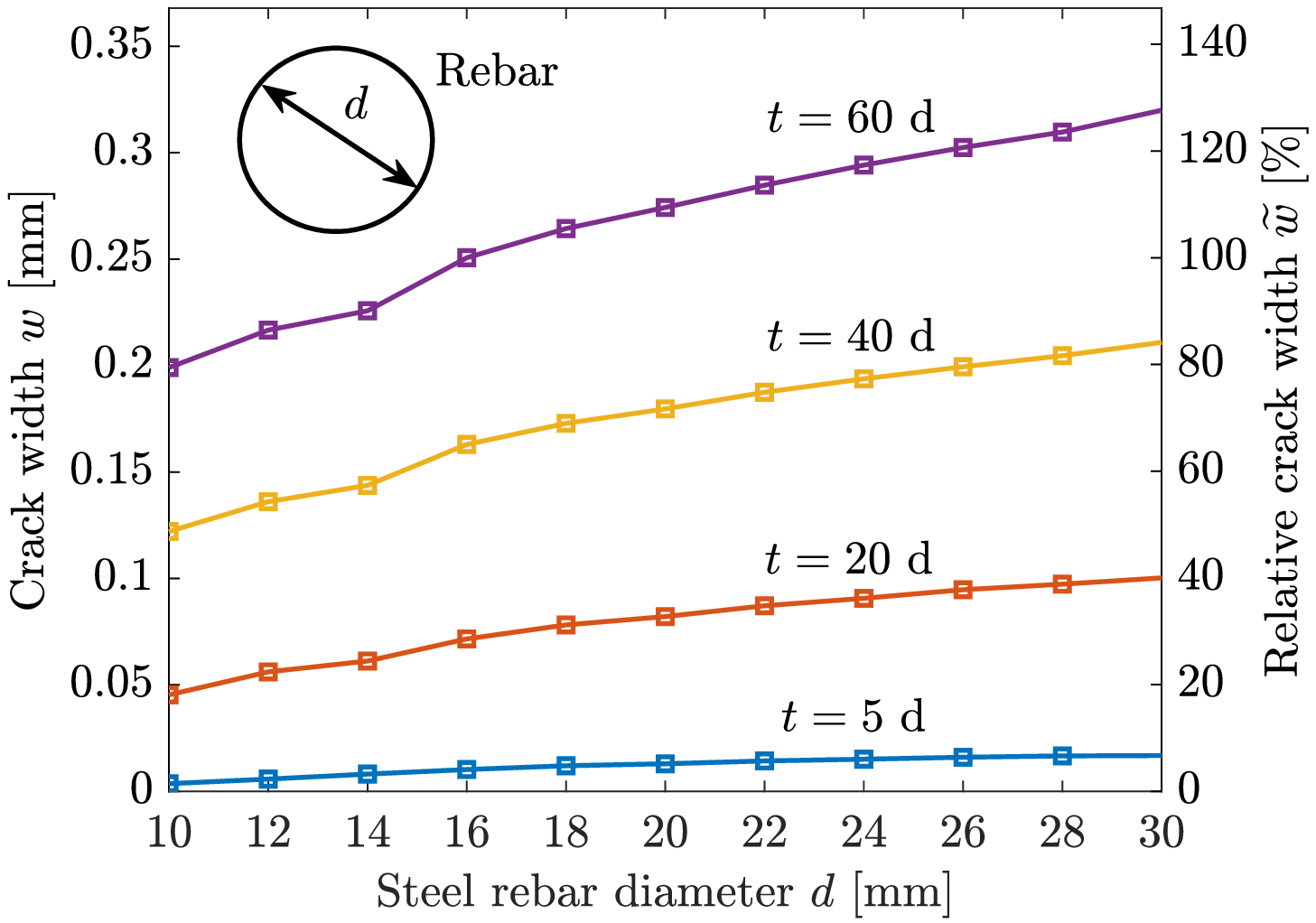}
    \caption{}
    \label{FigSweepBarDiam}    
    \end{subfigure}
    \hfill    
    \begin{subfigure}[!htb]{0.49\textwidth}
    \centering
    \includegraphics[width=\textwidth]{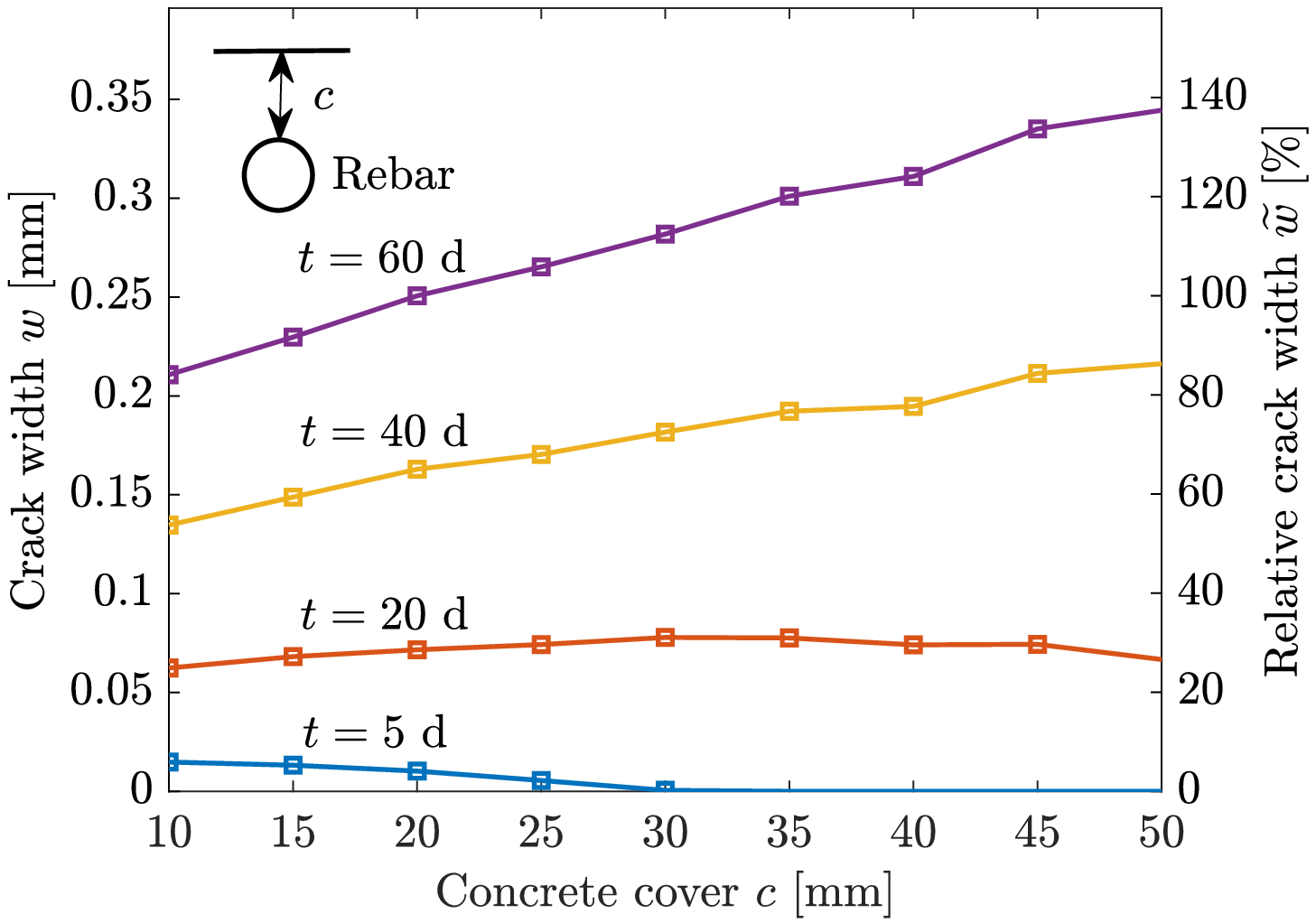}
    \caption{}
    \label{FigSweepCover}    
    \end{subfigure} 
    \hfill
    \begin{subfigure}[!htb]{0.49\textwidth}
    \centering
    \includegraphics[width=\textwidth]{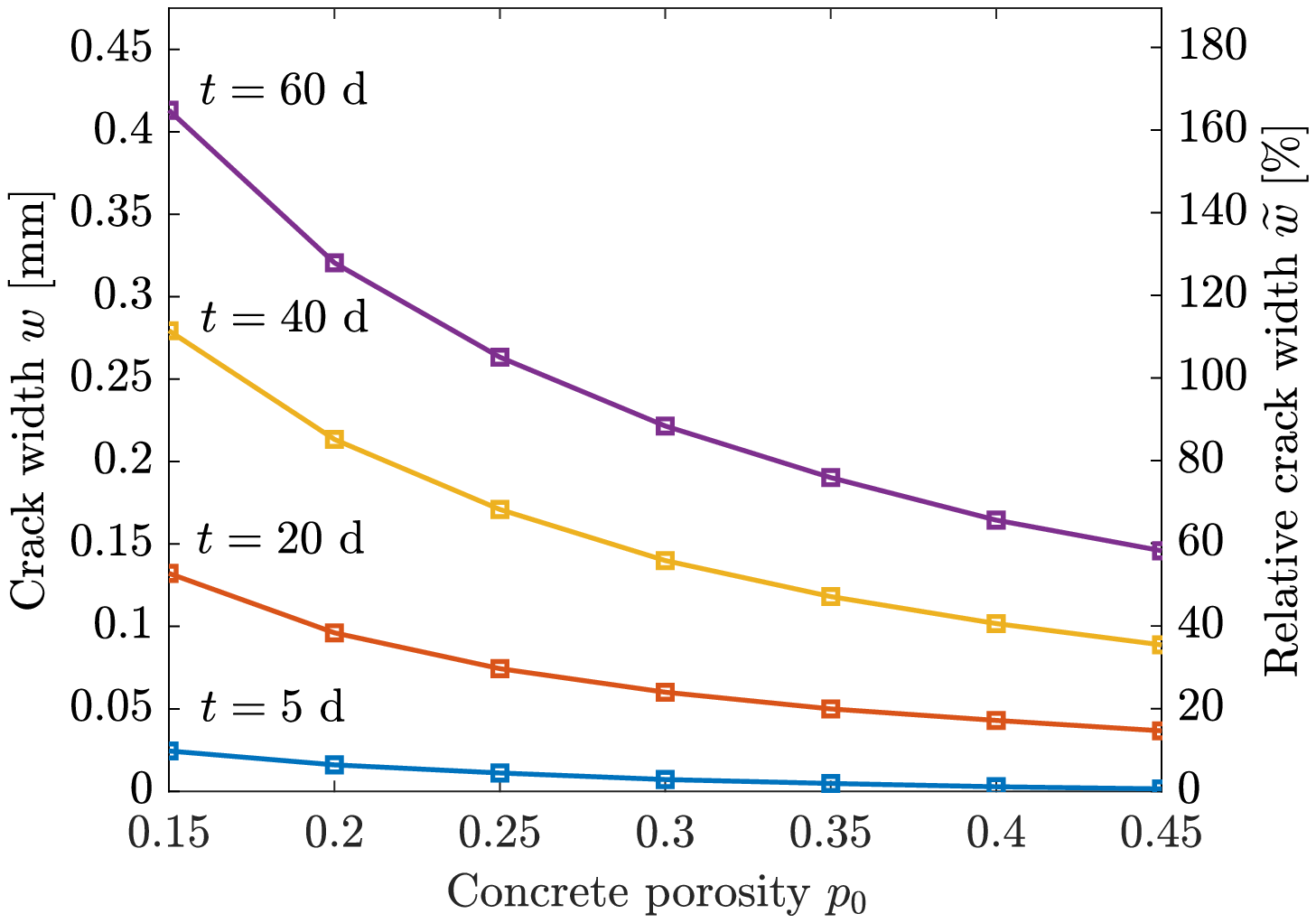}
    \caption{}
    \label{FigSweepPor}   
    \end{subfigure}
    \hfill
    \begin{subfigure}[!htb]{0.49\textwidth}
    \centering
    \includegraphics[width=\textwidth]{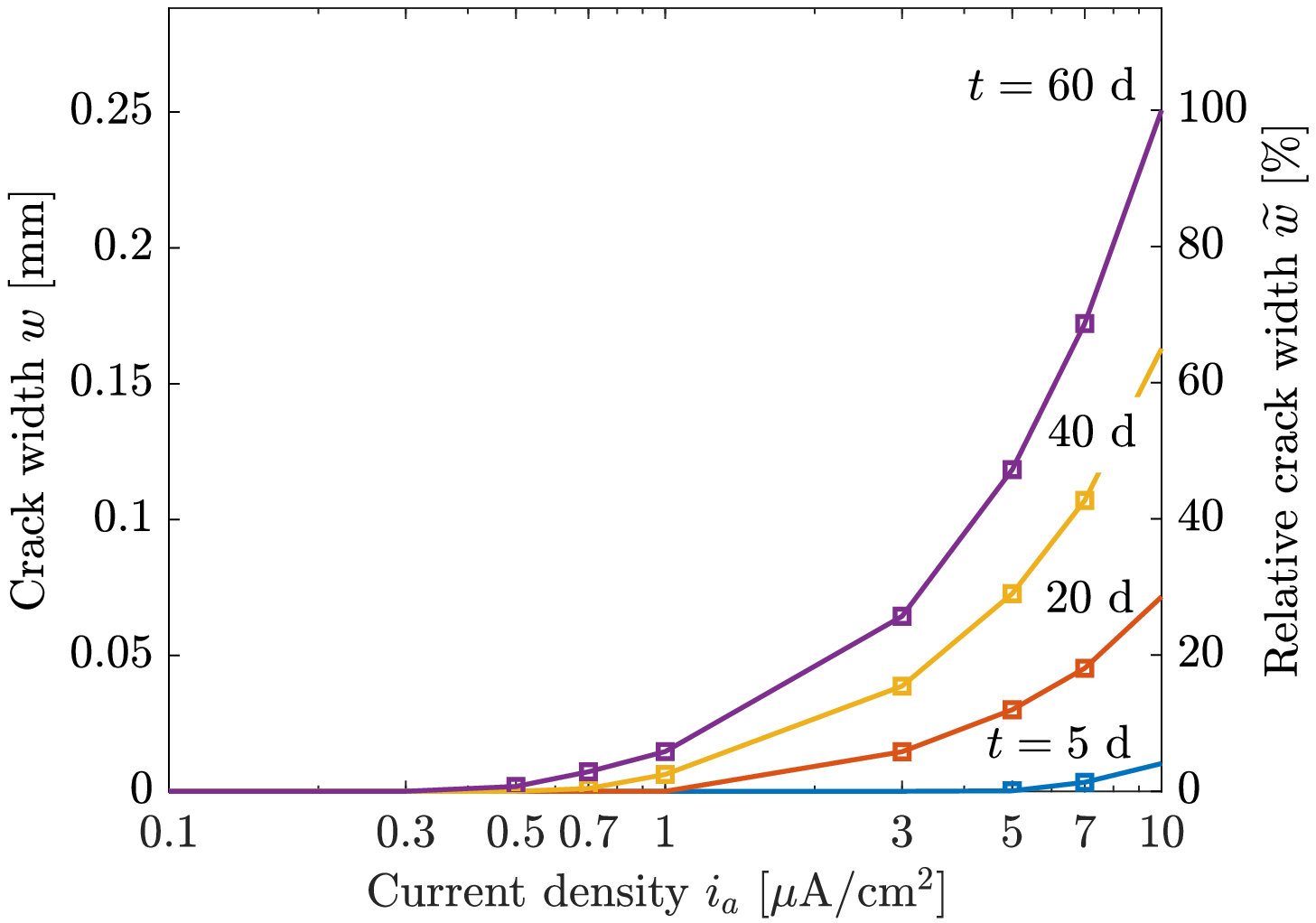}
    \caption{}
    \label{FigSweepCorrCurrDen}    
    \end{subfigure}
    \hfill    
    \begin{subfigure}[!htb]{0.49\textwidth}
    \centering
    \includegraphics[width=\textwidth]{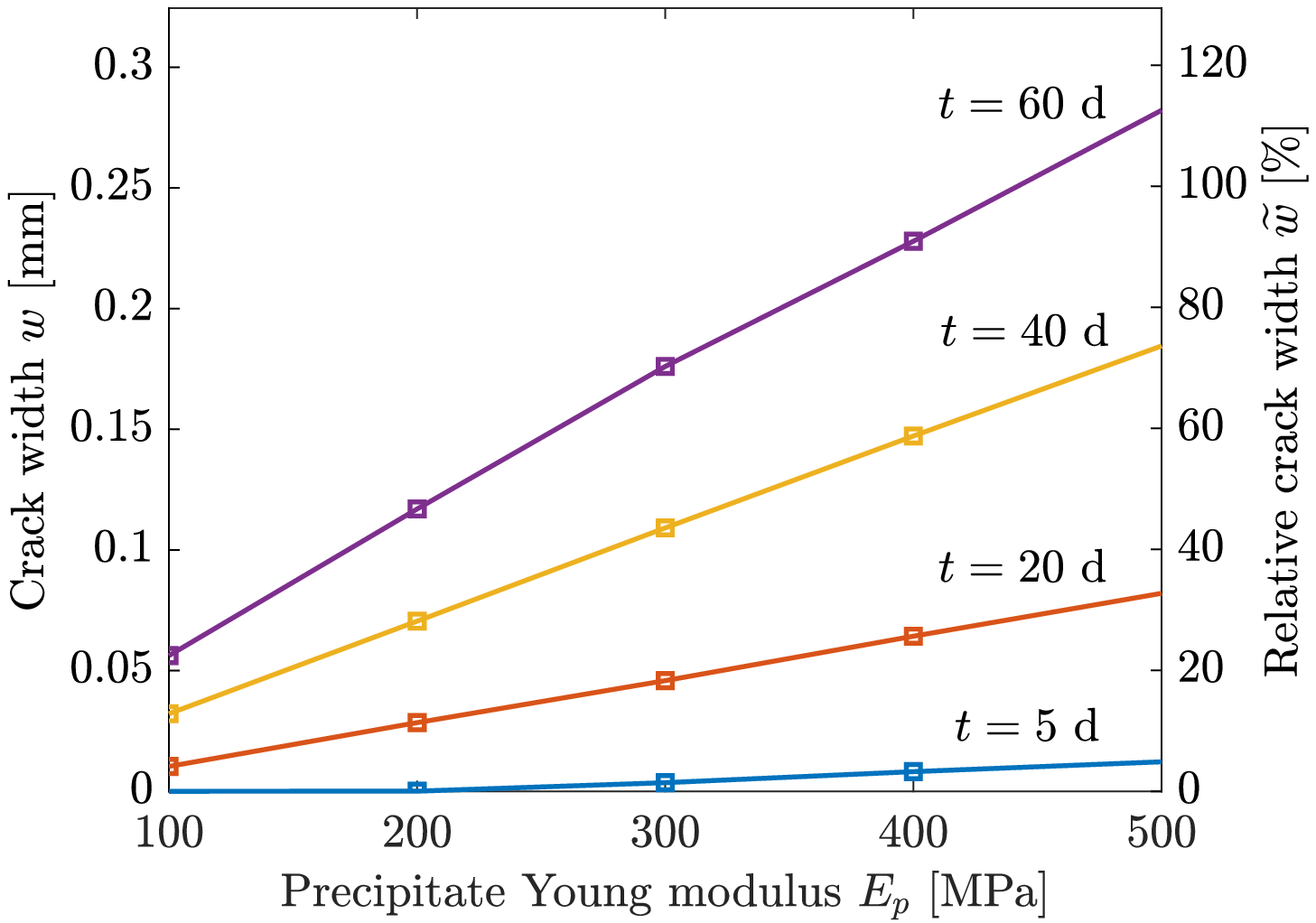}
    \caption{}
    \label{FigSweepRustYoungMod}    
    \end{subfigure} 
    \hfill    
    \begin{subfigure}[!htb]{0.49\textwidth}
    \centering
    \includegraphics[width=\textwidth]{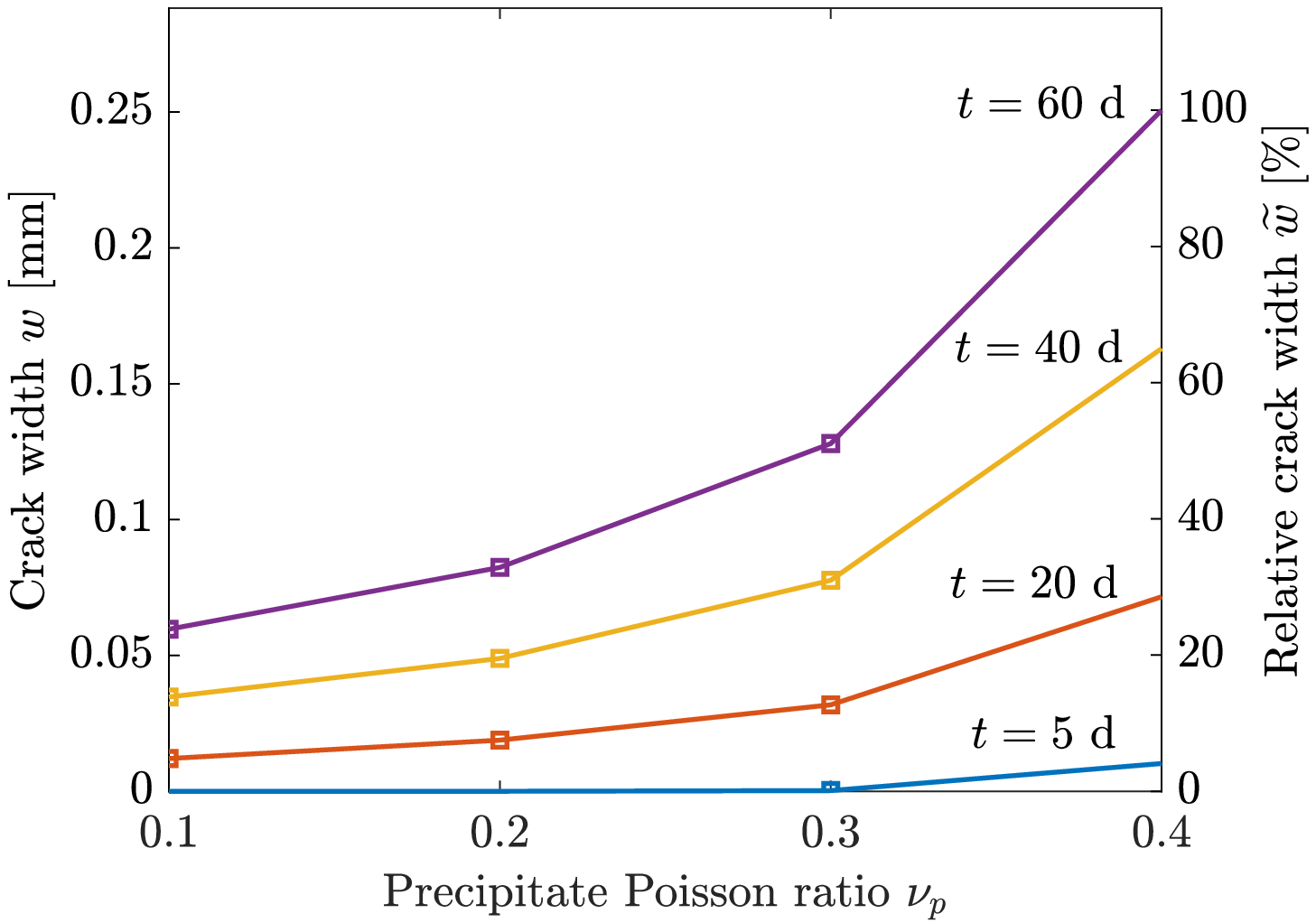}
    \caption{}
    \label{FigSweepRustPissonRat}    
    \end{subfigure}
    \caption{Parametric study: absolute and relative crack width after 5, 20, 40 and 60 days for varying (a) bar diameter $d$, (b) concrete cover $c$, (c) concrete porosity $p_{0}$, (d) corrosion current density $i_{a}$ (with the $x$-axis in log scale), (e) Young modulus of rust $E_{p}$, and (f) Poisson's ratio of rust $\nu_{p}$.}
    \label{Fig:ParametricStudy}
\end{figure}

The validated model is subsequently used to explore the sensitivity of the results to changes in model parameters. The surface crack width $w$ is evaluated for every parameter value at 5, 20, 40 and 60 days. Only the value of the parameter being investigated is varied and the remaining parameters are taken to be those relevant to test 2 ($i_{a} = 10$ \unit{\micro\ampere\per\centi\metre^2}, $c = 20$ mm), a representative case study. An exception to this is the concrete's tensile strength $ f_{t} $ and fracture energy $G_{f}$, which depend on each other and on Young's modulus as per the Eurocode standards \cite{standard2004eurocode}. For the sake of brevity, only the results for the most relevant parameters are discussed in this section, and the reader is referred to \ref{Sec:C} for the remainder. Specifically, crack width predictions as a function of time are here provided for varying values of the steel rebar diameter $d$, the concrete cover distance $c$, the concrete porosity $p_0$, the corrosion current density $i_a$, and the precipitate mechanical properties (Young's modulus $E_p$ and Poisson's ratio $\nu_p$) - see Fig. \ref{Fig:ParametricStudy}. To facilitate interpretation, the crack width is displayed both in its absolute value $ w $ (in mm) and in its relative value $\widetilde{w}$ (in $\%$). The relative crack width $\widetilde{w}$ is calculated as the ratio of the crack width $ w $ to 0.25 mm, which was the maximum crack width for test 2 ($i_{a} = 10$ \unit{\micro\ampere\per\centi\metre^2} and $c = 20$ mm) at 60 days.

Consider first the results obtained for a varying steel rebar size, Fig. \ref{FigSweepBarDiam}. Increasing the rebar diameter $d$ implies having a larger corroding surface and thus more released $ \mathrm{Fe}^{2+} $ ions gradually turning into rust. Hence, as $d$ increases, more precipitates distribute over a greater region, leading to a more profound expansive pressure and thus a greater crack width. The results relevant to a varying concrete cover depth $c$ are given in Fig. \ref{FigSweepCover}. Initially, a larger concrete cover causes the crack to start growing later, as it can be seen in the 5 days curve. However, the predicted rate of surface crack width growth is higher for larger covers, as it can be seen from the simulation at 40 and 60 days. This result can be thought of as counter-intuitive, given that design codes recommend increasing the concrete cover depth to improve corrosion resistance. However, this is aimed at delaying the transport of aggressive chemical species such as chlorides or carbon dioxide, while in this set of experiments chlorides are present near the rebar from the beginning of the test. In the scenario considered here, the larger surface crack width obtained with a larger concrete cover is the result of mechanical (geometric) effects. This has also been reported in computational studies \cite{Chen2015,Chen2020}, and \citet{Alonso1996} showed experimentally that for larger covers the cracking process is delayed but the rate of the crack width in time is not smaller.

\FloatBarrier

The sensitivity of the surface crack width to the concrete porosity is shown in Fig. \ref{FigSweepPor}. Here, we vary the bulk porosity $p_0$ but leave the SCI porosity constant at 0.52. The results show how the crack width rapidly decreases with increasing porosity. This is because precipitates occupy the pore space surrounding the rebar much quicker for low porosities, inducing large pressures at an earlier stage. The results show that the explicit modelling of the reactive transport and precipitation of iron species allows the profound impact of porosity on corrosion-induced cracking to be captured, without the need to employ the simplified concept of an artificial corrosion accommodation region. However, these predictions only reflect the impact of porosity on transport properties while in reality porosity would have an impact on the strength and fracture energy of concrete, potentially counterbalancing the effect described in Fig. \ref{FigSweepPor}. The extension of the model to capture such dependencies is straightforward, provided that the strength and fracture energy sensitivities to $p_0$ are known. 

The impact of the corrosion current density on the surface crack width is given in Fig. \ref{FigSweepCorrCurrDen}. The results show very significant sensitivity to $i_a$ within the range of applied currents evaluated (0.1 to 10 \unit{\micro\ampere\per\centi\metre^2}). This strong dependence is to be expected as it is commonly utilised in experimental accelerated corrosion studies to shorten testing times. The model captures this dependency by increasing the amount of $ \mathrm{Fe}^{2+} $ ions that can be released from the steel surface with increasing applied current, which then results in a higher content of $ \mathrm{Fe}^{3+} $ ions via oxidation and subsequent precipitation.\\ 

Finally, Figs. \ref{FigSweepRustYoungMod} and \ref{FigSweepRustPissonRat} show, respectively, the crack width sensitivity to the Young's modulus $E_{p}$ and Poisson's ratio $\nu_{p}$ of rust. The results show significant sensitivity to these two parameters, whose values carry a relatively high degree of uncertainty \cite{ZHAO201619}. Thus, these results highlight the need for careful characterisation studies of rust behaviour to enable mechanistic modelling of corrosion-induced cracking.

\FloatBarrier
\subsection{Simulation of spalling and delamination as a function of the reinforcement configuration}
\label{Sec:ResultsDelamSpalling}

\begin{figure}[!htb]
    \centering
    \begin{subfigure}[!htb]{0.49\textwidth}
    \centering
    \includegraphics[width=\textwidth]{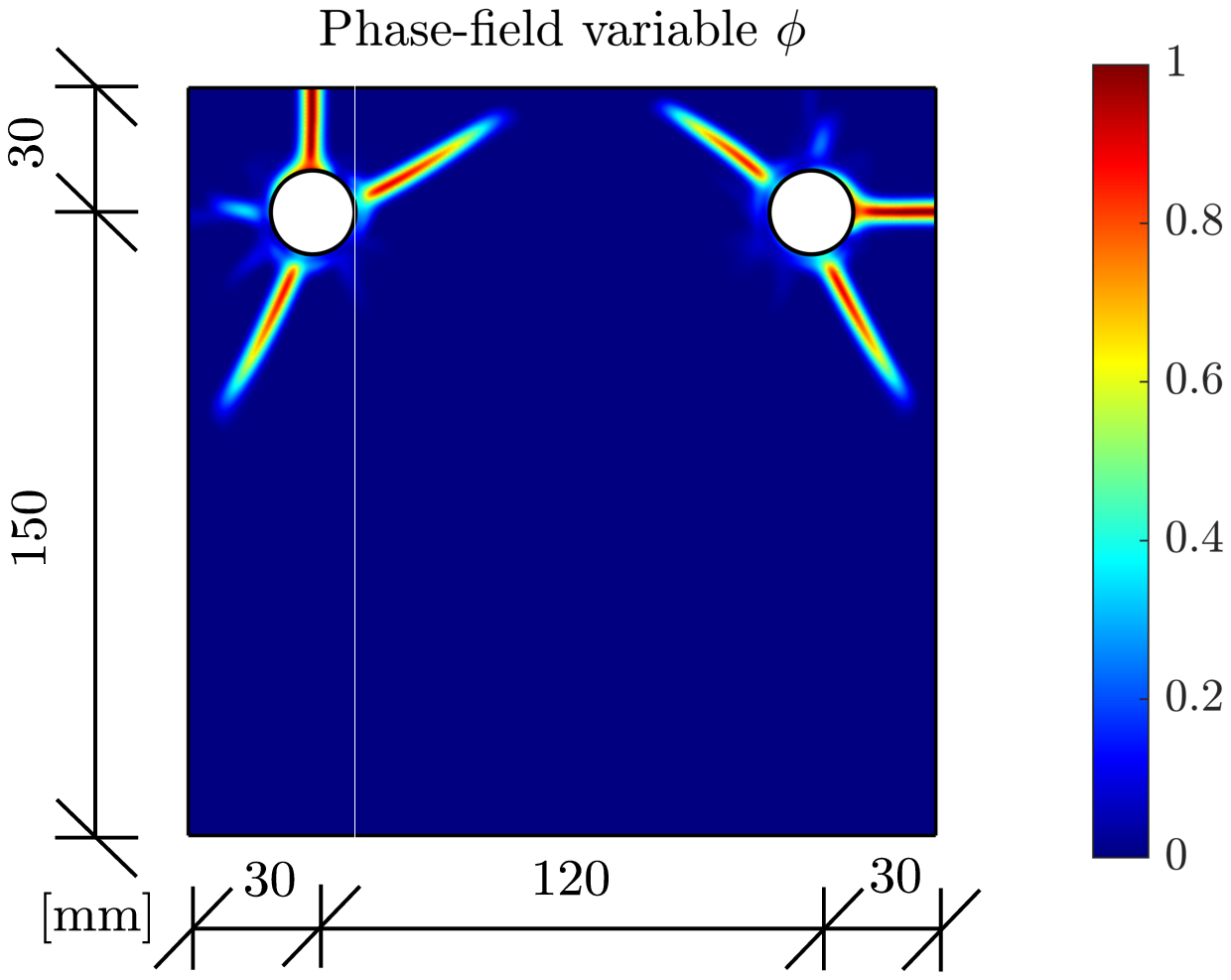}
    \caption{}
    \label{FigSpall2reb}    
    \end{subfigure}
    \hfill
    \begin{subfigure}[!htb]{0.49\textwidth}
    \centering
    \includegraphics[width=\textwidth]{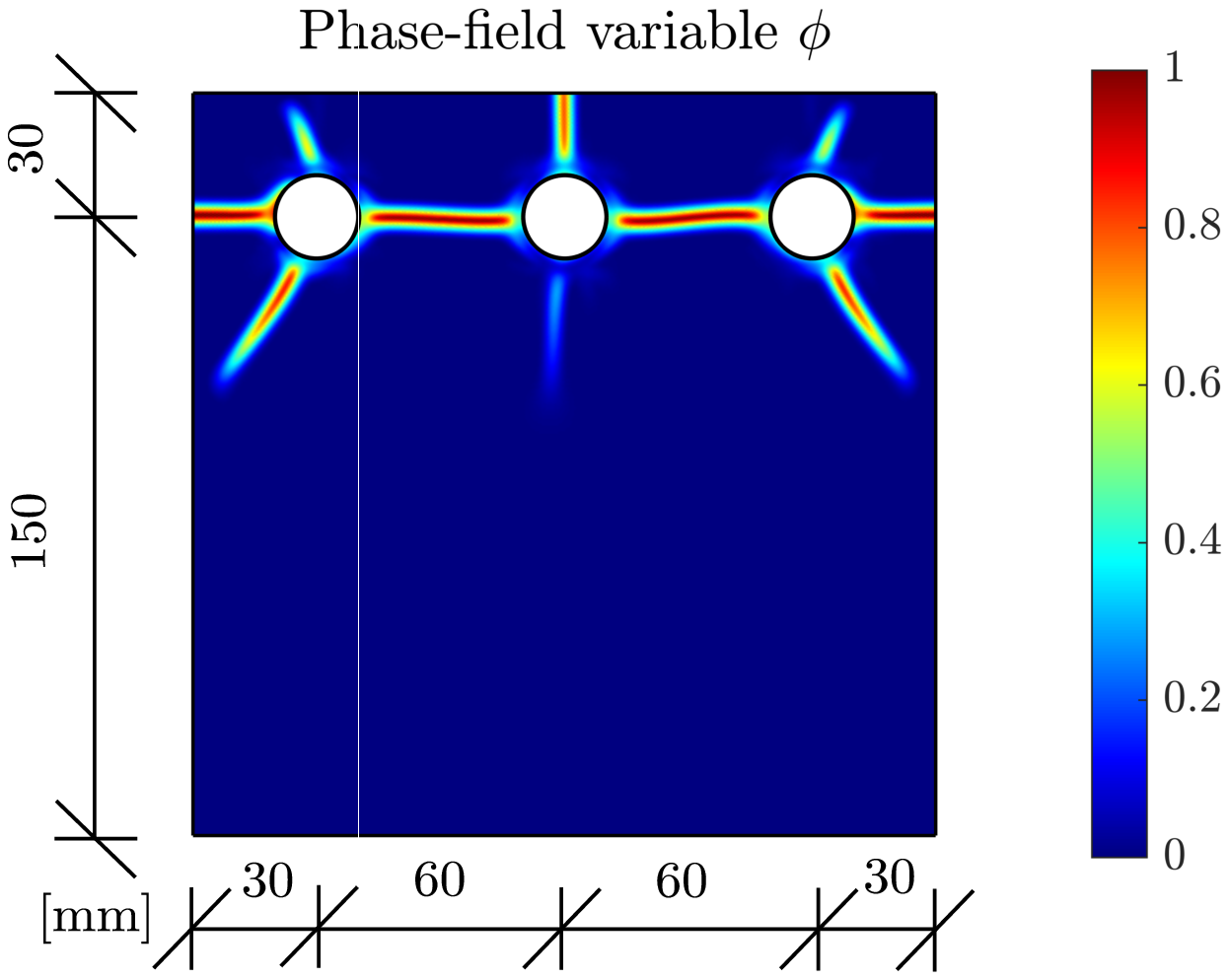}
    \caption{}
    \label{FigSpall3reb}    
    \end{subfigure}
    \hfill
    \begin{subfigure}[!htb]{0.49\textwidth}
    \centering
    \includegraphics[width=\textwidth]{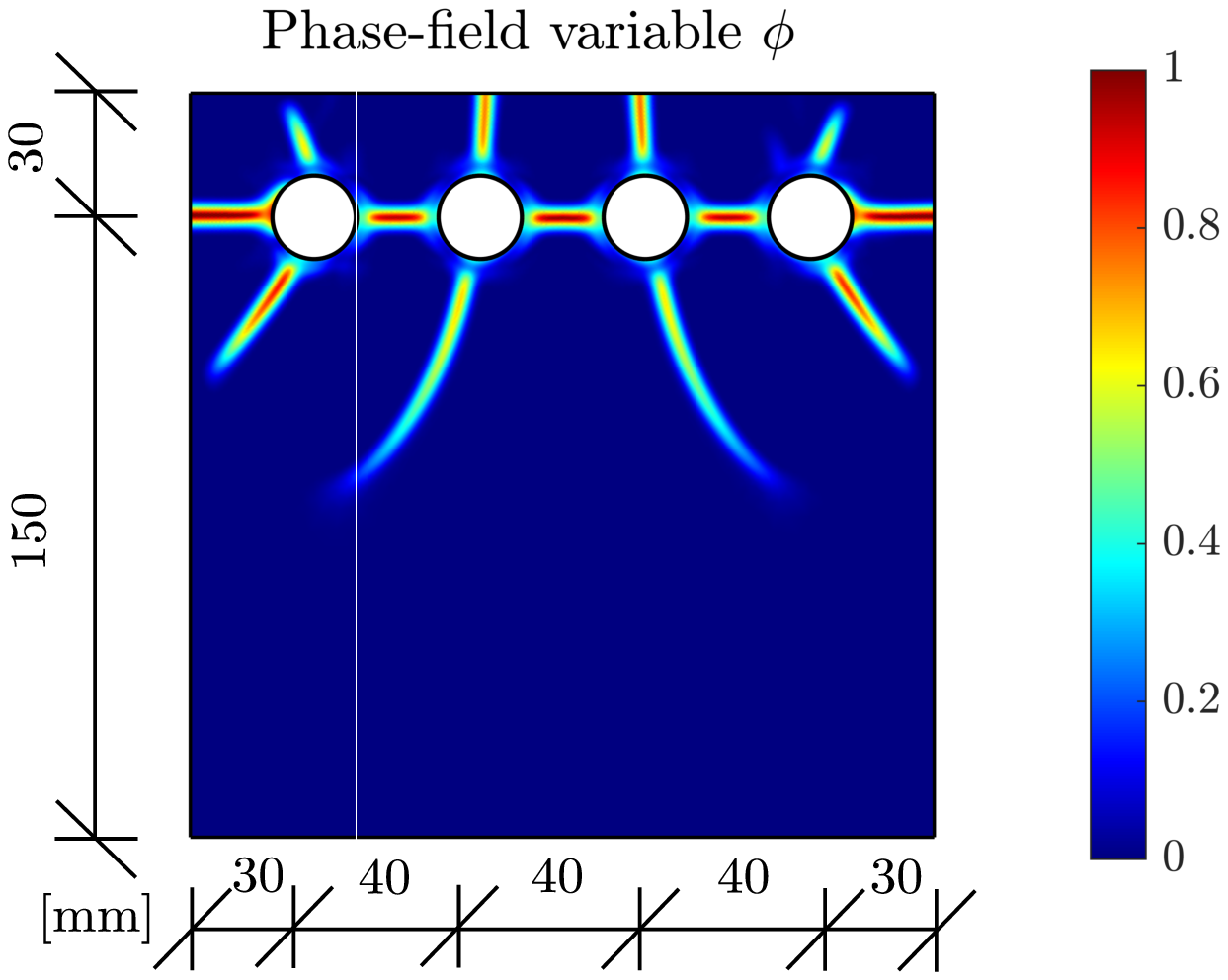}
    \caption{}
    \label{FigSpall4reb}    
    \end{subfigure}
    \hfill
    \begin{subfigure}[!htb]{0.49\textwidth}
    \centering
    \includegraphics[width=\textwidth]{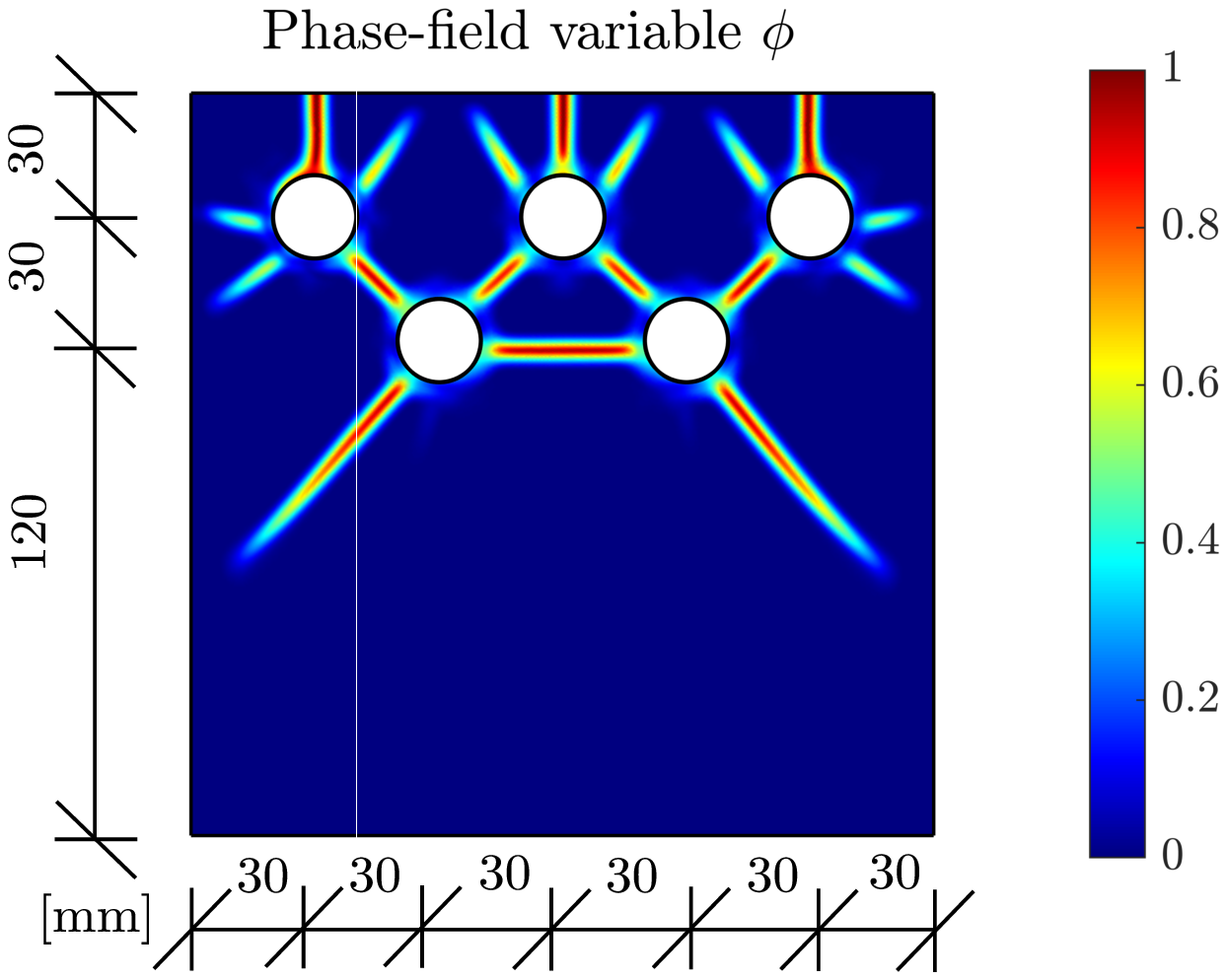}
    \caption{}
    \label{FigSpall5reb}    
    \end{subfigure}
    \caption{Phase-field variable $\phi$ contours after 60 days for (a) 2 embedded rebars in one row, (b) 3 embedded rebars in one row, (c) 4 embedded rebars in one row, and (d) 5 embedded rebars in two rows. In all cases, $i_{a} = 10$ \unit{\micro\ampere\per\centi\metre^2} and $d = 20$ mm.}
    \label{fig:ReinforcementConfig}
\end{figure}

We subsequently exploit the abilities of the model in capturing complex cracking phenomena to gain insight into the role of the reinforcement configuration and rebar interactions. As shown in Fig. \ref{fig:ReinforcementConfig}, simulations are conducted for a 180x180 mm concrete sample reinforced with a varying number of rebars: two (Fig. \ref{FigSpall2reb}), three (Fig. \ref{FigSpall3reb}), four (Fig. \ref{FigSpall4reb}) and five (Fig. \ref{FigSpall5reb}). The rebar diameter is 20 mm and in all but the five-rebar case the rebars are located at the same distance from the outer surface. In the case study involving five rebars, the configuration involves two rows of rebars, with the upper one having three. The model parameters employed correspond to those used in the previous section to validate the conditions of test 2. All rebars are corroding uniformly and the corrosion current density is $i_{a} = 10$ \unit{\micro\ampere\per\centi\metre^2}. 

The simulation results showcase the ability of the model to predict spalling and delamination events. Cracks typically originate between the rebars, and between the rebars and the closest free surface, highlighting their interaction. These interactions are governed by the rebar spacing. If the spacing is large relative to the concrete cover depth, spalling patterns are observed in the corners of the concrete sample (Fig. \ref{FigSpall2reb}). However, if the rebars are spaced more closely, lateral cracks interact and form a delamination pattern (Figs. \ref{FigSpall3reb} and \ref{FigSpall4reb}). This behaviour is typically observed in reinforced concrete structures \cite{ZHAO201619}. In the two-rebar case (Fig. \ref{FigSpall2reb}), non-symmetric behaviour can be observed, with a vertical crack being the most developed in the left rebar and a horizontal crack being dominant in the right rebar. This is not surprising since both rebars are equally distant from the vertical and horizontal concrete surfaces and thus small numerical differences can decide which crack will prevail. In practice, material heterogeneities will decide the competition between two equally probable cracking scenarios. Increasing the number of rebars increases the number of cracks, in agreement with expectations. Adding a new row of rebars (Fig. \ref{FigSpall5reb}) results in additional crack interactions, with a zigzag cracking pattern between rebar rows emerging as a result of the tendency of cracks to follow the shortest path between rebars. In addition, the comparison with Fig. \ref{FigSpall3reb} reveals that these 45$^\circ$-angle cracks dominate over the horizontal cracks that connect rebars in the absence of a second row that is sufficiently close. 

\FloatBarrier
\subsection{Corrosion-induced cracking of a 3D bent rebar}
\label{Sec:ResultsThreeDim}

\begin{figure}[htp]
    \centering
    \begin{subfigure}[!htb]{0.49\textwidth}
    \centering
    \includegraphics[width=\textwidth]{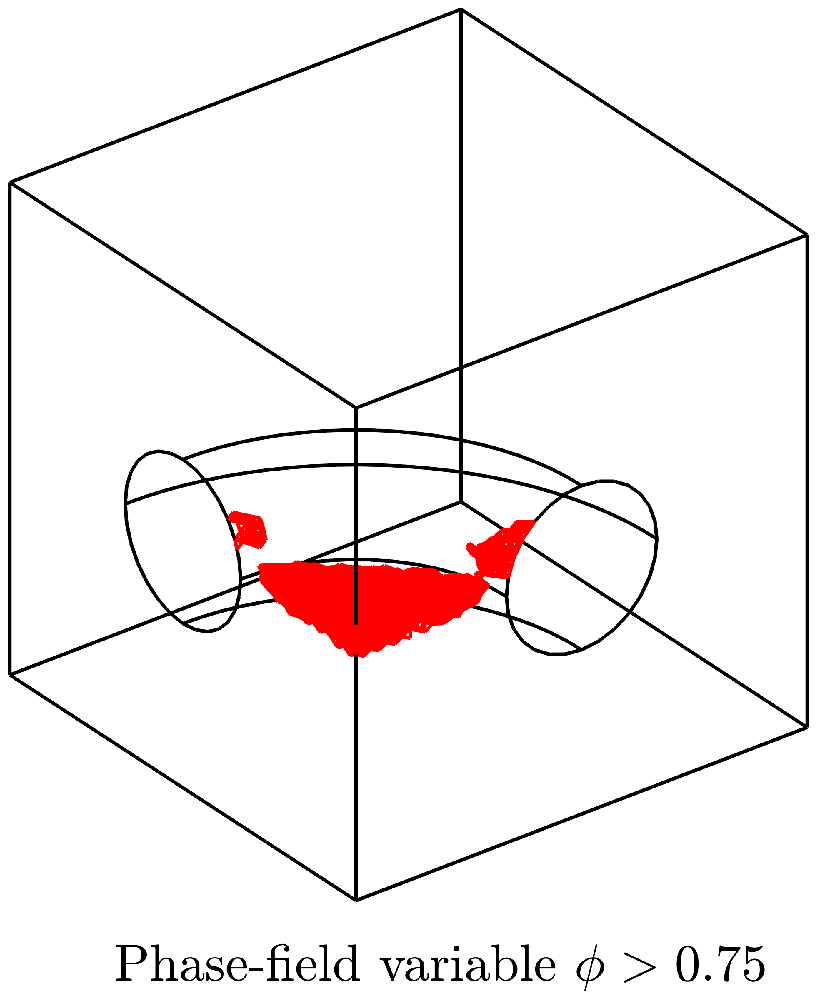}
    \caption{}
    \label{FigPF90days3DBendReb2}    
    \end{subfigure}
    \hfill
    \begin{subfigure}[!htb]{0.49\textwidth}
    \centering
    \includegraphics[width=\textwidth]{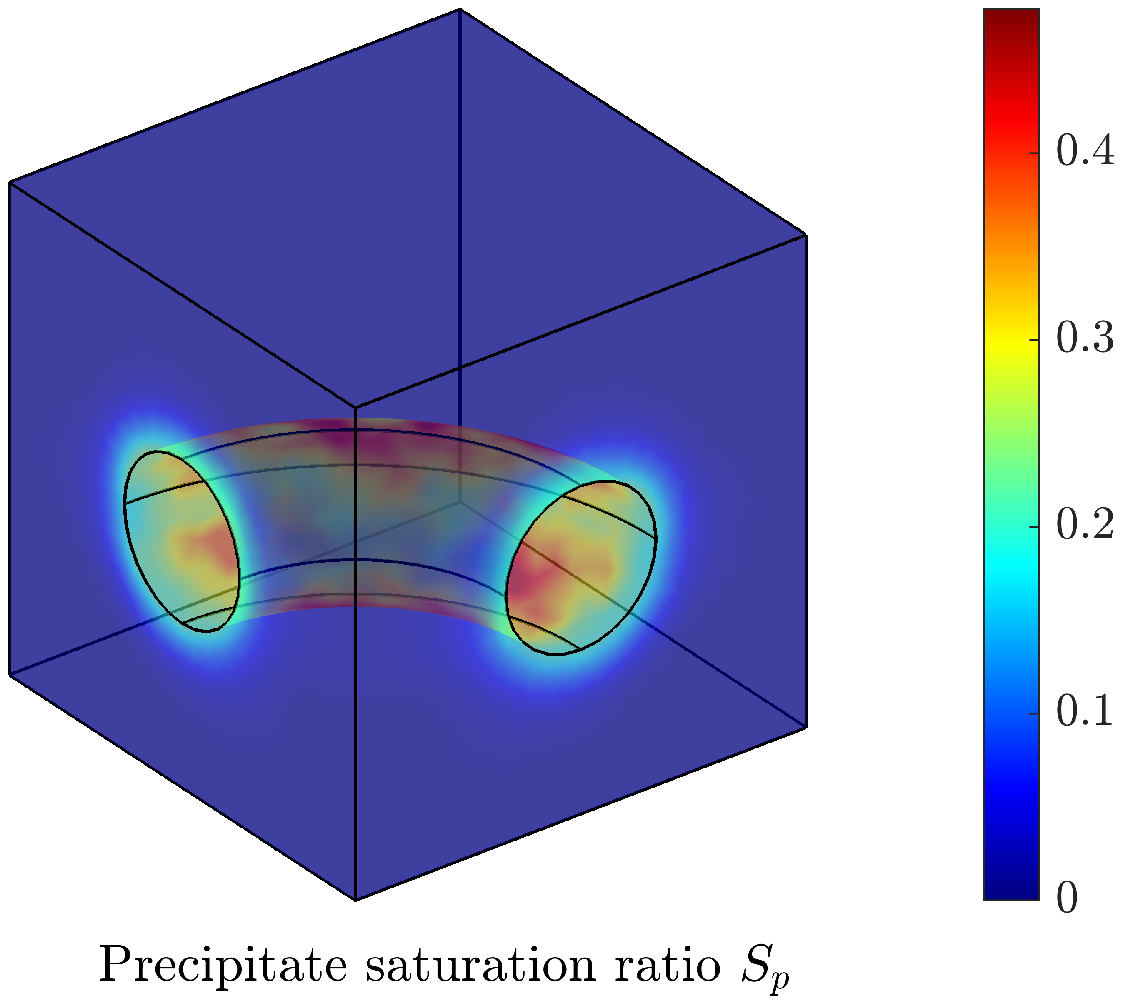}
    \caption{}
    \label{FigSP90days3DBendReb}  
    \end{subfigure} 
    \caption{Contours after 50 days of (a) evolving cracks, as characterised by contours of the phase-field variable above 0.75 ($\phi>0.75$), and (b) precipitate saturation ratio $S_{p}$ for a 30x30x30 mm concrete cube reinforced with a bent rebar. Here, $i_{a}$, $f_{t}$ and $G_{f}$ have been randomised following a continuous uniform probability distribution.}  
    \label{fig:3Dstudy}
\end{figure}

Finally, a three-dimensional case study is addressed to showcase the ability of the computational framework presented to predict corrosion-induced cracking in complex geometries and over technologically-relevant length scales. The boundary value problem under consideration is a 30x30x30 mm concrete cube reinforced with a bent rebar of 5 mm radius - see Fig. \ref{fig:3Dstudy}. The displacement is constrained in the vertical direction at the bottom surface (i.e., in the direction normal to the plane). Also, rigid body motion is prevented by prescribing the displacement in all directions at one of the corners of the cube. Regarding the transport boundary conditions, zero flux is assumed at the free surfaces except for the application of an inward flux for the concentration of $\mathrm{Fe}^{2+}$ ions, as given by Faraday's law (\ref{new_FarLaw}). We also utilise this case study to investigate the role of assigning a random variation to relevant parameters, so as to mimic the heterogeneous conditions expected in realistic scenarios. Thus, the corrosion current density $ i_{a} $ acting on the steel surface is varied between 1 and 9  \unit{\micro\ampere\per\centi\metre^2} using a continuous uniform probability distribution. In addition, the tensile strength and fracture energy of concrete are also randomised by employing a continuous uniform probability distribution such that that the maximum difference from the values given in Table \ref{tab:tableMechRust1} is 2.5 $\%$. Otherwise, the model parameters resemble those of test 2 in the validation study (see Sections \ref{Sec:modelParam} and \ref{Sec:ResultsValidationAndrade}). 

The results obtained are shown in Fig. \ref{fig:3Dstudy} in terms of the cracking patterns, as characterised by the domains where the phase field exceeds 0.75 (Fig. \ref{FigPF90days3DBendReb2}), and contours of the precipitate saturation ratio $S_p$ (Fig. \ref{FigSP90days3DBendReb}). It can be observed that the randomisation of the corrosion current density results in a non-uniform distribution of the precipitate saturation ratio $S_{p}$ but that the cracking behaviour shows a relatively symmetric pattern. Two sharp cracks form at the rebar edges closest to the free surface and grow perpendicularly to the vertical edges of the concrete cube. Then, damage expands and the cracked region eventually fills the space between the corner and the rebar, with some cracks appearing at other locations within the surface of the rebar. Thus, the cracking pattern is relatively insensitive to the randomisation of properties considered and is driven by geometrical effects. 

\section{Conclusions}
\label{Sec:Conclusions}

We have presented a new phase-field-based chemo-mechanical model capable of predicting corrosion-induced cracking of reinforced concrete under natural-like corrosion current densities. The theoretical framework is built upon three coupled attributes: (i) a model for the reactive transport and precipitation of $\mathrm{Fe}^{2+}$ and $\mathrm{Fe}^{3+}$ ions, (ii) an eigenstrain-based model to capture the pressure buildup resulting from the accumulation of precipitates under confined conditions in the concrete pore space, and (iii) a quasi-brittle phase-field description of fracture that accounts for the enhanced transport of relevant species through the crack network. The combination of these elements brings, for the first time, an understanding and modelling framework for corrosion-induced cracking driven by the growth of precipitates (rust) under confined conditions. This enables overcoming the limitations of expansion layer descriptions, which implicitly assumes the growth of incompressible precipitates from the steel surface, and the assumption of a stress-free accumulation of precipitates in the concrete pore space. Contrarily to this simplified approach, the proposed model:
\begin{itemize}
\item Resolves the evolution of the distribution of compressible precipitates in the concrete pore space in time. Thus, the model naturally captures the delaying effects of concrete porosity surrounding rebar and cracks on corrosion-induced cracking without necessity to consider an artificial  corrosion accommodation region (CAR) around the rebar which size and capacity is very hard to estimate. Even if the  corrosion accommodation region (CAR) would be accurately considered, it neglects the pressure of gradually forming rust in concrete porosity which is considered by the proposed model. 
\item Takes into consideration the compressibility and elastic properties of the rust which were found to importantly affect the corrosion-induced crack width.  
\item Allows to investigate the impact of varying current density on the composition of rust which is the objective of an ongoing research effort.
\end{itemize}
The resulting model consists of five differential equations and associated field quantities, which are solved by using the finite element method. The computational model is validated against the impressed current tests by \citet{Pedrosa2017}, showing a good agreement between the predicted and the experimentally measured crack widths during the early propagation stage, and subsequently used to gain insight into the role of relevant parameters and their interactions in 2D and 3D boundary value problems involving different rebar configurations, geometries and environmental conditions. The following key physical phenomena are captured: (i) the transport of $\mathrm{Fe}^{2+}$ and $\mathrm{Fe}^{3+}$ through the concrete matrix, (ii) the oxidation of $\mathrm{Fe}^{2+}$ to $\mathrm{Fe}^{3+}$, (iii) the precipitation of $\mathrm{Fe}^{2+}$ and $\mathrm{Fe}^{3+}$ in the concrete pore space, (iv) the clogging of the pore system by precipitates (rust), (v) the pressure resulting from the accumulation of precipitates (rust) under confined conditions in the concrete pore space, (vi) the progressive damage and fracture of the concrete matrix due to a precipitation pressure, and (vii) the enhanced transport of relevant chemical species through cracks. In addition, the results reveal the following main findings:
\begin{itemize}
    \item The model can accurately capture the evolution of surface crack width for different concrete cover depths and applied currents but the heterogeneous nature of concrete must be accounted for to accurately predict the onset of crack growth.
    \item Precipitates are found to be largely near the rebar but also spread up to millimetres away from the steel surface, reducing the precipitation-induced pressure and delaying cracking.
    \item Fracture proceeds due to pressure induced by precipitate growth and this can happen with a partial saturation of the concrete pore space.
    \item In regions with locally growing cracks, dissolved iron species are preferentially transported deeper into the cracks and precipitate there or in areas away from the steel rebar. 
    \item Surface crack width increases with rebar diameter (due to a larger precipitate distribution), concrete cover (due to geometrical effects) and current density (due to a higher release of $\mathrm{Fe}^{2+}$ ions).
    \item Predictions appear to be sensitive to the mechanical properties of rust, highlighting the need for appropriate characterisation studies
    \item Spalling and delamination events are predicted when multiple rebars are considered, showing that the rebar spacing plays a dominant role.
    \item While the nucleation of cracks is governed by local material heterogeneities, the final crack patterns are relatively insensitive to these as they are dominated by geometrical effects. 
\end{itemize}

Potential avenues for future work include the extension of the model to encompass non-uniform chloride-induced corrosion, carbonation-induced corrosion, and late-stage corrosion-induced cracking.

\section{Acknowledgements}
\label{Acknowledge of funding}

The authors gratefully acknowledge stimulating discussions with Prof Nick Buenfeld and Tanmay Ubgade (Imperial College London), and Prof Milan Kouril (University of Chemistry and Technology, Prague). E. Korec acknowledges financial support from the Imperial College President’s PhD Scholarships. M. Jirásek acknowledges the support of the European Regional Development Fund (Center of Advanced Applied Sciences, project CZ.02.1.01/0.0/0.0/16\_19/0000778). E. Mart\'{\i}nez-Pa\~neda was supported by an UKRI Future Leaders Fellowship [grant MR/V024124/1]. We additionally acknowledge computational resources and support provided by the Imperial College Research Computing Service (http://doi.org/10.14469/hpc/2232). 


\appendix
\FloatBarrier
\section{On the ability of various phase-field fracture models in reproducing quasi-brittle behaviour}
\label{Sec:A0}
\setcounter{figure}{0}

\begin{figure}[!htb]
    \centering
    \begin{subfigure}{0.48\textwidth}
    \centering
    \includegraphics[width=\textwidth]{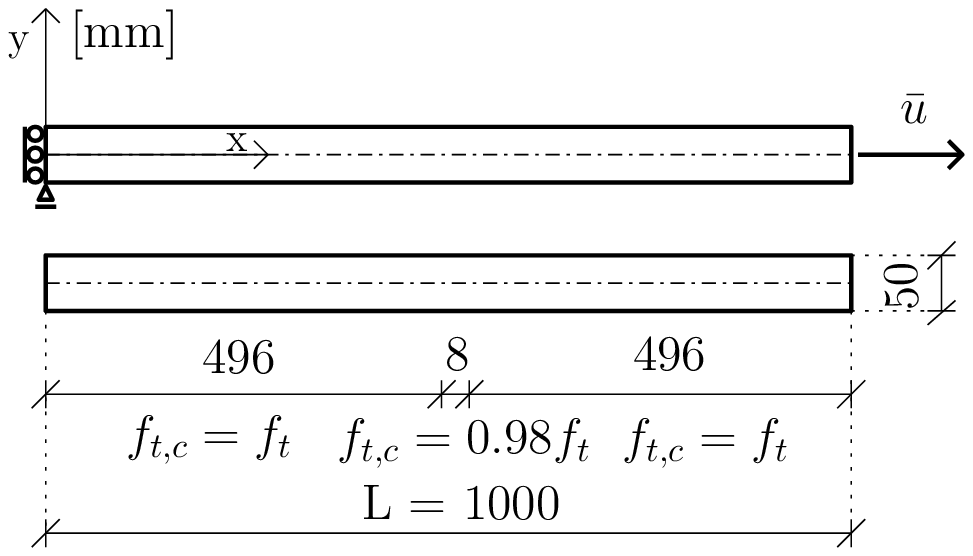}
    \caption{}
    \label{FigQuasiBrittle}  
    \end{subfigure} 
    \hfill
    \begin{subfigure}{0.48\textwidth}
    \centering
    \includegraphics[width=\textwidth]{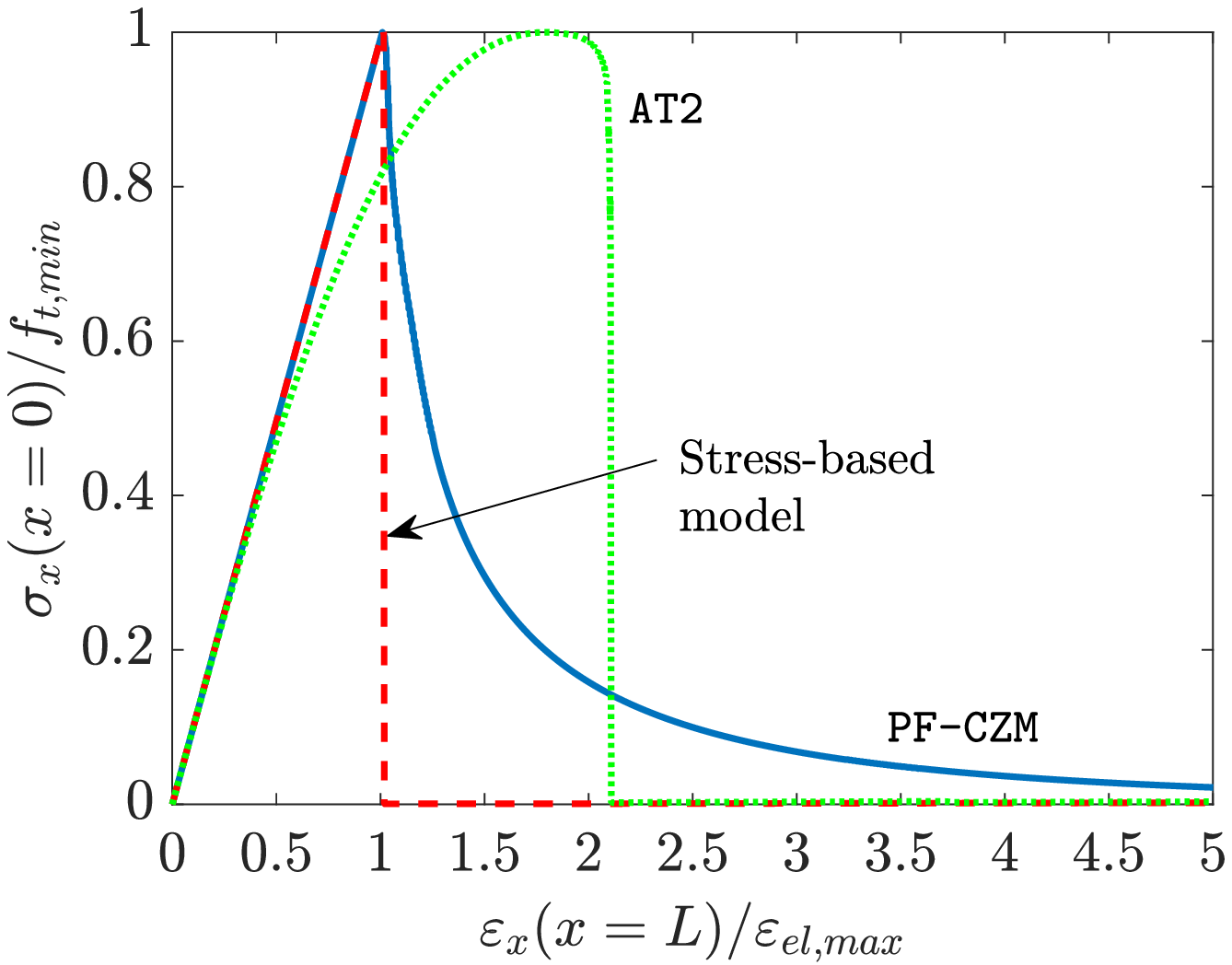}
    \caption{}
    \label{FigQuasiBrittle2}
    \end{subfigure}
    \caption{Elongated concrete bar benchmark -- (a) geometry (in mm), strength distribution and the boundary conditions, (b) comparison of stress-strain curves obtained with the conventional (\texttt{AT2}) phase field model \cite{Bourdin2000}, the stress-based model by Miehe et al. \cite{Miehe2015a}, and the quasi-brittle \texttt{PF-CZM} model \cite{Wu2017,Wu2018}. The stress $\sigma_{x}(0)$ is normalised with $f_{t,min}$ (the tensile strength of the sample in the weakened middle section), and the strain $\varepsilon_{x}(L)$ is normalised with $ \varepsilon_{el,max} = f_{t,min}/E_{c}$ (the maximum value of the $x$-component of the elastic strain.)}
    \label{fig:AppAPFcomparison}
\end{figure}

In the past year, studies have been conducted involving the application of phase field models to corrosion-induced cracking in concrete \cite{Wei2021,Hu2022,Freddi2022}. However, these studies employ phase-field fracture models originally designed for brittle fracture. To capture the quasi-brittle behaviour of concrete, we build our formulation upon the so-called phase-field cohesive zone model (\texttt{PF-CZM}) by Wu and co-workers \cite{Wu2017, Wu2018}. The differences between the three widely used phase-field models are highlighted in Fig. \ref{fig:AppAPFcomparison}. Specifically, the mechanical response of a two-dimensional bar is investigated by means of the \texttt{PF-CZM} model \cite{Wu2017,Wu2018}, the conventional or so-called \texttt{AT2} model \citet{Bourdin2000}, and the stress-based phase field model by Miehe et al. \cite{Miehe2015a,Clayton2022}. The values of Young's modulus $E_{c}$, Poisson's ratio $\nu_{c}$, tensile strength $f_{t}$ and fracture energy $G_{f}$ are the same as for the 147 days cured concrete samples (see Table \ref{tab:tableMechRust1}).  

The \texttt{PF-CZM} model by \citet{Wu2017, Wu2018} allows us to directly set both tensile strength and fracture energy as model parameters. The stress-based model of \citet{Miehe2015a} takes only tensile strength into consideration. The input to the \texttt{AT2} model \cite{Bourdin2000} is the fracture energy $G_f$ but the tensile strength can be set implicitly through the value of characteristic length $\ell$, following the relation:
\begin{equation}\label{charLengthTenStrengthForm}
f_{t} = \dfrac{9}{16}\sqrt{\dfrac{E_{c}G_{f}}{3\ell}}
\end{equation}

In order to trigger a localised fracture, the tensile strength in an 8 mm long section in the middle of the sample is reduced to $ f_{t,min} = 0.98f_{t} $ for the \texttt{PF-CZM} and stress-based models. In the \texttt{AT2} case, the fracture energy in the same section of the sample is reduced to 98$\%$ of its value. The phase-field characteristic length is chosen to be $ \ell  = 3 $ mm for the \texttt{PF-CZM} and stress-based models while $ \ell  = 72.4 $ mm for the \texttt{AT2} model to obtain the desired tensile strength when using (\ref{charLengthTenStrengthForm}). Lastly, the stress-based model by \citet{Miehe2015a} contains the dimensionless parameter $ \xi>0$ which affects the post peak slope of the local stress-strain diagram. We assume $ \xi=1$ but calculations conducted with other values do not appear to influence the results in this boundary value problem. The resulting stress-strain responses are shown in Fig. \ref{FigQuasiBrittle2}. Both the stress-based model by \citet{Miehe2015a} and the \texttt{AT2} model by \citet{Bourdin2000} predict a sudden drop in the load carrying capacity, indicative of brittle failure. Also, in the \texttt{AT2} case, damage starts to grow immediately after loading the sample and this leads to a concave stress-strain prior to sudden failure. In contrast, the \texttt{PF-CZM} model by \citet{Wu2017,Wu2018} predicts a long convex post-peak softening regime characteristic of quasi-brittle materials such as concrete.
 
\section{Evolution of the $\mathrm{Fe}^{2+}$ and $\mathrm{Fe}^{3+}$ concentrations}
\label{Sec:A}
\setcounter{figure}{0}

\begin{figure}[!htb]
    \centering
    \begin{subfigure}{0.48\textwidth}
    \centering
    \includegraphics[width=\textwidth]{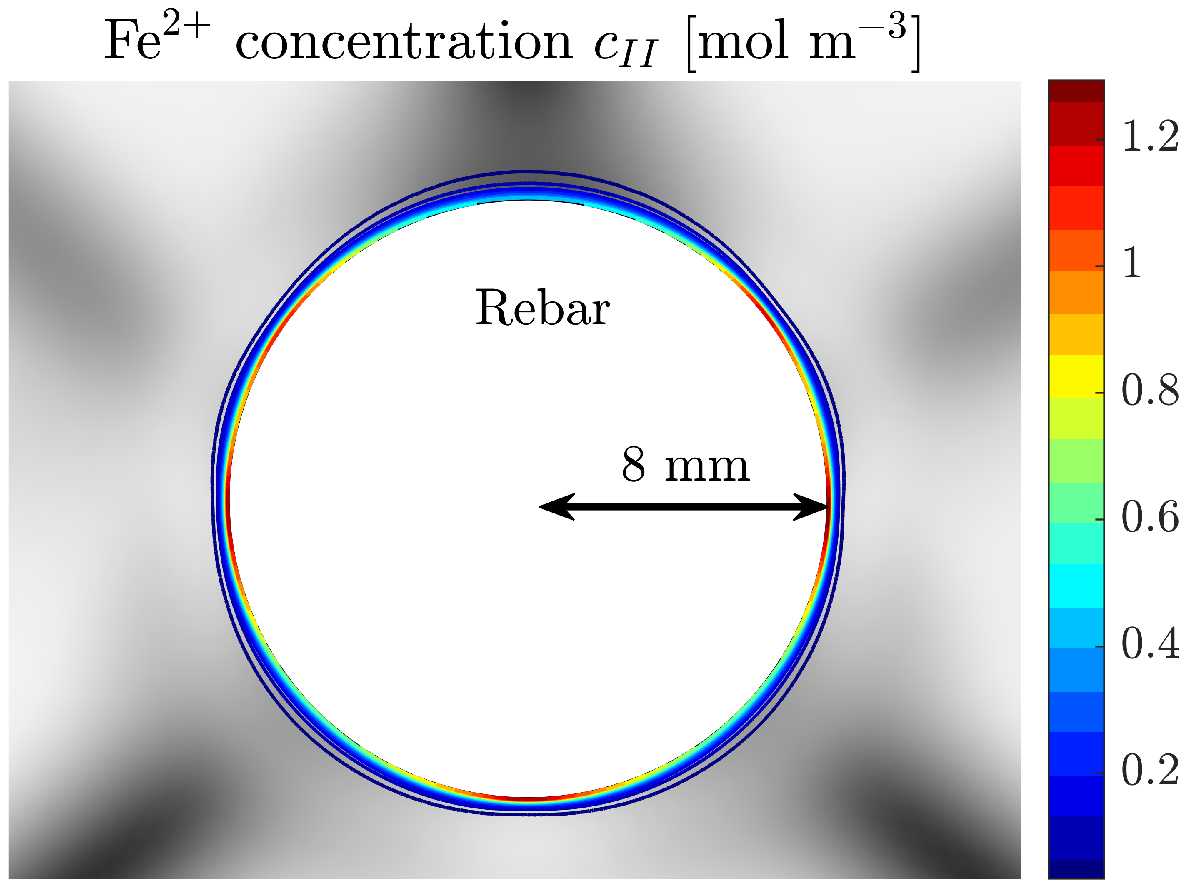}
    \caption{}
    \label{FigCIIcont60d}    
    \end{subfigure}
    \hfill    
    \begin{subfigure}{0.48\textwidth}
    \centering
    \includegraphics[width=\textwidth]{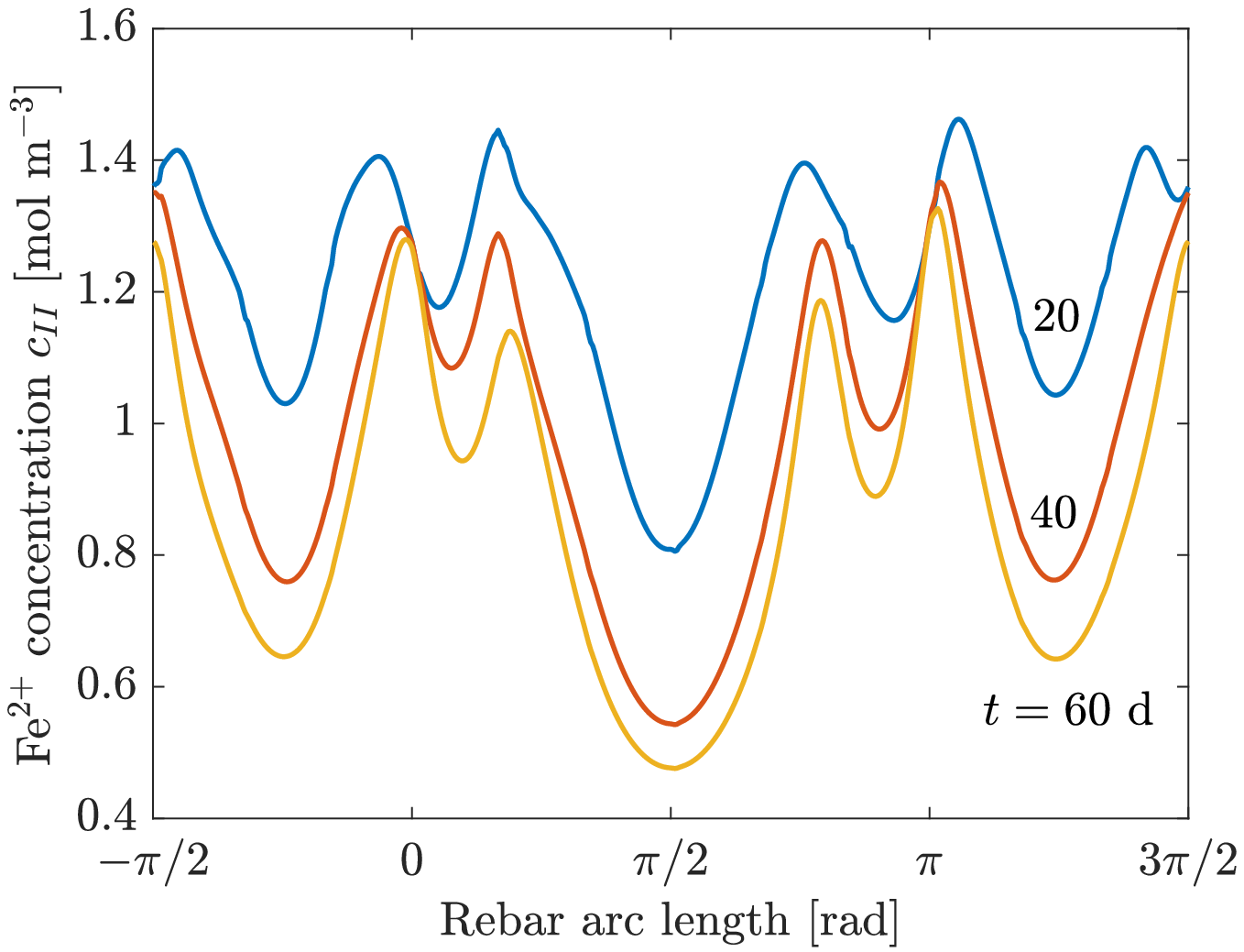}
    \caption{}
    \label{FigAlongRebarFe2}  
    \end{subfigure} 
    \caption{$\mathrm{Fe}^{2+}$ concentration $c_{II}$ for test 2 ($i_{a} = 10$ \unit{\micro\ampere\per\centi\metre^2}, $c = 20$ mm) -- (a) contours of $c_{II}$  in the vicinity of steel rebar at 60 days, phase-field variable $\phi$ in the shades of grey (0 -- white, 1 -- black), and (b) evolution of $c_{II}$ around the circumference of steel rebar.} 
    \label{fig:AppCII}
\end{figure}

\begin{figure}[!htb]
    \centering    
    \begin{subfigure}{0.48\textwidth}
    \centering
    \includegraphics[width=\textwidth]{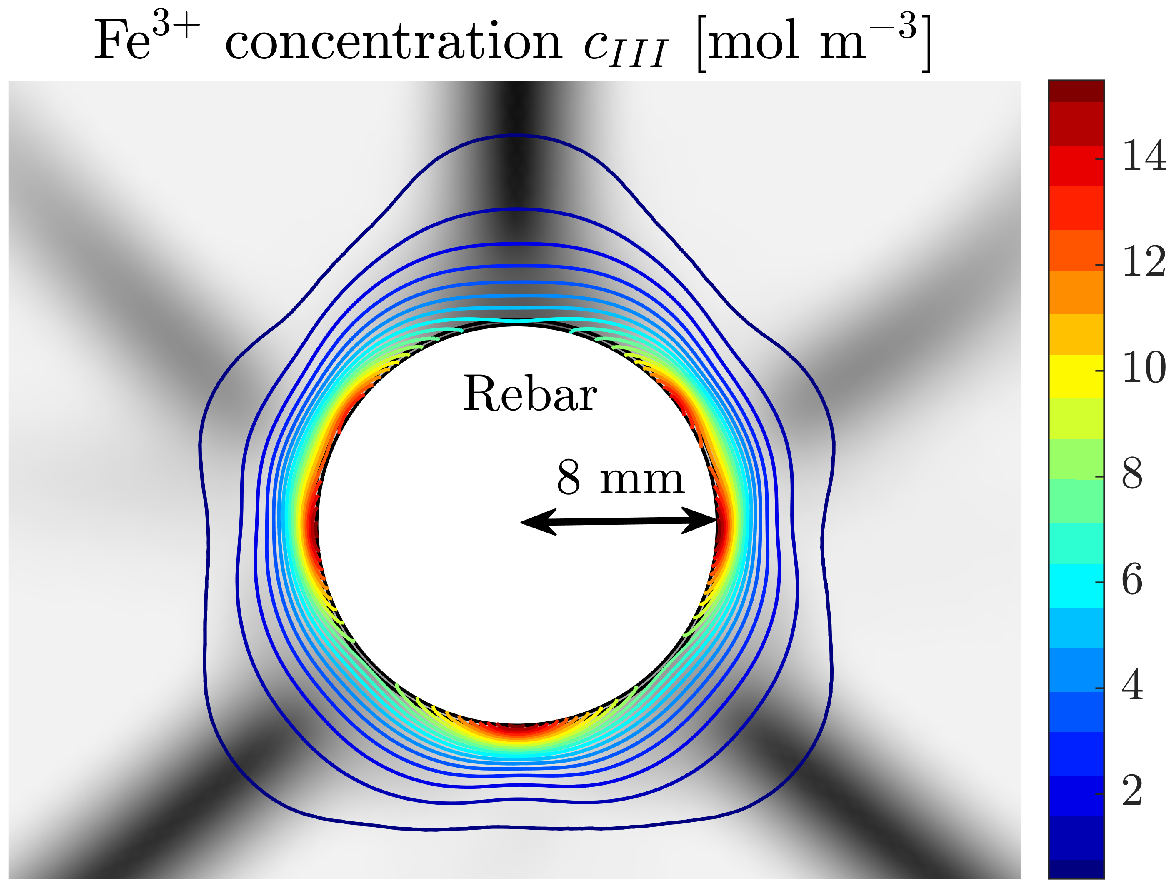}
    \caption{}
    \label{FigCIIIcont60d}    
    \end{subfigure}
    \hfill    
    \begin{subfigure}{0.48\textwidth}
    \centering
    \includegraphics[width=\textwidth]{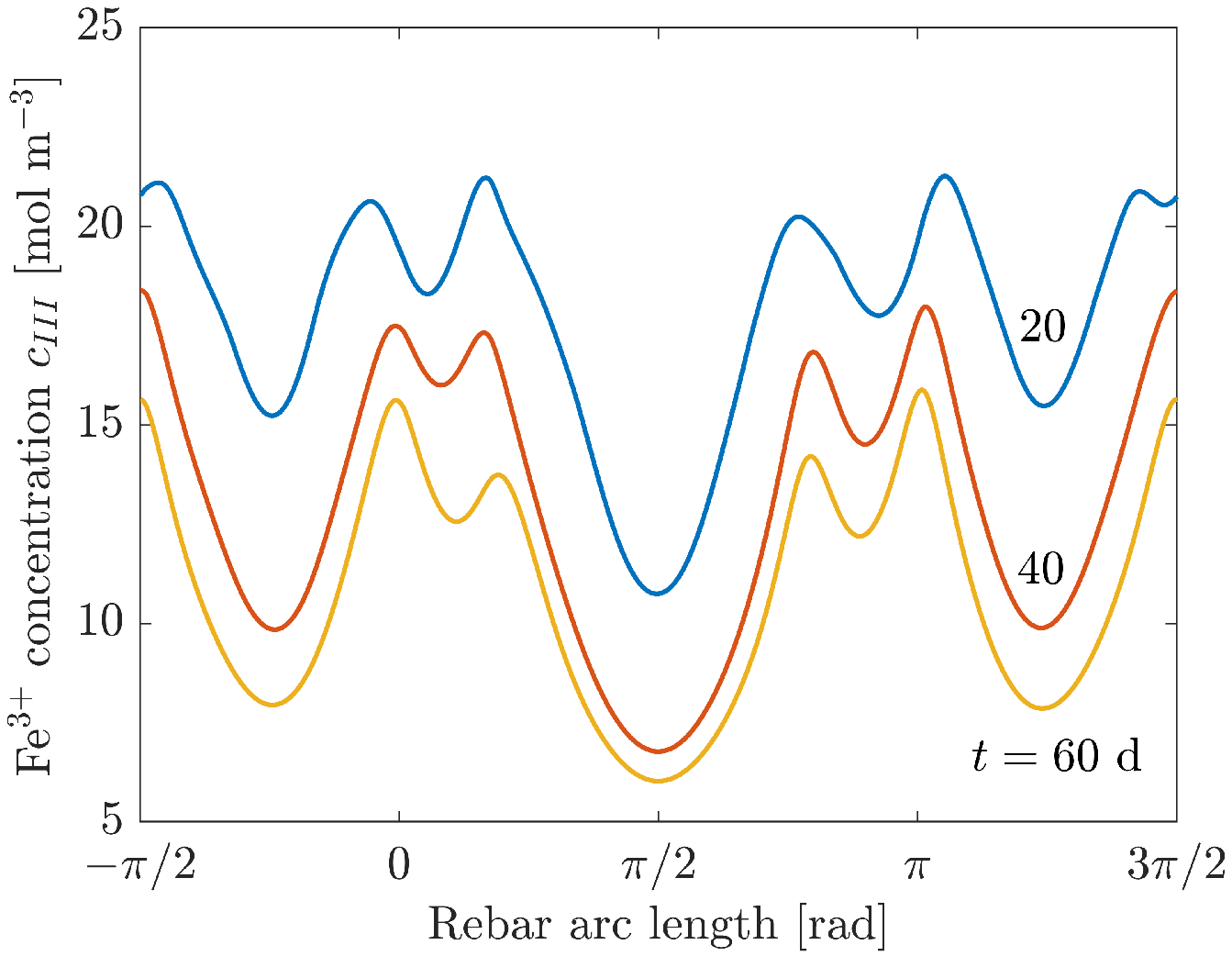}
    \caption{}
    \label{FigAlongRebarFe3}  
    \end{subfigure} 
    \caption{$\mathrm{Fe}^{3+}$ concentration $c_{III}$ for test 2 ($i_{a} = 10$ \unit{\micro\ampere\per\centi\metre^2}, $c = 20$ mm) -- (a) contours of $\mathrm{Fe}^{3+}$ in the vicinity of steel rebar at 60 days, phase-field variable $\phi$ in the shades of grey (0 -- white, 1 -- black), and (b) evolution of $c_{III}$ around the circumference of steel rebar.}
    \label{fig:AppCIII}
\end{figure}

The distribution of $\mathrm{Fe}^{2+}$ and $\mathrm{Fe}^{3+}$ concentrations is respectively shown in Figs. \ref{fig:AppCII} and \ref{fig:AppCIII}, for the conditions relevant to the validation test 2 (see Section \ref{Sec:Andrade}). The results show that the concentration of $\mathrm{Fe}^{2+}$ is at least one order of magnitude smaller than that of $\mathrm{Fe}^{3+}$. This is because the concentration of $\mathrm{Fe}^{2+}$ is continuously diluted by diffusion and fast oxidation to $\mathrm{Fe}^{3+}$. Hence, the assumption of negligible precipitation of $\mathrm{Fe}^{2+}$ under low, natural-like current densities appears to be sensible. Also, the results show that both concentrations decrease in time, with a particularly noticeable minima in the region with the highest damage, as a result of the damage-enhancement of local diffusivity (the distribution of phase-field variable around the circumference of steel rebar in time is depicted in Fig. \ref{FigAlongRebarPFvsSp}).     

\FloatBarrier

\section{Additional parametric studies}
\label{Sec:C}
\setcounter{figure}{0}

\begin{figure}[htp!]
    \begin{adjustbox}{minipage=\textwidth,scale=0.97}
    \centering
    \begin{subfigure}[!htb]{0.49\textwidth}
    \centering
    \includegraphics[width=\textwidth]{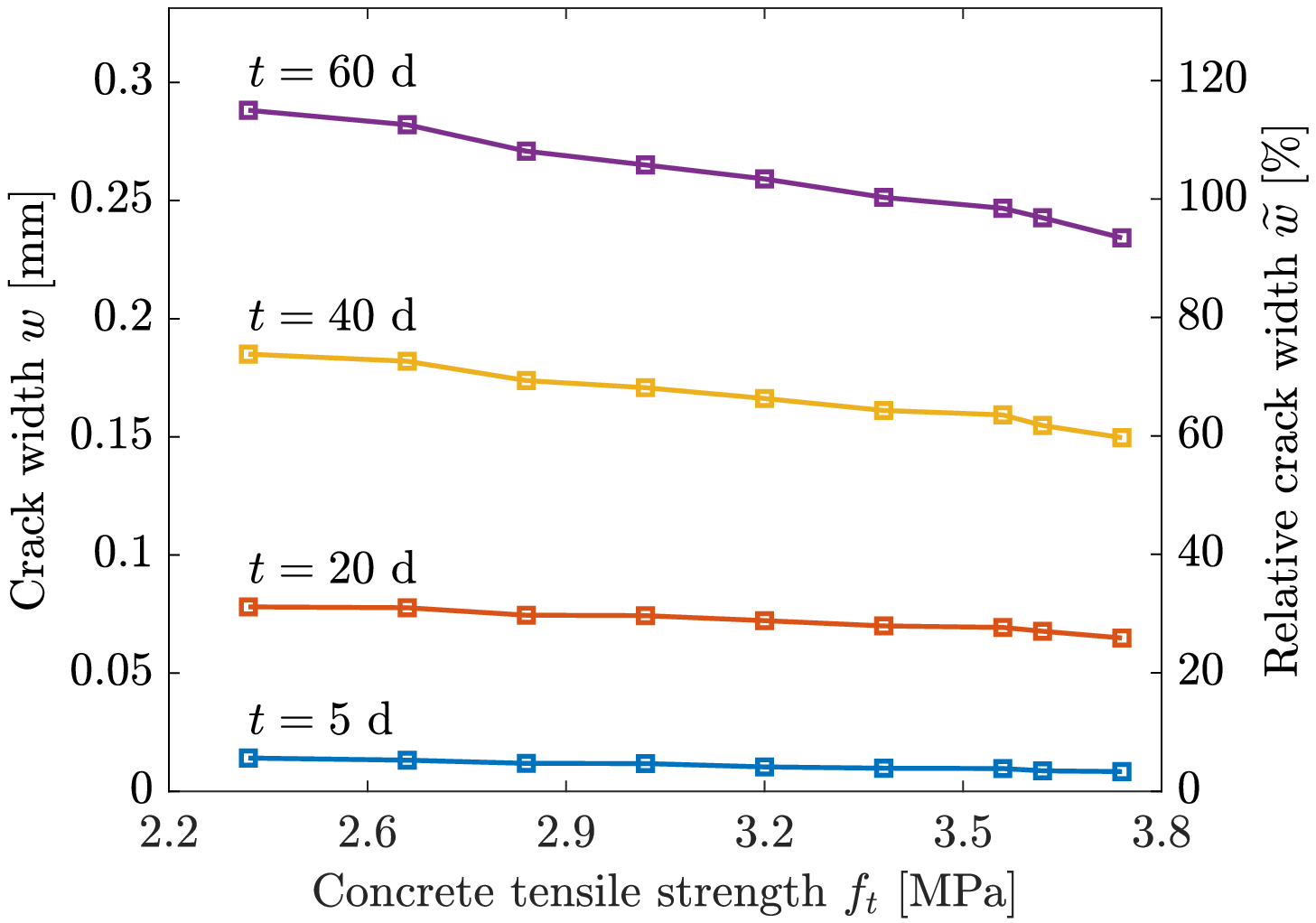}
    \caption{}
    \label{FigSweepTenStren} 
    \end{subfigure}   
    \hfill
    \begin{subfigure}[!htb]{0.49\textwidth}
    \centering
    \includegraphics[width=\textwidth]{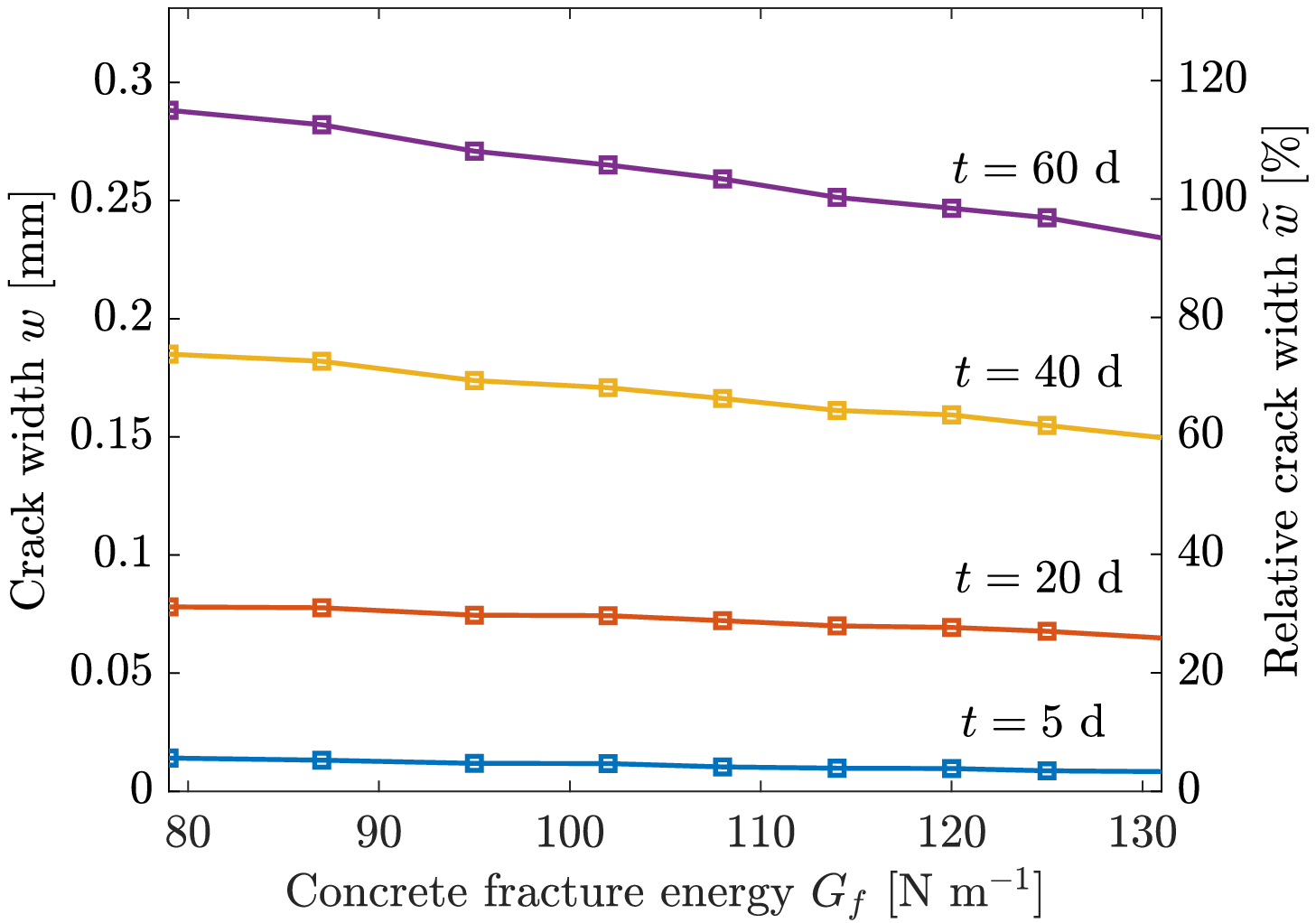}
    \caption{}
    \label{FigSweepFrEnrg}
    \end{subfigure}
    \begin{subfigure}[!htb]{0.49\textwidth}
    \centering
    \includegraphics[width=\textwidth]{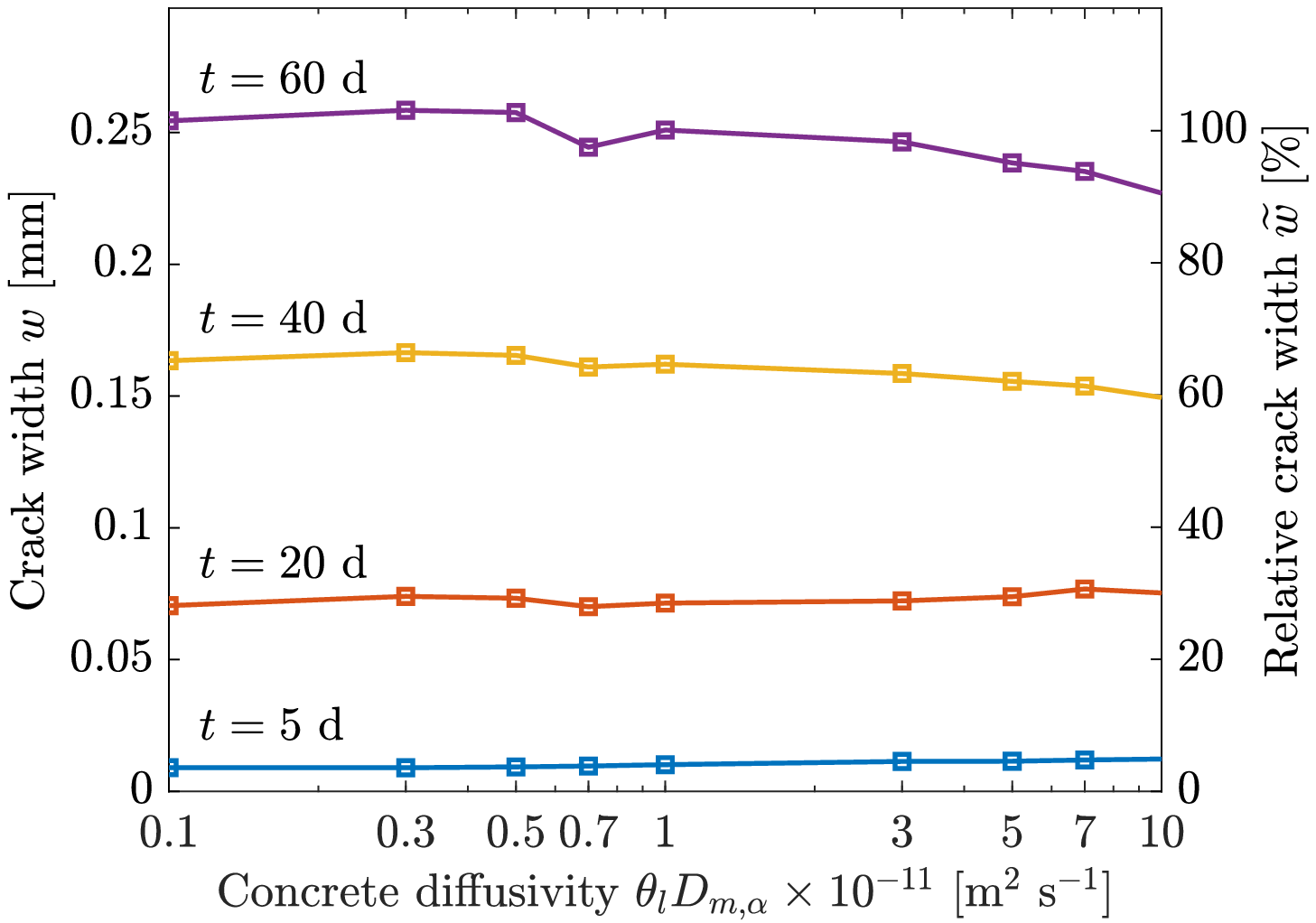}
    \caption{}
    \label{FigSweepDiff}
    \end{subfigure} 
    \begin{subfigure}[!htb]{0.49\textwidth}
    \centering
    \includegraphics[width=\textwidth]{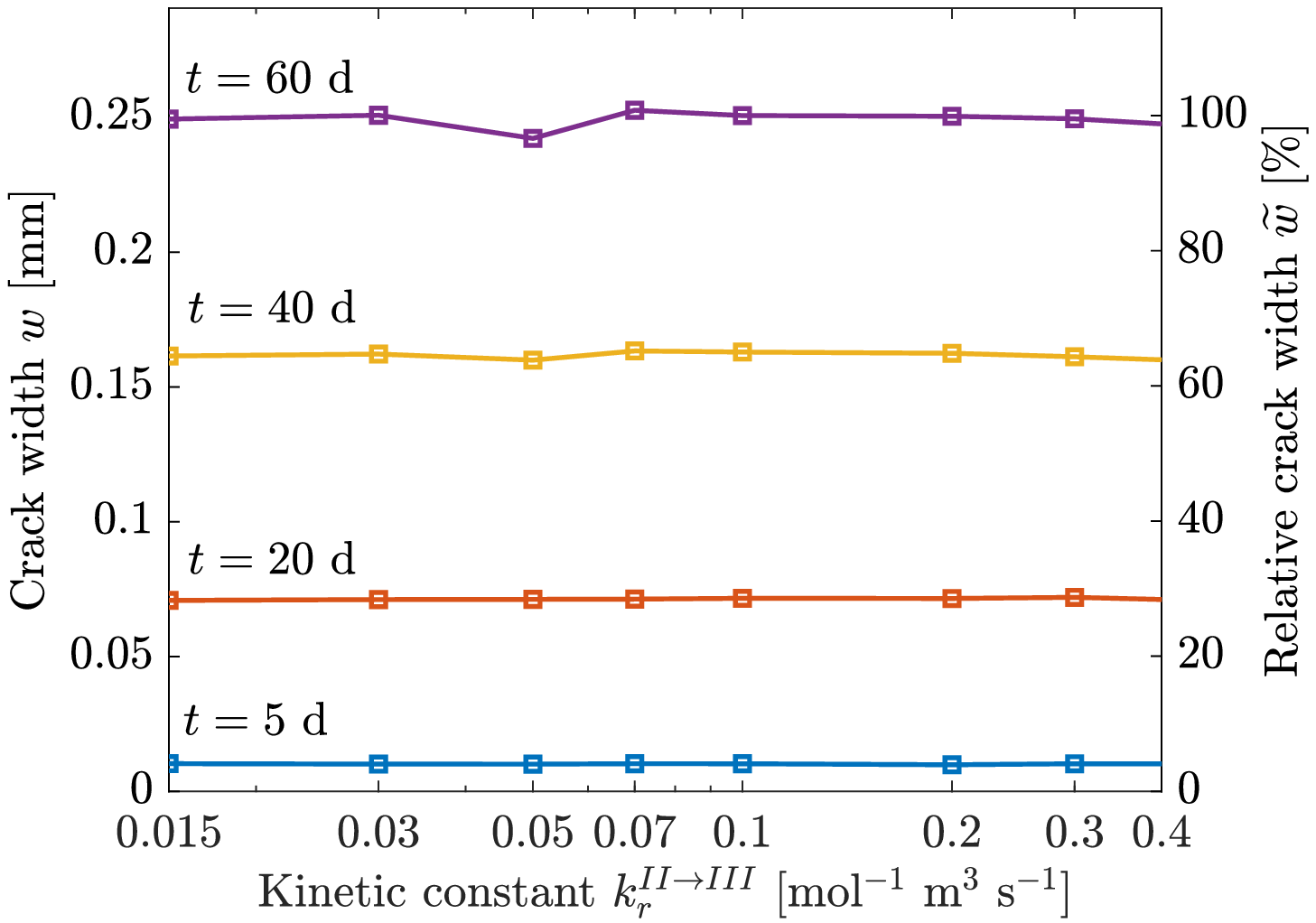}
    \caption{}
    \label{FigSweepRateConstII}  
    \end{subfigure}
    \begin{subfigure}[!htb]{0.49\textwidth}
    \centering
    \includegraphics[width=\textwidth]{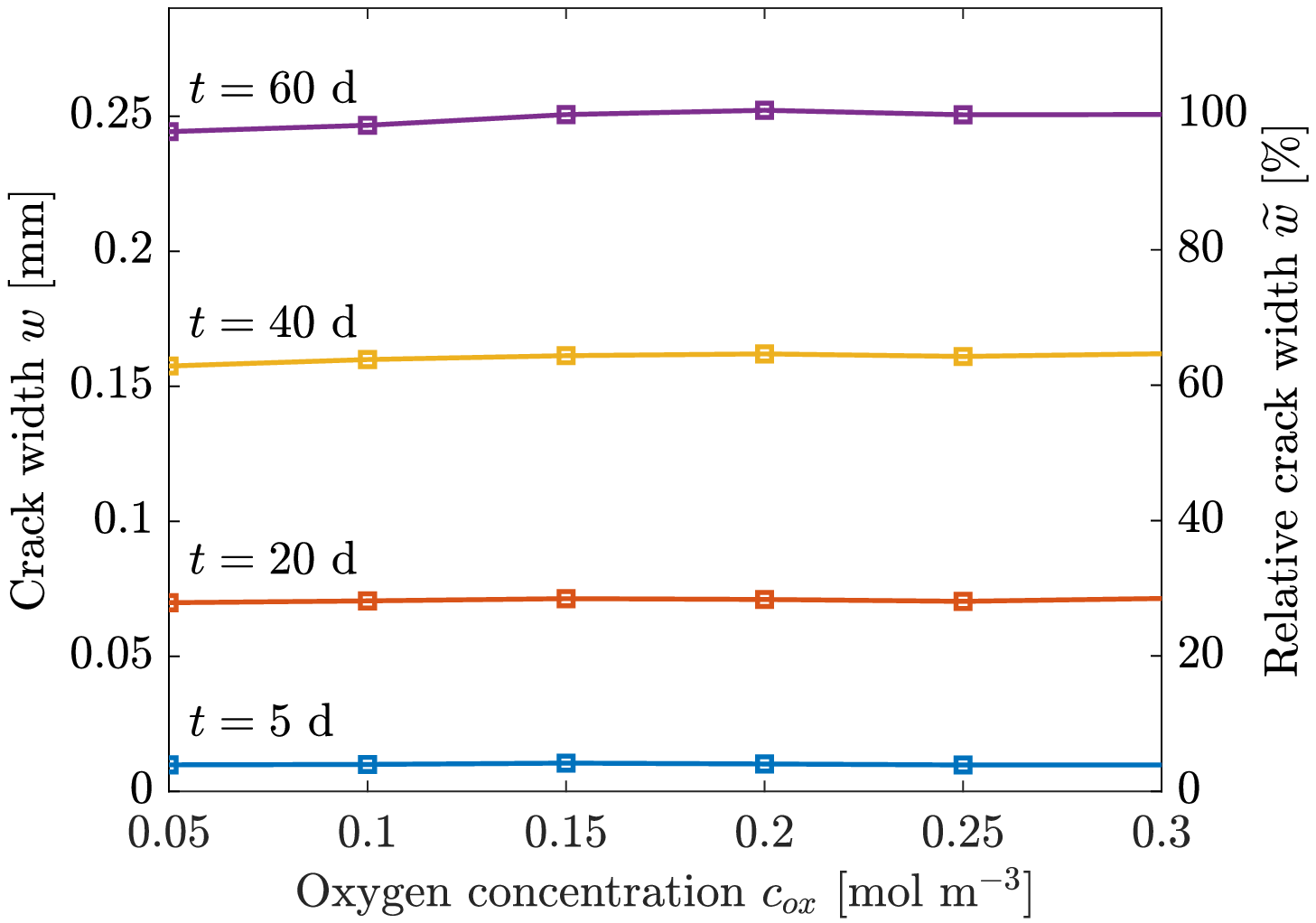}
    \caption{}
    \label{FigSweepRateOxCon}    
    \end{subfigure}
    \begin{subfigure}[!htb]{0.49\textwidth}
    \centering
    \includegraphics[width=\textwidth]{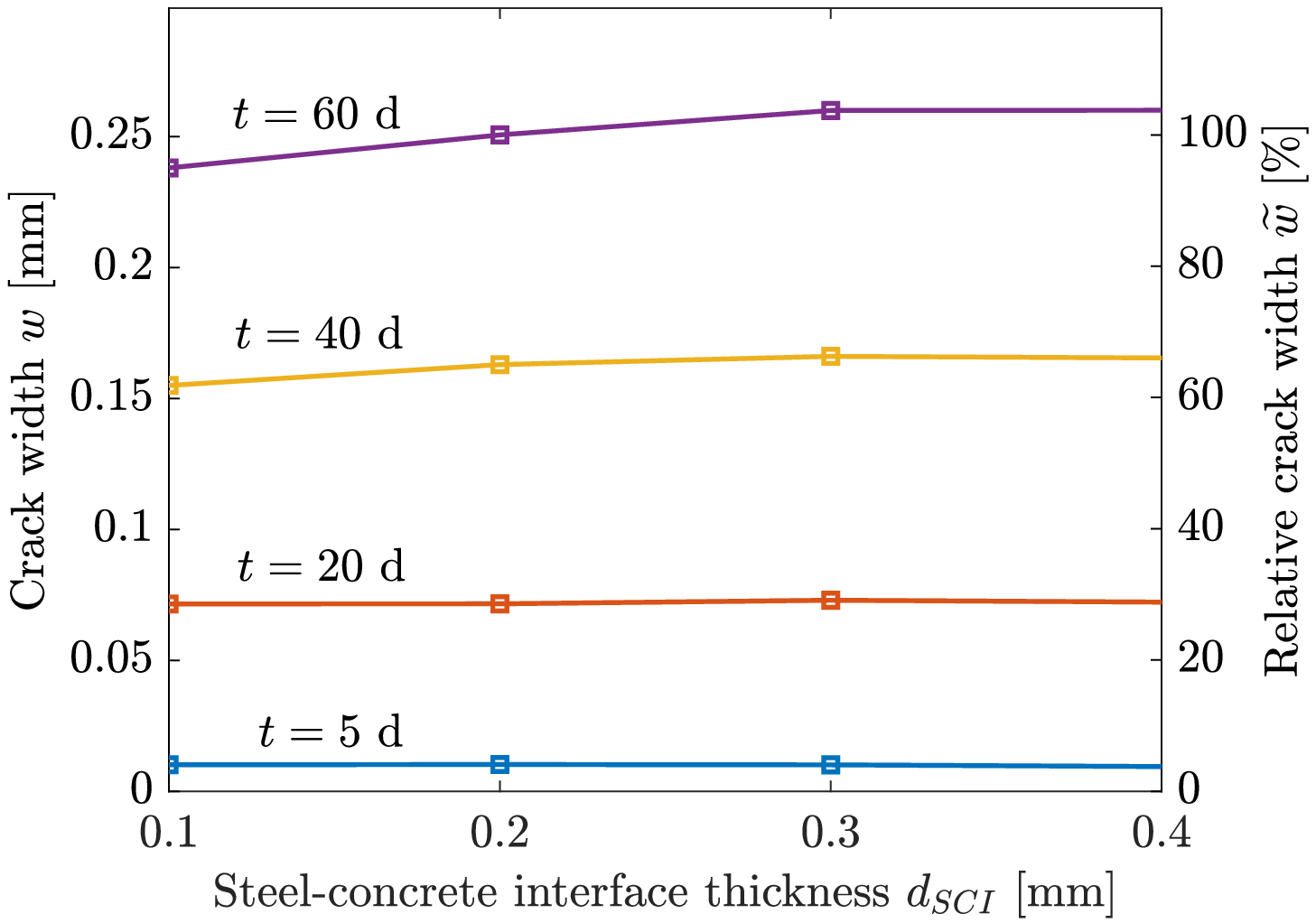}
    \caption{}
    \label{FigSweepSCIthc}    
    \end{subfigure}
    \caption{Parametric study: predicted absolute and relative crack width after 5, 20, 40 and 60 days for (a) varying tensile strength $f_{t}$, (b) fracture energy $G_{f}$, (c) initial (undamaged) concrete diffusivity $\theta_{l}D_{m,\alpha} = (1 - p_{0})D_{m,\alpha}$ (with the $x$-axis in log scale), (d) rate constant $k^{II \rightarrow III}_{r}$ (with the $x$-axis in log scale), (e) oxygen concentration $c_{ox}$, and (f) steel-concrete interface thickness $d_{SCI}$.}
    \label{Fig:ParametricStudy2}
    \end{adjustbox}
\end{figure}

\begin{figure}[H]
    \centering
    \begin{subfigure}[!htb]{0.49\textwidth}
    \centering
    \includegraphics[width=\textwidth]{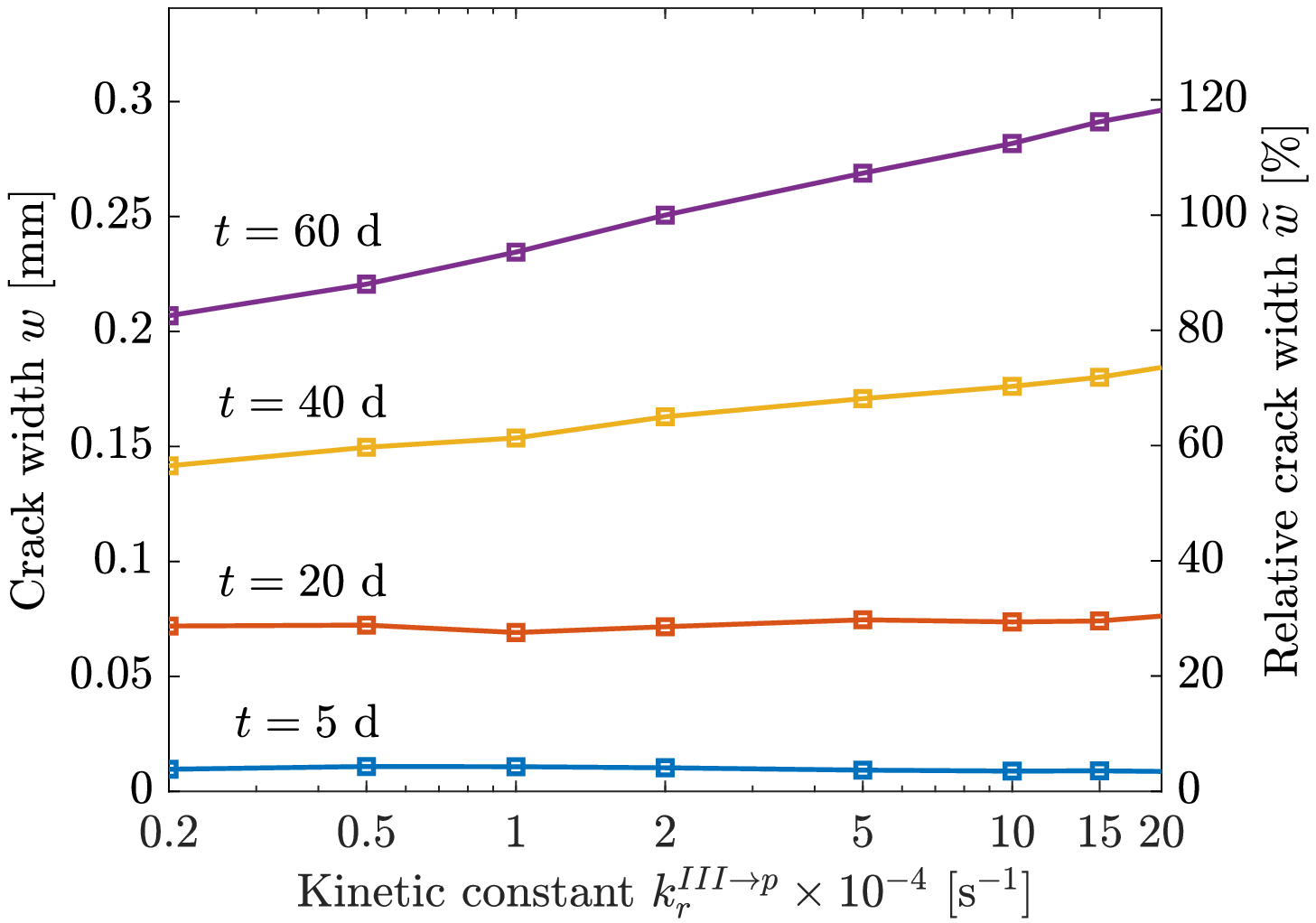}
    \caption{}
    \label{FigSweepRateConstIII}    
    \end{subfigure}
    \hfill
    \begin{subfigure}[!htb]{0.49\textwidth}
    \centering
    \includegraphics[width=\textwidth]{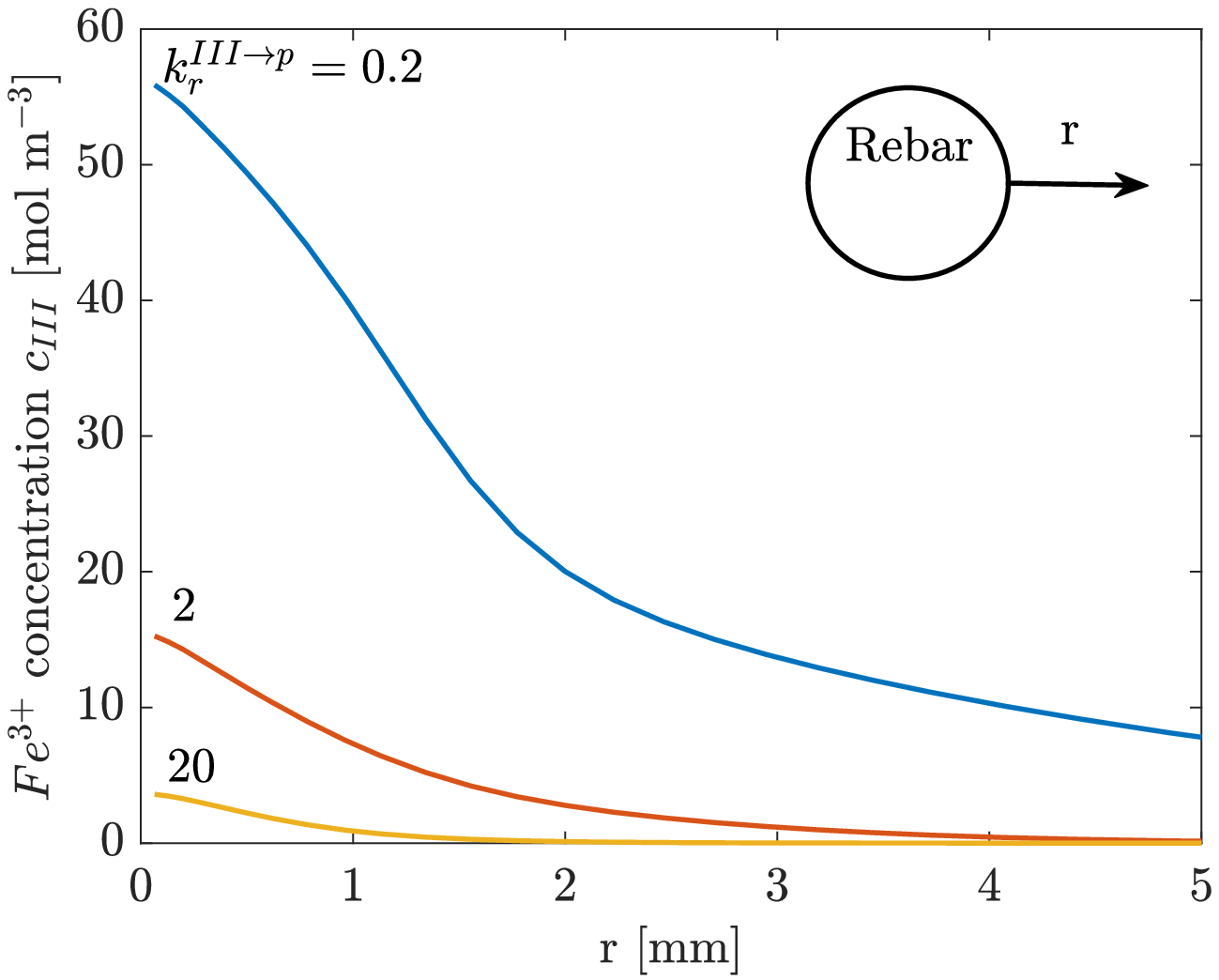}
    \caption{}
    \label{FigSweepExplkIIIFe3}    
    \end{subfigure}
    \caption{Influence of the rate constant $k^{III \rightarrow p}_{r}$: (a) absolute and relative crack width at 5, 20, 40 and 60 days for varying $k^{III \rightarrow p}_{r}$ ($k^{III \rightarrow p}_{r}$ is in log scale), and (b) $\mathrm{Fe}^{3+}$ concentration $c_{III}$ for varying $k_{r}^{III \rightarrow p}$ in the vicinity of steel rebar at 60 days.}
    \label{fig:AppCkIII}
\end{figure}

Complementing the parametric study discussed in Section \ref{Sec:ResultsParamStudy}, we proceed here to assess the sensitivity of surface crack width $w$ predictions to the remaining parameters of the model. Specifically, as shown in Figs. \ref{Fig:ParametricStudy2} and \ref{fig:AppCkIII}, results are obtained as a function of the concrete tensile strength $f_t$ (Fig. \ref{FigSweepTenStren}), the concrete fracture energy $G_f$ (Fig. \ref{FigSweepFrEnrg}), the initial (undamaged) concrete diffusivity $\theta_{l}D_{m,\alpha}$ (Fig. \ref{FigSweepDiff}), the rate constant $k_{r}^{II \rightarrow III}$ (Fig. \ref{FigSweepRateConstII}), the oxygen concentration $c_{ox}$ (Fig. \ref{FigSweepRateOxCon}), the steel-concrete interface thickness $d_{SCI}$ (Fig. \ref{FigSweepSCIthc}) and the rate constant $k_{r}^{III \rightarrow p}$ (Fig. \ref{fig:AppCkIII}). 

Consider first the role of the tensile strength $f_{t}$ (Fig. \ref{FigSweepTenStren}) and fracture energy $G_{f}$ (Fig. \ref{FigSweepFrEnrg}) of concrete. For this set of results, the chosen values of Young's modulus, fracture energy and tensile strength are correlated. The estimated values of Young's modulus have been adopted from Eurocode 2 \citep{standard2004eurocode} and the fracture energy has been estimated with the formula of \citet{Bazant2002}, assuming crushed aggregates. The results show that, in agreement with expectations, the crack width decreases with increasing $f_{t}$ and $G_{f}$. Now consider the role of the initial (undamaged) concrete diffusivity $\theta_{l}D_{m,\alpha} = (1 - p_{0})D_{m,\alpha}$, Fig. \ref{FigSweepDiff}. The crack width is seen to decrease with increasing diffusivity values, as dissolved iron species can more easily escape from the steel surface for higher diffusivities, precipitating deeper into the concrete space and reducing damage. The relatively slow decrease of the crack width with growing diffusivity could be possibly attributed to the large contribution of damaged diffusivity (see equation \ref{diffusivity_tensor_2}) in the vicinity of steel surface, diminishing the role of the original undamaged diffusivity. The influence of the rate constant $k_{r}^{II \rightarrow III}$ is given in Fig. \ref{FigSweepRateConstII}. The surface crack width appears to be largely insensitive to this parameter when varied along the range of values reported in the literature, which span two orders of magnitude \cite{Leupin2021, stefanoni_kinetic_2018}. Even though a lower $k_{r}^{II \rightarrow III}$ leads to a slower oxidation reaction and thus to a higher concentration of $\mathrm{Fe}^{2+}$, the resulting reaction rate in all tested cases is so quick that the final concentration of $\mathrm{Fe}^{3+}$ is nearly not affected. These results suggest that even though the value of $k_{r}^{II \rightarrow III}$ is relatively uncertain, the resulting error in terms of crack width is negligible. The oxygen concentration $c_{ox}$ (Fig. \ref{FigSweepRateOxCon}) also shows little influence on surface crack width predictions, unless oxygen is nearly entirely depleted and the reaction stops. The final concentration of $\mathrm{Fe}^{2+}$ is higher for smaller oxygen concentrations because the transformation to $\mathrm{Fe}^{3+}$ proceeds more slowly. However, the resulting reaction rate is still so quick that the final concentration of $\mathrm{Fe}^{3+}$ (and thus the distribution of precipitates) is nearly not affected. Next, Fig. \ref{FigSweepSCIthc} shows the sensitivity of the crack width to the steel-concrete interface thickness $d_{SCI}$, which is found to be small. The difference between crack width predicted for $d_{SCI} = 0.1 $ mm and $d_{SCI} = 0.4 $ mm is approximately 9 $\%$ in terms of relative crack width. Thus, even though there is some uncertainty associated with the choice of $d_{SCI}$, the associated error appears to be small. Similarly to the parametric study of the influence of the initial (undamaged) diffusivity, this result is possibly influenced by the large contribution of damaged part of the diffusivity tensor \ref{diffusivity_tensor_2}.

The influence of the rate constant $k_{r}^{III \rightarrow p}$ is explored separately in Fig. \ref{fig:AppCkIII}. Its value is varied over two orders of magnitude from $2\cdot10^{-5} $ to $2\cdot10^{-3} $, and this leads to a ~35$\%$ difference in relative crack width. As shown in Fig. \ref{FigSweepExplkIIIFe3}, with growing $k_{r}^{III \rightarrow p}$, $\mathrm{Fe}^{3+}$ precipitates more quickly, accumulating rust and increasing damage. Hence, our analysis suggests that this rate constant plays a non-negligible role and thus should be carefully characterised. 

\FloatBarrier
\section{Evolution of the distribution of cracks}
\label{Sec:D}
\setcounter{figure}{0}
\noindent In order to better understand the evolution of cracking during the performed numerical simulations, let us compare the distribution of phase-field variable $\phi$ at 2 days (figure \ref{FigPFsmall2days}) and 60 days (figure \ref{FigPFsmall60days}) of 50 by 50 mm concrete sample with the rebar of 16 mm diameter which centre is shifted by 5 mm horizontally towards the upper concrete surface from the centre of the square. The material parameters are chosen the same as for 28 day cured samples in table \ref{tab:tableMechRust1}. Corrosion current density is $i_{a} = 10$ \unit{\micro\ampere\per\centi\metre^2}. In figure \ref{FigPFsmall2days} we can see there is an initial damaged layer around the rebar which is slightly offsetted from the steel surface. This is related to the spatial distribution of precipitates in the pore space, as can be observed on the isolines of precipitate saturation ratio $S_{p} = 0.01$ (gray) and $S_{p} = 0.001$ (green). The first crack starts to nucleate vertically from the surface of the sample at the shortest distance between the concrete surface and the surface of the rebar. In figure \ref{FigPFsmall60days}, we can see that at 60 days, the vertical crack is fully developed and several lateral cracks start to nucleate. The more rapidly developing lower lateral cracks are slightly offset from the steel surface which is again corresponding to the spatial distribution of precipitates. The computational reason for the connection between the offset of damage from the steel surface and the distribution of precipitates is that corrosion-induced pressure of the rust accumulating under confined conditions in concrete porosity is simulated as a precipitation eigenstrain. Expanding precipitates accumulate in a thin concrete region adjacent to the steel rebar causing tensile first principle effective stress further in the concrete which eventually leads to crack nucleation. However, the thin rust-filled region itself is locked between steel and remaining concrete causing the first effective principal stress there to be initially negative which prevents the local damage onset. Thus, only further in the concrete, where the first principle effective stress is positive, cracks can initiate. More experimental data are needed to clarify whether the described damage offset effect is accurate or whether it is artificially caused by the eigenstrain-based simulation technique.  

\begin{figure}[!htb]
    \centering
    \begin{subfigure}[!htb]{0.49\textwidth}
    \centering
    \includegraphics[width=\textwidth]{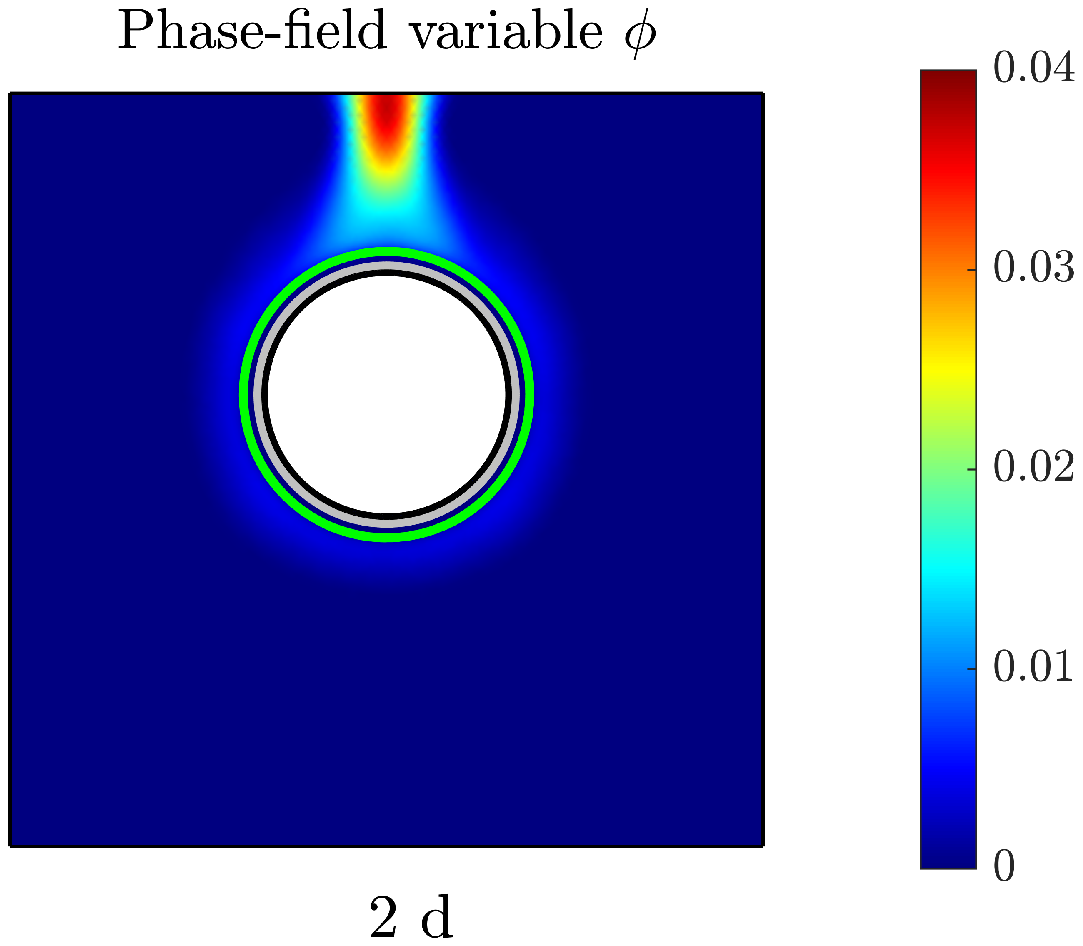}
    \caption{}
    \label{FigPFsmall2days}    
    \end{subfigure}
    \hfill
    \begin{subfigure}[!htb]{0.49\textwidth}
    \centering
    \includegraphics[width=\textwidth]{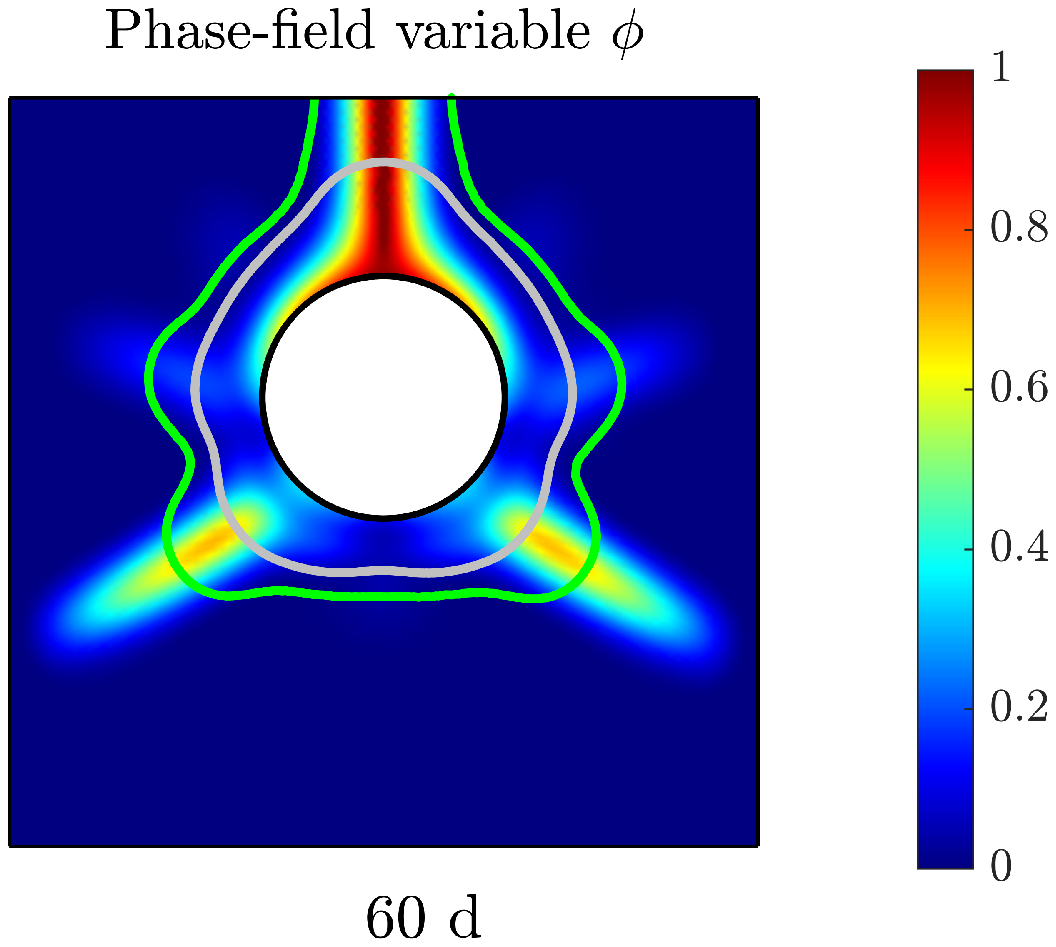}
    \caption{}
    \label{FigPFsmall60days}    
    \end{subfigure}
    \caption{Distribution of phase-field variable $\phi$ at (a) 2 days (b) 60 days. Isolines of precipitate saturation ratio in gray ($S_{p} = 0.01$) and green ($S_{p} = 0.001$), $i_{a} = 10$ \unit{\micro\ampere\per\centi\metre^2}, $c = 12$ mm}
\end{figure}


\bibliographystyle{elsarticle-num-names}
\bibliography{paper2Bib.bib}
\end{document}